\documentclass[a4paper]{amsart}
\usepackage{amsmath}
\usepackage{stmaryrd} 
\usepackage{xypic}
\usepackage{amsaddr}
\usepackage{geometry}
\allowdisplaybreaks 

\usepackage{hyperref}
\hypersetup{
  bookmarks=true,                
  unicode=false,                 
  pdftoolbar=true,               
  pdfmenubar=true,               
  pdffitwindow=false,            
  pdfstartview={FitH},           
  pdftitle={Fermionic systems for quantum information people},    
  pdfauthor={Szil{\'a}rd Szalay} {Zolt{\'a}n Zimbor{\'a}s} {Mih{\'a}ly M{\'a}t{\'e}} {Gergely Barcza} {Christian Schilling} {{\"O}rs Legeza},        
  pdfsubject={},                 
  pdfcreator={pdflatex},         
  pdfproducer={vim},             
  pdfkeywords={fermion} {fermionic second quantization} {quantum information theory} {reduced density matrix} {CAR algebra} {Jordan-Wigner representation} {fermionic correlation and entanglement} {fermionic parity superselection}
  pdfnewwindow=true,             
  colorlinks=true,               
  linktoc=page,                  
  linkcolor=blue,                
  citecolor=blue,                
  filecolor=blue,                
  urlcolor=blue                  
}
\usepackage[numbers,sort&compress]{natbib}

\usepackage{graphicx,accents}

\makeatletter
\DeclareRobustCommand{\cev}[1]{%
  \mathpalette\do@cev{#1}%
}
\newcommand{\do@cev}[2]{%
  \fix@cev{#1}{+}%
  \reflectbox{$\m@th#1\vec{\reflectbox{$\fix@cev{#1}{-}\m@th#1#2\fix@cev{#1}{+}$}}$}%
  \fix@cev{#1}{-}%
}
\newcommand{\fix@cev}[2]{%
  \ifx#1\displaystyle
    \mkern#23mu
  \else
    \ifx#1\textstyle
      \mkern#23mu
    \else
      \ifx#1\scriptstyle
        \mkern#22mu
      \else
        \mkern#22mu
      \fi
    \fi
  \fi
}

\makeatother

\makeatletter
\AtBeginDocument{%
  \check@mathfonts
  \fontdimen10\textfont3=0.2pt
  \fontdimen12\textfont3=7.2pt
  }
\makeatother

\newcommand{\equals}{{\;\;=\;\;}}
\newcommand{\equalsref}[1]{\overset{\eqref{#1}}{\equals}}
\newcommand{\equalsreff}[2]{\overset{\substack{\eqref{#1}\\\eqref{#2}}}{\equals}}

\newcommand{\arrthen}{{\quad\Longrightarrow\quad}}
\newcommand{\arrthenref}[1]{\overset{\eqref{#1}}{\arrthen}}

\newcommand{\arriff}{{\quad\Longleftrightarrow\quad}}

\newcommand{\restrict}[1]{|_{#1}}
\newcommand{\bigrestrict}[1]{\big|_{#1}}

\newcommand{\LieGrp}[1]{\mathrm{#1}} 

\newcommand{\field}[1]{\mathbb{#1}}
\newcommand{\vs}[1]{\boldsymbol{#1}}

\DeclareMathOperator{\Span}{Span}

\DeclareMathOperator{\Lin}{Lin}
\DeclareMathOperator{\Conv}{Conv}

\newcommand{\f}[1]{\widetilde{#1}}
\newcommand{\fp}[1]{\widehat{#1}}

\DeclareMathOperator{\Ad}{Ad}
\DeclareMathOperator{\Tr}{Tr}
\DeclareMathOperator{\Trf}{\f{\Tr}}
\DeclareMathOperator{\otimesf}{\f{\otimes}}
\DeclareMathOperator{\otimesfp}{\fp{\otimes}}
\DeclareMathOperator*{\bigotimesf}{\f{\bigotimes}}
\DeclareMathOperator*{\bigotimesfp}{\fp{\bigotimes}}

\newcommand{\iotaf}{\f{\iota}}
\newcommand{\iotabb}{\iota\hspace{-2.4pt}\iota}
\newcommand{\iotabbf}{\f{\iotabb}}
\newcommand{\iotabbfp}{\fp{\iotabb}}
\newcommand{\Gammaf}{\f{\Gamma}}

\newcommand{\alleven}{{\#}} 

\DeclareMathOperator*{\prodord}{\prod^\rightarrow\limits}
\DeclareMathOperator*{\prodordback}{\prod^\leftarrow\limits}

\newcommand{\ord}[1]{\vec{#1}}
\newcommand{\ordback}[1]{\cev{#1}}

\newcommand{\esum}{\sum\limits}

\newcommand{\isom}{\cong}

\newcommand{\indexdu}[2]{^{#1,#2}}
\newcommand{\indexddu}[3]{_{#1}^{#2,#3}}
\newcommand{\indexud}[2]{^{#1,#2}}
\newcommand{\indexdud}[3]{_{#1}^{#2,#3}}

\newcommand{\cket}[1]{\vert #1 \rangle}

\newcommand{\bra}[1]{\langle #1 \vert}

\newcommand{\bracket}[1]{\langle #1 \rangle}
\newcommand{\bigbracket}[1]{\bigl\langle #1 \bigr\rangle}

\newcommand{\skalp}[2]{\langle #1 \,\vert\, #2 \rangle}

\newcommand{\skalpHS}[2]{( #1 \,\vert\, #2 )}
\newcommand{\bigskalpHS}[2]{\bigl( #1 \,\big\vert\, #2 \bigr)}
\newcommand{\BigskalpHS}[2]{\Bigl( #1 \,\Big\vert\, #2 \Bigr)}

\newcommand{\daggerHS}{{\boldsymbol{\dagger}}}

\providecommand{\abs}[1]{\lvert#1\rvert}
\providecommand{\bigabs}[1]{{\bigl\lvert#1\bigr\rvert}}

\providecommand{\norm}[1]{\lVert#1\rVert}

\newcommand{\set}[1]{\{ #1 \}}
\newcommand{\bigset}[1]{\bigl\{ #1 \bigr\}}
\newcommand{\Bigset}[1]{\Bigl\{ #1 \Bigr\}}

\newcommand{\sset}[2]{\set{ #1 \;\vert\; #2 }}
\newcommand{\bigsset}[2]{\bigset{ #1 \;\big\vert\; #2 }}
\newcommand{\Bigsset}[2]{\Bigset{ #1 \;\Big\vert\; #2 }}

\newcommand{\tuple}[1]{( #1 )}

\DeclareMathOperator{\Alg}{Alg}

\newcommand{\alg}[2]{\Alg( #1, #2 )}

\newcommand{\id}{\mathrm{id}}
\newcommand{\Id}{I}
\newcommand{\IId}{\mathcal{I}}

\newcommand{\Idf}{\f{I}}
\newcommand{\IIdf}{\f{\mathcal{I}}}
\newcommand{\IIIdf}{\f{\mathbb{I}}}

\begin{document}
\newgeometry{top=3cm}
\title{Fermionic systems for quantum information people}

\author{Szil{\'a}rd Szalay$^1$,
Zolt{\'a}n Zimbor{\'a}s$^{2,3,4}$,
Mih{\'a}ly M{\'a}t{\'e}$^{1,5}$,
Gergely Barcza$^{1,6}$,
Christian Schilling$^{7,8,9}$,
{\"O}rs Legeza$^{1,10}$}
\email{\href{mailto:szalay.szilard@wigner.hu}{szalay.szilard@wigner.hu}}

\address{$^1$Strongly Correlated Systems ``Lend{\"u}let'' Research Group,\\
Institute for Solid State Physics and Optics,
Wigner Research Centre for Physics,\\
29-33, Konkoly-Thege Mikl{\'o}s str., H-1121, Budapest, Hungary}

\address{$^2$Quantum Computing and Information Research Group,\\
Institute for Particle and Nuclear Physics,
Wigner Research Centre for Physics,\\
29-33, Konkoly-Thege Mikl{\'o}s str., H-1121, Budapest, Hungary}

\address{$^3$MTA-BME Lend{\"u}let Quantum Information Theory Research Group, Budapest, Hungary}

\address{$^4$Mathematical Institute, Budapest University of Technology and Economics,\\
P.O.~Box 91, H-1111, Budapest, Hungary}

\address{$^5$Department of Physics of Complex Systems, E{\"o}tv{\"o}s Lor{\'a}nd University, P.O.~Box 32, H-1518 Budapest, Hungary }

\address{$^6$J.~Heyrovsk\'{y} Institute of Physical Chemistry, Academy of Sciences of the Czech Republic, v.v.i., Dolejškova 3, 18223 Prague 8, Czech Republic}

\address{$^7$Faculty of Physics, Arnold Sommerfeld Centre for Theoretical Physics (ASC),\\
Ludwig-Maximilians-Universit{\"a}t M{\"u}nchen, Theresienstr.~37, 80333 M{\"u}nchen, Germany}

\address{$^8$Munich Center for Quantum Science and Technology (MCQST),\\
Schellingstrasse 4, 80799 M{\"u}nchen, Germany}

\address{$^9$Wolfson College, University of Oxford, Linton Rd, Oxford OX2 6UD, United Kingdom}

\address{$^{10}$Fachbereich Physik, Philipps Universit{\"a}t Marburg, 35032 Marburg, Germany}

\date{December 10, 2021}

\begin{abstract}
The operator algebra of fermionic modes is isomorphic to that of qubits, 
the difference between them is twofold: 
the embedding of subalgebras corresponding to mode subsets and multiqubit subsystems on the one hand, 
and the parity superselection in the fermionic case on the other. 
We discuss these two fundamental differences extensively, 
and illustrate these through the Jordan--Wigner representation 
in a coherent, self-contained, pedagogical way, from the point of view of quantum information theory.
Our perspective leads us to develop useful new tools for the treatment of fermionic systems,
such as the fermionic (quasi-) tensor product, fermionic canonical embedding, fermionic partial trace,
fermionic products of maps and fermionic embeddings of maps.
We formulate these by direct, easily applicable formulas, without mode permutations, for arbitrary partitionings of the modes.
It is also shown that fermionic reduced states can be calculated by the fermionic partial trace, containing the proper phase factors.
We also consider variants of the notions of fermionic mode correlation and entanglement,
which can be endowed with the usual, local operation based motivation,
if the parity superselection rule is imposed.
We also elucidate some other fundamental points, 
related to joint map extensions,
which make the parity superselection inevitable in the description of fermionic systems.
\end{abstract}

\maketitle

\newpage
\restoregeometry
\tableofcontents

\section{Introduction}
\label{sec:Intro}

\emph{Fermionic systems in the second quantized form} 
are described by the algebra of canonical anticommutation relations,
also called \emph{CAR algebra} \cite{Jordan-1928,Bratteli-s}.
In order to be able to handle
notions and quantities important also in quantum information theory,
we have to use
the abstract terms of \emph{algebraic quantum mechanics} \cite{Bratteli-s,Alicki-2001,Beny-2015}.
\emph{The standard quantum information theory of qubits}
is described by a natural 
\emph{tensor product structure}, 
which makes possible the use of the terms of
the simpler language of \emph{Hilbert space quantum mechanics,}
used by a much wider and still rapidly growing community \cite{Nielsen-2000,Keyl-2002,Petz-2008,Wilde-2013}.
In the CAR algebra, there is no tensor product structure which fits naturally to the structure of the fermionic modes.
Imposing the \emph{(fermion number) parity superselection rule} \cite{Wick-1952,Hegerfeldt-1968},
however, makes possible to use the standard concepts of quantum information theory, 
besides issues concerning the definitions of correlation and entanglement 
\cite{Verstraete-2003,Wiseman-2003,Bartlett-2003,Schuch-2004a,Schuch-2004b,Wiseman-2004,Banuls-2007,Benatti-2014,DAriano-2014,Benatti-2020}.
Even with superselection, however, 
if we go down to the level of a concrete representation
(e.g., Jordan--Wigner representation of the CAR algebra on qubit- or spin-chains \cite{Jordan-1928})
for performing numerical \cite{Barthel-2009,Kraus-2010,Pineda-2010,Corboz-2010,Murg-2010,Legeza-2014,Szalay-2015a,Chan-2016,Szalay-2017,Orus-2019,Baiardi-2020}, 
or even symbolical calculations \cite{Banuls-2007,Friis-2013,Amosov-2016,Eisler-2015,Spee-2018,Ding-2020},
the quantities become involved. 
This is mainly because phase factors arise due to the fixed ordering of the Jordan--Wigner representation.

The main motivation of this work is
to illustrate the abstract notions 
of the representation of a \emph{finite} number of fermionic modes,
by showing the structures explicitly.
For this, we construct a toolbox for the book-keeping of fermionic phase factors,
by which the calculations become straightforward.
We present this in a coherent, self-contained, pedagogical way, from the point of view of quantum information.
After recalling the Jordan--Wigner representation (Section~\ref{sec:JW}),
 we introduce fermionic versions of 
\emph{tensor product}, \emph{partial trace} and \emph{canonical embedding} (Section~\ref{sec:Tensors}), 
by the use of which we formulate
\emph{subsets of modes}, \emph{state reduction} and \emph{state extension} in a simple and painless way (Section~\ref{sec:States}).
Then we turn to the notions of \emph{correlation} and \emph{entanglement} (Section~\ref{sec:CorrEnt}),
which turn out to be physically motivated by the notion of \emph{locality} (Section~\ref{sec:ParityPhys})
if the \emph{parity superselection rule} is imposed (Section~\ref{sec:ParityMath}),
being the main difference compared to the usual correlation and entanglement theory of qubits.

The Jordan--Wigner representation is always given with respect to a fixed linear ordering of the fermionic modes.
This is particularly important on the level of concrete matrix algebras.
Then, with our formalism presented,
it is simple to treat correlation and entanglement
with respect to any different partitionings of the modes
\emph{in a unified way, without the need for ad hoc reordering of the modes.}
Avoiding mode permutations is highly useful in multipartite correlation and entanglement theory \cite{Szalay-2015b,Szalay-2017,Szalay-2018,Szalay-2019},
serving as the other motivation of this work.
Such treatment is made possible by that
our formalism provides
reordering-free way of writing 
state extension and state reduction (calculating reduced density matrices), as well as 
map (e.g., quantum channel) extension.
The resulting formulas are easy to implement also in numerical program packages.

The third motivation of the present work is to emphasize the importance and context 
of the \emph{parity superselection rule} \cite{Wick-1952,Hegerfeldt-1968}.
Although reduced states can be defined and reduced density matrices can be calculated without the superselection,
this leads to ambiguities \cite{Friis-2013}, for example, 
unusual behavior of entropic quantities \cite{Montero-2011,Bradler-2012,Montero-2012,Bradler-2011},
such as failure of entropic inequalities \cite{Moriya-2002,Araki-2003a,Moriya-2005},
and different (non-zero part of) spectra of reduced states of pure bipartite states \cite{Moriya-2002,Amosov-2016}.
These can be resolved by the imposition of the parity superselection rule \cite{Araki-2003a,Araki-2003b,Moriya-2005,Moriya-2006,Bradler-2011,Friis-2013,Amosov-2016}.
The deeper cause of these ambiguities
is the \emph{lack of the statistical independence}
(lack of existence of joint state extension) 
of disjoint mode subsets without parity superselection \cite{Moriya-2002,Araki-2003a,Araki-2003b,Moriya-2006}.
The \emph{lack of the stronger, algebraic independence} (commutativity) leads to that the locality of maps cannot even be formulated,
on which the mere definition of separability is based from the physical point of view.
Again, these are resolved by the imposition of the parity superselection rule,
establishing commutativity.
Although only parity superselection is considered here,
since it is the one being necessary for local operations and entanglement,
we note that there are several other superselection rules, discussed in a nice, general framework \cite{Bartlett-2003,Wiseman-2004,Bartlett-2007}.
The different superselection rules
constrain the possible operations differently \cite{Bartlett-2003},
making possible different applications \cite{Verstraete-2003},
and affecting differently the notion and measures of entanglement \cite{Wiseman-2003,Schuch-2004a,Schuch-2004b,Wiseman-2004}. 

An index of the different notations is given for convenience in Table \ref{tab:notations}. 

\vfill

\begin{table}[h]
\caption{Index of the different notations}
\label{tab:notations}
\renewcommand{\arraystretch}{1.4}
\begin{tabular}{l l}
\hline 
notation & name and reference \\
\hline 
$E\indexddu{i}{\nu}{\nu'}$, $\Gamma_Y$, $\Gammaf_Y$ &
elementary basis \eqref{eq:Ei}, standard and fermionic extension \eqref{eq:Gammas}  \\
$E\indexddu{Y}{\vs{\nu}}{\vs{\nu}'}$, $\f{E}\indexddu{Y}{\vs{\nu}}{\vs{\nu}'}$ &
standard and fermionic bases \eqref{eq:bases}   \\
$\Phi_Y$, $f\indexddu{Y}{\vs{\nu}}{\vs{\nu}'}$ &
qubit-fermion mapping and phase factors \eqref{eq:PhiY} \\
$\otimesf$, $\otimesfp$, $\f{\Psi}_\xi$, $\f{\Lambda}_{\ord{\xi}}$ &
fermionic products \eqref{eq:EYfXs}, \eqref{eq:TPfp} and related maps \eqref{eq:TPf}, \eqref{eq:Lambda}  \\
$\iota_{X,Y}$, $\iotaf_{X,Y}$, $U_{X\bar{X}}$ &
std./ferm.~canonical embedding \eqref{eq:cEmbs}, unitary connecting them \eqref{eq:iotaU}  \\
$\Tr_{Y,X}$, $\Trf_{Y,X} $ &
std./ferm.~partial trace \eqref{eq:PT}, \eqref{eq:PTf} \\
$\otimesf$, $\otimesfp$ &
fermionic products for maps \eqref{eq:mTPTPs} \\
$\iotabb_{X,Y}$, $\iotabbf_{X,Y}$, $\iotabbfp_{X,Y}$ &
std./ferm.~canonical embeddings of maps \eqref{eq:mapInjs} \\
$\f{T}_Y$, $\f{P}_Y^\pm$, $\f{P}_\xi^{\vs{\epsilon}}$, $\mathcal{H}_Y^\pm$, $\mathcal{H}_\xi^{\vs{\epsilon}}$ &
parity, projections, subspaces for the Hilbert space, Sect.~\ref{sec:ParityMath.H} \\
$\f{\Theta}_Y$, $\f{\Pi}_Y^\pm$, $\f{\Pi}_\xi^{\vs{\epsilon}}$, $\f{\mathcal{A}}_Y^\pm$, $\f{\mathcal{A}}_\xi^{\vs{\epsilon}}$ &
parity, projections, subspaces for the operators, Sect.~\ref{sec:ParityMath.A} \\
$\f{\mathbb{T}}_Y$, $\f{\mathbb{P}}_Y^\pm$, $\f{\mathbb{P}}_\xi^{\vs{\epsilon}}$, $\f{\mathcal{B}}_Y^\pm$, $\f{\mathcal{B}}_\xi^{\vs{\epsilon}}$ &
parity, projections, subspaces for the maps, Sect.~\ref{sec:ParityMath.B}  \\
$\mathcal{D}_X$, $\f{\mathcal{D}}_X$ &
sets of qubit/fermionic states, inline before \eqref{eq:defreduced}, \eqref{eq:defreducedf}\\
$\f{\mathcal{D}}_Y^+$, $\f{\mathcal{D}}_\xi^\alleven$ &
sets of physical/locally physical ferm.~states, beginning of Sect.~\ref{sec:ParityPhys.Corrent}\\
$\mathcal{D}_{\text{$\xi$-unc}}$, $\mathcal{D}_{\text{$\xi$-sep}}$ &
sets of uncorrelated/separable qubit states \eqref{eq:Dunc}, \eqref{eq:Dsep} \\
$\f{\mathcal{D}}_{\text{$\xi$-unc}}$, $\f{\mathcal{D}}_{\text{$\xi$-sep}}$ &
sets of uncorrelated/separable ferm.~states, without SSR \eqref{eq:DfpupsuncwoSSR}, \eqref{eq:DfpupssepwoSSR} \\
$\f{\mathcal{D}}_{\text{$\xi$-unc}}^+$ &
sets of uncorrelated physical ferm.~states \eqref{eq:Dfpupsunc} \\
$\f{\mathcal{D}}_{\text{$\xi$-unc}}^\alleven $ &
sets of product (strongly uncorrelated) physical ferm.~states \eqref{eq:Dupsprodfp} \\
$\f{\mathcal{D}}_{\text{$\xi$-sep}}^+ $ &
sets of weakly separable physical ferm.~states \eqref{eq:Dfpupsconv} \\
$\f{\mathcal{D}}_{\text{$\xi$-sep}}^\alleven$ &
sets of separable physical ferm.~states \eqref{eq:Dfpupssep}\\
\hline 
\end{tabular}
\end{table}

\section{Jordan--Wigner representation}
\label{sec:JW}

In this section,
we introduce the tools for the description of fermionic occupations
by qubits through the Jordan--Wigner representation of the CAR algebra \cite{Jordan-1928}.

Since we consider a \emph{finite} number of fermionic modes only,
we have finite dimensional Hilbert spaces $\mathcal{H}$,
and subalgebras of the finite dimensional algebras $\Lin\mathcal{H}$.
Then, on $\Lin\mathcal{H}\isom\mathcal{H}\otimes\mathcal{H}^*$, we have 
the \emph{trace} map,
given by the invariant definition 
as $\Tr\bigl(\cket{\psi}\bra{\phi}\bigr):=\skalp{\phi}{\psi}$,
and the \emph{Hilbert--Schmidt inner product} \cite{Petz-2008,Wilde-2013,Beny-2015},
given as
\begin{equation}
\label{eq:HS}
A,B\in\Lin\mathcal{H},\qquad \skalpHS{A}{B}:=\Tr(A^\dagger B),
\end{equation}
which are important tools in the sequel.
Recall also that
the adjoint $A^\dagger$ of operators $A\in\Lin\mathcal{H}$ is given 
with respect to the inner product of the Hilbert space as
$\skalp{A^\dagger\phi}{\psi}=\skalp{\phi}{A\psi}$,
for all $\cket{\psi},\cket{\phi}\in\mathcal{H}$; 
as well as
the adjoint $\Omega^\daggerHS$ of maps $\Omega\in\Lin\Lin\mathcal{H}$ is given 
with respect to the Hilbert--Schmidt inner product as
$\bigskalpHS{\Omega^\daggerHS(A)}{B}=\bigskalpHS{A}{\Omega(B)}$ 
for all $A,B\in\Lin\mathcal{H}$ \cite{Petz-2008,Wilde-2013,Beny-2015}.

\subsection{Qubits}
\label{sec:JW.qubits}

Let us consider an $L$-qubit system for $L\geq1$,
labeled by the elements of the index set $M:=\{1,2,\dots,L\}$. 
For each qubit, $i\in M$,
we have a two-dimensional Hilbert space $\mathcal{H}_i:=\Span_\field{C}\bigsset{\cket{\phi_i^\nu}}{\nu\in\set{0,1}}$ associated to it,
where $\cket{\phi_i^0}$ and $\cket{\phi_i^1}$ are orthonormal basis vectors, 
$\skalp{\phi_i^\mu}{\phi_i^\nu}=\delta\indexud{\mu}{\nu}$.
If we consider a given qubit $i\in M$ \emph{by itself,}
then the projectors $\cket{\phi_i^0}\bra{\phi_i^0}$ and $\cket{\phi_i^1}\bra{\phi_i^1}$,
or the subspaces they project onto,
represent two mutually exclusive elementary ``quantum events''.

We also have the algebra $\mathcal{A}_i:=\Lin\mathcal{H}_i$ of linear operators acting on $\mathcal{H}_i$,
with the standard basis
\begin{equation}
\label{eq:Ei}
\bigsset{E\indexddu{i}{\nu}{\nu'} := \cket{\phi_i^\nu}\bra{\phi_i^{\nu'}}\in\mathcal{A}_i}{\nu,\nu'\in\set{0,1}},
\end{equation}
which is orthonormal, 
$\bigskalpHS{E\indexddu{i}{\mu}{\mu'}}{E\indexddu{i}{\nu}{\nu'}}=\delta\indexud{\mu}{\nu}\delta\indexud{\mu'}{\nu'}$.
The orthonormality of the standard basis in $\Lin\mathcal{H}_i$ arises from the
orthonormality of the basis in $\mathcal{H}_i$.

\subsection{Jordan--Wigner representation of fermionic occupations}
\label{sec:JW.fermions}

In the Jordan--Wigner representation, 
the qubits are used to store the quantum information
about the occupation of the fermionic modes.
Let us consider a system of $L$ fermionic modes, 
labeled by the index set $M=\{1,2,\dots,L\}$.
If we consider a given mode $i\in M$ \emph{by itself,}
then let the projectors $\cket{\phi_i^0}\bra{\phi_i^0}$ and $\cket{\phi_i^1}\bra{\phi_i^1}$ 
represent the elementary ``quantum events'' of unoccupied and occupied modes, respectively.
We also have some standard linear operators in hand, as
the \emph{creation operator} $a_i^\dagger := \cket{\phi_i^1}\bra{\phi_i^0}$,
the \emph{annihilation operator} $a_i := \cket{\phi_i^0}\bra{\phi_i^1}$,
the \emph{particle number operator} $n_i := a_i^\dagger a_i = \cket{\phi_i^1}\bra{\phi_i^1}$,
the \emph{``no particle'' operator} $\Id_i-n_i := a_i a_i^\dagger = \cket{\phi_i^0}\bra{\phi_i^0}$
(these four are also the elements of the standard basis \eqref{eq:Ei}),
the \emph{identity operator} $\Id_i=\cket{\phi_i^0}\bra{\phi_i^0} + \cket{\phi_i^1}\bra{\phi_i^1}$ and
the \emph{phase operator} $p_i := \cket{\phi_i^0}\bra{\phi_i^0} - \cket{\phi_i^1}\bra{\phi_i^1}$.
We emphasize that these operators act on different Hilbert spaces for different values of the index $i$.

Now, let us ``join'' some fermionic modes (as well as the corresponding qubits), 
and consider them as a \emph{(sub)set of modes} (as well as \emph{composite (sub)system} of qubits).
That is, for each \emph{subsystem} (or \emph{mode subset}) $Y\subseteq M$, 
we have the Hilbert space $\mathcal{H}_Y := \bigotimes_{i\in Y}\mathcal{H}_i$
(Fock space of mode subset $Y$),
together with the standard tensor product basis
\begin{equation}
\label{eq:phiY}
\Bigsset{\cket{\phi_Y^{\vs{\nu}}} := \bigotimes_{i\in Y}\cket{\phi_i^{\nu_i}}}{\forall\vs{\nu}: Y\to\{0,1\}},
\end{equation}
with the \emph{multi-index} $\vs{\nu}: Y\to\{0,1\}$, $i\mapsto\nu_i$.
(For $Y=\emptyset$, we define the one-dimensional Hilbert space $\mathcal{H}_\emptyset\isom\field{C}$
with the empty tensor product.)
This basis is also orthonormal, $\skalp{\phi_Y^{\vs{\mu}}}{\phi_Y^{\vs{\nu}}}=\delta\indexud{\vs{\mu}}{\vs{\nu}}=\prod_{i\in Y}\delta\indexud{\mu_i}{\nu_i}$.

Now we extend the operators of each mode $i\in M$ linearly
to $\Lin\mathcal{H}_Y\isom\bigotimes_{i\in Y}\mathcal{A}_i$
in two different ways, called ``standard'' and ``fermionic'', as
\begin{subequations}
\label{eq:Gammas}
\begin{align}
\label{eq:Gamma}
\begin{split}
\Gamma_Y: \quad
\mathcal{A}_i \quad&\longrightarrow\quad
\bigotimes_{j\in Y}\mathcal{A}_j,\\
E\indexddu{i}{\nu}{\nu'} \quad&\longmapsto\quad
 \Bigl(\bigotimes_{k\in Y, k<i}\Id_k\Bigr)
 \otimes E\indexddu{i}{\nu}{\nu'} 
 \otimes \Bigl(\bigotimes_{k\in Y, i<k}\Id_k\Bigr),
\end{split}\\
\label{eq:Gammaf}
\begin{split}
\Gammaf_Y: \quad
\mathcal{A}_i \quad&\longrightarrow\quad
\bigotimes_{j\in Y}\mathcal{A}_j,\\
E\indexddu{i}{\nu}{\nu'} \quad&\longmapsto\quad
 \Bigl(\bigotimes_{k\in Y, k<i} p_k^{\nu+\nu'}\Bigr)
 \otimes E\indexddu{i}{\nu}{\nu'}
 \otimes \Bigl(\bigotimes_{k\in Y, i<k}\Id_k\Bigr).
\end{split}
\end{align}
\end{subequations}
(For $Y=\emptyset$, we define the one-dimensional algebra
$\bigotimes_{j\in \emptyset}\mathcal{A}_j\isom\Lin\mathcal{H}_\emptyset\isom\field{C}$ 
with the empty tensor product.)
With a slight abuse of notation, we use the same letter for all modes,
i.e., we do not denote the mode index $i$ on $\Gamma_Y$ and $\Gammaf_Y$,
since this will not cause confusion.
(Note that the ordering in the writing of the tensor product is illustrative only:
the product of two elementary tensors is 
the elementary tensor formed by the products of operators \emph{of the same modes}.)
These maps are $*$-homomorphisms, that is,
\begin{subequations}
\begin{align}
\label{eq:GammaYHom}
\Gamma_Y(A_i) \Gamma_Y(B_i) &=\Gamma_Y(A_iB_i),& \qquad \Gamma_Y(A_i^\dagger) &=\Gamma_Y(A_i)^\dagger,\\ 
\label{eq:GammafYHom}
\Gammaf_Y(A_i)\Gammaf_Y(B_i)&=\Gammaf_Y(A_iB_i),&\qquad \Gammaf_Y(A_i^\dagger)&=\Gammaf_Y(A_i)^\dagger.
\end{align}
\end{subequations}
The \emph{fermionic extensions of the $a_i^\dagger$ and $a_i$ creation and annihilation operators,} 
$\tilde{a}_{i,Y}^\dagger := \Gammaf_Y(a_i^\dagger)$ and
$\tilde{a}_{i,Y} := \Gammaf_Y(a_i)$,
are then the \emph{Jordan--Wigner representations} of the \emph{fermionic} creation and annihilation operators \cite{Jordan-1928},
which are the generators of the CAR algebra, obeying the \emph{canonical anticommutation relations}
\begin{equation}
\label{eq:modeacommf}
\{\tilde{a}_{i,Y},\tilde{a}_{j,Y}\}=\{\tilde{a}_{i,Y}^\dagger,\tilde{a}_{j,Y}^\dagger\}=0,\qquad
\{\tilde{a}_{i,Y},\tilde{a}_{j,Y}^\dagger\}=\delta_{i,j} \tilde{\Id}_Y,
\end{equation}
for all $i,j\in Y$,
and $\Idf_Y=\prodord_{i\in Y}\Gammaf_Y(\Id_i)=\prod_{i\in Y}\Gamma_Y(\Id_i)$
is the identity operator.
(Here $\prodord$ denotes
that the factors in the product are ordered for increasing values of the mode indices.
Although it is not necessary for the identity operators, we use it to emphasize
that the fermionic extensions of operators of different modes do not commute.)
Note the elegant way 
how $p_k^{\nu+\nu'}$ inserts the phase operators in \eqref{eq:Gammaf} exactly for the cases when $\nu\neq\nu'$,
which coincides with the case of products of odd number of creation and annihilation operators, see \eqref{eq:GammafYHom}.
Also, the \emph{standard extensions of the $a_i^\dagger$ and $a_i$ creation and annihilation operators},
$a_{i,Y}^\dagger := \Gamma_Y(a_i^\dagger)$ and 
$a_{i,Y} := \Gamma_Y(a_i)$,
are the usual representations of the \emph{hardcore boson} creation and annihilation operators \cite{Matsubara-1956,Matsubara-1957,Tennie-2017},
which we do not consider here.

Using the two extensions above, we define two bases in $\bigotimes_{i\in Y}\mathcal{A}_i$,
called ``standard'' and ``fermionic'',
\begin{subequations}
\label{eq:bases}
\begin{align}
\label{eq:EY}
\Bigsset{E\indexddu{Y}{\vs{\nu}}{\vs{\nu}'}
 &:= \prod_{i\in Y}\Gamma_Y(E\indexddu{i}{\nu_i}{\nu_i'}) \in \bigotimes_{i\in Y}\mathcal{A}_i}{\forall\vs{\nu},\vs{\nu}':Y\to\{0,1\}},\\
\label{eq:EYf}
\Bigsset{\f{E}\indexddu{Y}{\vs{\nu}}{\vs{\nu}'}
 &:= \prodord_{i\in Y} \Gammaf_Y(E\indexddu{i}{\nu_i}{\nu_i'}) \in \bigotimes_{i\in Y}\mathcal{A}_i}{\forall\vs{\nu},\vs{\nu}':Y\to\{0,1\}}.
\end{align}
\end{subequations}
Both of these are orthonormal with respect to the Hilbert--Schmidt inner product \eqref{eq:HS}, 
that is,
$\bigskalpHS{E\indexddu{Y}{\vs{\mu}}{\vs{\mu'}}}{E\indexddu{Y}{\vs{\nu}}{\vs{\nu}'}}
=\bigskalpHS{\f{E}\indexddu{Y}{\vs{\mu}}{\vs{\mu'}}}{\f{E}\indexddu{Y}{\vs{\nu}}{\vs{\nu}'}}
=\delta\indexud{\vs{\mu}}{\vs{\nu}}\delta\indexud{\vs{\mu}'}{\vs{\nu}'}$,
see later in \eqref{eq:PhiU}.
Note that the standard basis is consistent 
with the tensor product in $\bigotimes_{i\in Y}\mathcal{A}_i$,
since
\begin{subequations}
\label{eq:basesexpl}
\begin{equation}
\label{eq:EYexpl}
E\indexddu{Y}{\vs{\nu}}{\vs{\nu}'}
=\prod_{i\in Y}\Gamma_Y(E\indexddu{i}{\nu_i}{\nu_i'})
=\bigotimes_{i\in Y} E\indexddu{i}{\nu_i}{\nu_i'},
\end{equation}
while the fermionic basis is consistent with the fermionic operators,
since
\begin{equation}
\label{eq:EYfexpl}
\f{E}\indexddu{Y}{\vs{\nu}}{\vs{\nu}'}
= \prodord_{i\in Y} \Gammaf_Y(E\indexddu{i}{\nu_i}{\nu_i'})
= \prodord_{i\in Y} \left\{\begin{aligned}
\tilde{a}_{i,Y}\tilde{a}_{i,Y}^\dagger \; &\text{if $\nu_i=0$ and $\nu_i'=0$}\\
\tilde{a}_{i,Y}                        \; &\text{if $\nu_i=0$ and $\nu_i'=1$}\\
\tilde{a}_{i,Y}^\dagger                \; &\text{if $\nu_i=1$ and $\nu_i'=0$}\\
\tilde{a}_{i,Y}^\dagger\tilde{a}_{i,Y} \; &\text{if $\nu_i=1$ and $\nu_i'=1$}
\end{aligned}\right\},
\end{equation}
\end{subequations}
the basis elements are ordered products of fermionic creation and annihilation operators.
(For explicit calculations, see Appendices~\ref{appsec:JW.identities} and \ref{appsec:JW.prod}.)

Let $\vs{0}:i\mapsto0$ be the zero multi-index, labeling the \emph{vacuum} state,
$\cket{\text{Vac}_Y}:=\cket{\phi_Y^{\vs{0}}}=\bigotimes_{i\in Y}\cket{\phi_i^0}$, see \eqref{eq:phiY}.
Then, on the one hand,
 $\cket{\phi_Y^{\vs{\nu}}}\bra{\phi_Y^{\vs{0}}} = E_Y^{\vs{\nu},\vs{0}}$ creates
the state vector of occupation $\vs{\nu}$ from the vacuum,
\begin{subequations}
\begin{equation}
E_Y^{\vs{\nu},\vs{0}}\cket{\phi_Y^{\vs{0}}}
= \Bigl(\bigotimes_{i\in Y} \cket{\phi_i^{\nu_i}}\bra{\phi_i^0}\Bigr) \Bigl(\bigotimes_{i\in Y} \cket{\phi_i^0}\Bigr)
= \cket{\phi_Y^{\vs{\nu}}}.
\end{equation}
On the other hand, $\f{E}_Y^{\vs{\nu},\vs{0}}$ does the same,
\begin{equation}
\f{E}_Y^{\vs{\nu},\vs{0}}\cket{\text{Vac}_Y}
= \prodord_{i\in Y} \left\{\begin{aligned}
\tilde{a}_{i,Y}\tilde{a}_{i,Y}^\dagger \; &\text{if $\nu_i=0$}\\
\tilde{a}_{i,Y}^\dagger                \; &\text{if $\nu_i=1$}
\end{aligned}\right\}\cket{\text{Vac}_Y}
= \prodord_{\substack{i\in Y:\\\nu_i=1}} \tilde{a}_{i,Y}^\dagger\cket{\text{Vac}_Y}
= \cket{\phi_Y^{\vs{\nu}}},
\end{equation}
\end{subequations}
where the last equality follows from 
that, because of the ordering of the operators,
the phase operators coming from \eqref{eq:Gammaf} act always on empty modes, $p_i\cket{\phi_i^0}=\cket{\phi_i^0}$.
Therefore \eqref{eq:phiY} is indeed the Fock basis (in concrete representation),
which is usually defined in fermionic second quantization by the usual ordering of the creation operators.

\subsection{Mapping between qubits and fermionic modes}
\label{sec:JW.Phi}

Using \eqref{eq:EY}-\eqref{eq:EYf}, let us define the 
two algebras endowed with canonical bases, 
$\mathcal{A}_Y$ and $\f{\mathcal{A}}_Y$, being two copies of $\bigotimes_{i\in Y}\mathcal{A}_i$ for a $Y\subseteq M$,
to which $\set{E_Y^{\vs{\nu},\vs{\nu}'}}$ and $\set{\f{E}_Y^{\vs{\nu},\vs{\nu}'}}$ are associated, respectively.
From now on, we denote the $Y\subseteq M$ subsystem/mode-subset index on the objects,
and, for example, writing $A_Y$ or $\f{A}_Y$, 
the statements are understood for all $A_Y\in\mathcal{A}_Y$ or $\f{A}_Y\in\f{\mathcal{A}}_Y$ automatically,
if not stated differently.
Let the identity operators be
$\prod_{i\in Y}\Gamma_Y(\Id_i)=\bigotimes_{i\in Y}\Id_i=\Id_Y\in\mathcal{A}_Y$ in the standard, and
$\prodord_{i\in Y}\Gammaf_Y(\Id_i)=\Idf_Y\in\f{\mathcal{A}}_Y$ in the fermionic case.
Now let us consider the linear map between the two algebras, given as
\begin{subequations}
\label{eq:PhiY}
\begin{equation}
\label{eq:EYEYftraf}
\begin{split}
\Phi_Y:\quad 
\mathcal{A}_Y \quad&\longrightarrow\quad
\f{\mathcal{A}}_Y,\\
E\indexddu{Y}{\vs{\nu}}{\vs{\nu}'} \quad&\longmapsto\quad 
\f{E}\indexddu{Y}{\vs{\nu}}{\vs{\nu}'}.
\end{split}
\end{equation}
Taking into account \eqref{eq:Gamma} and \eqref{eq:Gammaf},
we can write out the effect of this transformation
\begin{equation}
\label{eq:PhiYEY}
\Phi_Y(E\indexddu{Y}{\vs{\nu}}{\vs{\nu}'}) 
= \f{E}\indexddu{Y}{\vs{\nu}}{\vs{\nu}'} 
= f\indexddu{Y}{\vs{\nu}}{\vs{\nu}'} E\indexddu{Y}{\vs{\nu}}{\vs{\nu}'},
\end{equation}
which is the elementwise product with the $\pm1$ phase factors
\begin{equation}
\label{eq:phasefY}
f\indexddu{Y}{\vs{\nu}}{\vs{\nu}'} = (-1)^{\esum_{i\in Y}\nu_i'\esum_{k\in Y, i<k}(\nu_k+\nu_k')}.
\end{equation}
\end{subequations}
(For the proof, see Appendix~\ref{appsec:JW.Phi}.
For the explicit form for few modes, see Appendix~\ref{appsec:JW.Explf}.)
Since this map just multiplies the basis elements \eqref{eq:EY} with phase factors,
it is clearly \emph{unitary} with respect to the Hilbert--Schmidt inner product \eqref{eq:HS},
\begin{equation}
\label{eq:PhiU}
\bigskalpHS{\Phi_Y(A_Y)}{\Phi_Y(B_Y)}=\bigskalpHS{A_Y}{B_Y},
\end{equation}
or, equivalently, 
$\bigskalpHS{\f{A}_Y}{\Phi_Y(B_Y)}=\bigskalpHS{\Phi_Y^{-1}(\f{A}_Y)}{B_Y}$.
(In particular, $\Phi_Y$ preserves the Hilbert--Schmidt norm $\sqrt{\skalpHS{A_Y}{A_Y}}=\sqrt{\Tr(A_Y^\dagger A_Y)}$,
but not the trace norm $\Tr\bigl(\sqrt{A_Y^\dagger A_Y}\bigr)$ \cite{Wilde-2013}.)
It also turns out that the same map can be used
for transforming between the two extensions in \eqref{eq:Gamma}-\eqref{eq:Gammaf}, that is,
$\Phi_Y: 
\Gamma_Y(E\indexddu{i}{\nu}{\nu'}) \mapsto \Gammaf_Y(E\indexddu{i}{\nu}{\nu'})$,
in other words, we have
\begin{equation}
\label{eq:PhiGamma}
\Gammaf_Y = \Phi_Y \circ \Gamma_Y
\end{equation}
for the standard \eqref{eq:Gamma} and the fermionic \eqref{eq:Gammaf} extension maps.
Moreover,
this, and also \eqref{eq:EYEYftraf}, are special cases of that,
for arbitrary $X\subseteq Y$,
\begin{equation}
\label{eq:EYXEYXftraf}
\Phi_Y:\quad \prod_{j\in X} \Gamma_Y (A_j)  \quad\longmapsto\quad 
\prodord_{j\in X} \Gammaf_Y (A_j).
\end{equation}
(For the proof, see Appendix~\ref{appsec:JW.morphism}.)
We also have $\Idf_Y:=\Phi_Y(\Id_Y)=\Id_Y$ for the identity operators $\mathcal{H}_Y\to\mathcal{H}_Y$,
and $\IIdf_Y:=\Phi_Y\circ\IId_Y\circ\Phi_Y^{-1}=\IId_Y$ for the identity maps 
$\IId_Y:\mathcal{A}_Y\to\mathcal{A}_Y$ and 
$\IIdf_Y:\f{\mathcal{A}}_Y\to\f{\mathcal{A}}_Y$.
Similarly, $\Trf=\Phi_\emptyset\circ\Tr\circ\Phi_Y^{-1}$ for the trace maps 
$\Tr:\mathcal{A}_Y\to\field{C}$ and
$\Trf:\f{\mathcal{A}}_Y\to\field{C}$.
Here $\Tr : \mathcal{A}_Y\to\field{C}$ is given by the invariant definition, then $\Trf$ turns out to be given in the same way
(the diagonal elements of the phase factors \eqref{eq:phasefY} are $+1$).
By this, since $\Tr \bigl(\Gamma_Y (E\indexddu{i}{\nu}{\nu'})\bigr)=2^{\abs{Y}-1}\delta^{\nu,\nu'}$,
we have        $\Trf\bigl(\Gammaf_Y(E\indexddu{i}{\nu}{\nu'})\bigr)=2^{\abs{Y}-1}\delta^{\nu,\nu'}$.
Also, since $\Tr (    E\indexddu{Y}{\vs{\nu}}{\vs{\nu}'})=\delta^{\vs{\nu},\vs{\nu}'}$,
we have     $\Trf(\f{E}\indexddu{Y}{\vs{\nu}}{\vs{\nu}'})=\delta^{\vs{\nu},\vs{\nu}'}$.

From the algebraic point of view,
the subalgebras describing the subsystems, or mode subsets $X\subseteq Y$ are generated as
$\alg{a_{j,Y}}{j\in X}\subseteq \mathcal{A}_Y = \alg{a_{i,Y}}{i\in Y}$ and
$\alg{\f{a}_{j,Y}}{j\in X}\subseteq \f{\mathcal{A}}_Y= \alg{\f{a}_{i,Y}}{i\in Y}$
in the qubit and fermionic cases, respectively.
(Here $\alg{a_{j,Y}}{j\in X}$ denotes the algebra generated by the set $\sset{a_{j,Y}}{j\in X}$.)
So \eqref{eq:EYXEYXftraf} tells us that
$\Phi_Y$ respects the subsystem structure,
$\Phi_Y: \alg{a_{j,Y}}{j\in X}\to \alg{\f{a}_{j,Y}}{j\in X}$.
Note, however, the special form of the product on the left-hand side of \eqref{eq:EYXEYXftraf},
because in general, there exists $A_Y,B_Y\in\mathcal{A}_Y$ for which
$\Phi_Y(A_YB_Y)\neq\Phi_Y(A_Y)\Phi_Y(B_Y)$,
as well as
$\Phi_Y(A_Y^\dagger)\neq\Phi_Y(A_Y)^\dagger$,
\emph{the linear isomorphism $\Phi_Y$ is not a $*$-algebra isomorphism.}
Of course, a $*$-algebra isomorphism in such a role would not be interesting at all,
as it would not provide an essentially different structure compared to the structure of qubits.

\section{Fermionic tensors}
\label{sec:Tensors}

In this section
we construct a kind of ``fermionic version'' of operations widely used in quantum information
for our convenience,
and we consider the similarities and differences compared to the original ones.

\subsection{Fermionic tensor product}
\label{sec:Tensors.TP}

Let us have a \emph{partition} 
$\xi = \set{X_1,X_2,\dots,X_{\abs{\xi}}}\in\Pi(Y)$ 
of subsystem $Y\subseteq M$, that is, 
the subsystems $X\in\xi$ are nonempty and disjoint,
with $\bigcup_{X\in \xi }X= Y$.
In the qubit case, we have the \emph{tensor product}
for forming the joint algebra $\mathcal{A}_Y$ corresponding to $Y$
from the algebras $\mathcal{A}_X$ corresponding to $X\in\xi$,
as $\mathcal{A}_Y=\bigotimes_{X\in\xi}\mathcal{A}_X$.
This is written in the standard basis \eqref{eq:EY} as
\begin{subequations}
\begin{equation}
\label{eq:EYXs}
E\indexddu{Y}{\vs{\nu}}{\vs{\nu}'} = \bigotimes_{X\in\xi} E\indexddu{X}{\vs{\nu}_X}{\vs{\nu}_X'},
\end{equation}
see also \eqref{eq:EYexpl},
where the multi-index is
$\vs{\nu}: Y\to\{0,1\}$ as usual, and
$\vs{\nu}_X:=\vs{\nu}\restrict{X}: X\to\{0,1\}$ are the restrictions to the subsystems $X\subseteq Y$.
We would like to have a similar tool in the fermionic case
for joining subsystems, that is, mode subsets.
(Having $\otimes$ is particularly enlightening, because for applying voltage, it emits light.)
The usual tensor product cannot work in general,
it does not respect the fermionic nature of the operators, for example,
there exist operators $A_j\in\mathcal{A}_j$ for which
$\bigotimes_{X\in\xi} \prodord_{j\in X}\Gammaf_X(A_j)\neq \prodord_{i\in Y}\Gammaf_Y(A_i)$,
while equality holds for the qubit case without the tildes (see in \eqref{eq:algstuff} later).
Instead, let us define the \emph{fermionic tensor product} in the fermionic basis \eqref{eq:EYf} as
\begin{equation}
\label{eq:EYfXs}
\f{E}\indexddu{Y}{\vs{\nu}}{\vs{\nu}'} =: \bigotimesf_{X\in\xi} \f{E}\indexddu{X}{\vs{\nu}_X}{\vs{\nu}_X'}.
\end{equation}
\end{subequations}
Since $\f{E}\indexddu{Y}{\vs{\nu}}{\vs{\nu}'}
=\Phi_Y(E\indexddu{Y}{\vs{\nu}}{\vs{\nu}'})
=\Phi_Y\bigl( \bigotimes_{X\in\xi} E\indexddu{X}{\vs{\nu}_X}{\vs{\nu}_X'} \bigr)
=\Phi_Y\bigl( \bigotimes_{X\in\xi} \Phi_X^{-1}(\f{E}\indexddu{X}{\vs{\nu}_X}{\vs{\nu}_X'} )\bigr)$
on the left-hand side
by using \eqref{eq:EYEYftraf} and \eqref{eq:EYXs},
one can read off the form
\begin{subequations}
\label{eq:TPf}
\begin{equation}
\label{eq:TPfPsi}
\bigotimesf_{X\in\xi}\f{A}_X
= \f{\Psi}_\xi\Bigl(\bigotimes_{X\in\xi}\f{A}_X\Bigr)
\end{equation}
by linearity, 
with the map
\begin{equation}
\label{eq:Psiupsilon}
\f{\Psi}_\xi := \Phi_Y \circ \bigotimes_{X\in\xi}\Phi_X^{-1} : \quad 
\f{\mathcal{A}}_Y \quad\longrightarrow\quad \f{\mathcal{A}}_Y.
\end{equation}
\end{subequations}
This map is also given by multiplication of the \eqref{eq:EYf} basis with $\pm1$ phase factors.
(For the phase factors, see Appendix~\ref{appsec:Tensors.TP}.
For the explicit form for few modes, see Appendix~\ref{appsec:Tensors.Explh}.)
Note again, that the ordering in this tensor product is illustrative only, 
$\f{A}_X\otimesf\f{B}_{\bar{X}}=\f{B}_{\bar{X}}\otimesf\f{A}_X$
(using the notation $\bar{X} = Y\setminus X$).

Both the usual and the fermionic tensor products are consistent with the refinement of the partitions.
Indeed, let us have $Z\subseteq M$, the partition $\upsilon\in\Pi(Z)$,
and for all parts $Y\in\upsilon$, let us have the partitions $\xi_Y\in\Pi(Y)$ of them,
by which let $\xi\in\Pi(Z)$ be the ``merging'' of the partitions $\xi_Y$,
$\xi:=\bigcup_{Y\in\upsilon}\xi_Y$;
then we have the \emph{associativity} for the tensor products as 
\begin{subequations}
\begin{align}
\label{eq:TPAssoc}
\bigotimes_{Y\in\upsilon} \Bigl( \bigotimes_{X_Y\in\xi_Y} A_{X_Y} \Bigr) &= \bigotimes_{X\in\xi} A_X,\\
\label{eq:TPfAssoc}
\bigotimesf_{Y\in\upsilon} \Bigl( \bigotimesf_{X_Y\in\xi_Y} \f{A}_{X_Y} \Bigr) &= \bigotimesf_{X\in\xi} \f{A}_X.
\end{align}
\end{subequations}
(The first one is a property of the tensor product;
the second one follows from this, by the special structure \eqref{eq:Psiupsilon} of $\f{\Psi}_\xi$.)
This justifies the use of the ``associative'' tensor product sign.
We also have that the identity operators are
$\bigotimes_{X\in\xi}\Id_X = \Id_Y$ and
$\bigotimesf_{X\in\xi}\Idf_X = \Idf_Y$.

Since the map $\f{\Psi}_\xi$ in \eqref{eq:Psiupsilon} is a composition of unitaries \eqref{eq:PhiU},
it is \emph{unitary} itself,
\begin{equation}
\label{eq:PsiU}
\bigskalpHS{\f{\Psi}_\xi(\f{A}_Y)}{\f{\Psi}_\xi(\f{B}_Y)}=\bigskalpHS{\f{A}_Y}{\f{B}_Y},
\end{equation}
and we also have the
\emph{compatibility with the Hilbert--Schmidt inner product} \eqref{eq:HS},
\begin{subequations}
\begin{align}
\label{eq:TPHS}
\BigskalpHS{\bigotimes_{X\in\xi} A_X}{\bigotimes_{X\in\xi} B_X}
&= \prod_{X\in\xi}\skalpHS{A_X}{B_X},\\
\label{eq:TPfHS}
\BigskalpHS{\bigotimesf_{X\in\xi} \f{A}_X}{\bigotimesf_{X\in\xi} \f{B}_X}
&= \prod_{X\in\xi}\skalpHS{\f{A}_X}{\f{B}_X}.
\end{align}
\end{subequations}
(The first one is the standard construction of the inner product in tensor product spaces;
the second one follows from this, by using \eqref{eq:TPfPsi} and \eqref{eq:PsiU}.)
From this, the product property of the trace follows also for fermionic elementary tensors,
$\Tr \bigl( \bigotimes_{X\in\xi}    B_X\bigr) = \prod_{X\in\xi} \Tr (B_X)$ and
$\Trf\bigl(\bigotimesf_{X\in\xi}\f{B}_X\bigr) = \prod_{X\in\xi} \Trf(\f{B}_X)$,
by setting $A_X=\Id_X$ and $\f{A}_X=\Idf_X$ in \eqref{eq:TPHS} and \eqref{eq:TPfHS}, respectively.
(A direct calculation would be in the second case to show that
the diagonal elements of the phase factors of $\f{\Psi}_\xi$ are $+1$, see Appendix \ref{appsec:Tensors.TP}.)

With the fermionic tensor product \eqref{eq:TPf}, now we have
\begin{subequations}
\begin{align}
\label{eq:algstuff}
\prod_{i\in Y}\Gamma_Y(A_i)
&= \bigotimes_{X\in\xi} \prod_{j\in X}\Gamma_X(A_j),\\
\label{eq:algfstuff}
\prodord_{i\in Y}\Gammaf_Y(A_i)
&= \bigotimesf_{X\in\xi} \prodord_{j\in X}\Gammaf_X(A_j).
\end{align}
\end{subequations}
(The first one follows from $\prod_{j\in X}\Gamma_X(A_j) = \bigotimes_{j\in X}A_j$,
which is a simple consequence of \eqref{eq:EYexpl} and \eqref{eq:Gamma} by linearity;
the second one follows from this, by using \eqref{eq:EYXEYXftraf} and \eqref{eq:TPf}.)
By linearity and \eqref{eq:EYfexpl}, one can see that 
\eqref{eq:algfstuff} is an equivalent form of \eqref{eq:EYfXs},
expressing a different aspect of the fermionic tensor product.

Although the fermionic tensor product \eqref{eq:EYfXs} shows the above convenient properties,
which are extremely useful in symbolical as well as numerical calculations,
it does not obey all the properties of the usual tensor product,
as we will immediately see.
The fermionic tensor product is clearly linear, and
if we consider only the \emph{linear structure},
we could even write $\f{\mathcal{A}}_Y=\bigotimesf_{X\in\xi}\f{\mathcal{A}}_X$
(and we will actually use this notation later),
as far as this is understood simply as the linear hull of fermionic elementary tensors $\bigotimesf_{X\in\xi}\f{A}_X$.
However, the fermionic tensor product fails to obey the properties
of the whole \emph{$*$-algebraic structure},
since 
there exist operators $\f{A}_X,\f{B}_X\in\f{\mathcal{A}}_X$ such that
$\bigl(\bigotimesf_{X\in\xi} \f{A}_X\bigr)\bigl(\bigotimesf_{X'\in\xi} \f{B}_{X'}\bigr)
\neq \bigotimesf_{X\in\xi} \bigl(\f{A}_X\f{B}_X\bigr)$
and $\bigotimesf_{X\in\xi} \f{A}_X^\dagger
\neq \bigl(\bigotimesf_{X\in\xi} \f{A}_X\bigr)^\dagger$,
as well as, 
if we consider them as \emph{operators} acting on Hilbert spaces,
$\bigl(\bigotimesf_{X\in\xi} \f{A}_X\bigr) \bigl(\bigotimes_{X'\in\xi} \cket{\psi_{X'}}\bigr)
\neq \bigotimes_{X\in\xi} \bigl(\f{A}_X\cket{\psi_X}\bigr)$.
(For explicit examples, coming from a wider context, see Appendix~\ref{appsec:Parity.examples}.)
We keep using the name ``fermionic tensor product'' for convenience,
because \emph{$\otimesf$ shows up in the same role in the fermionic case
as $\otimes$ in the standard (qubit) case.}
However, we emphasize that it should be considered only as a shorthand notation 
for the ``$\Phi$-adjoined version'' \eqref{eq:TPf} of the usual tensor product.
Note also that 
this (and also \eqref{eq:Lambdaboth} later)
can be regarded as an inner definition 
in the concrete representation
(for the case of many mode subsets)
of the abstract construction
of the fermionic tensor product,
given usually through an outer definition
\cite{Shapourian-2019,Crismale-2020,Brunetti-2021}.

Following these lines,
we also have in general that
there exist operators $\f{A}_X\in\f{\mathcal{A}}_X$ and $\f{B}_{\bar{X}}\in\f{\mathcal{A}}_{\bar{X}}$ such that
$\bigl(\f{A}_X\otimesf\Idf_{\bar{X}}\bigr)\bigl( \Idf_X\otimesf \f{B}_{\bar{X}}\bigr)\neq \f{A}_X\otimesf \f{B}_{\bar{X}}$
(using the notation $\bar{X} = Y\setminus X$),
moreover,
$\bigl[\f{A}_X\otimesf\Idf_{\bar{X}}, \Idf_X\otimesf \f{B}_{\bar{X}} \bigr] \neq 0$,
although the equalities hold in these formulas in the standard (qubit) case, with $\otimes$ instead of $\otimesf$.
(We will see later that at least 
$\bigl(\f{A}_X\otimesf\Idf_{\bar{X}}\bigr)\bigl( \f{B}_X\otimesf\Idf_{\bar{X}}\bigr) = (\f{A}_X\f{B}_X)\otimesf \Idf_{\bar{X}}$
holds, see \eqref{eq:cEmbfHom}.
Also, in the so called physical subspace of the algebras, 
$\bigl[\f{A}_X\otimesf\Idf_{\bar{X}}, \Idf_X\otimesf \f{B}_{\bar{X}} \bigr] = 0$,
see \eqref{eq:Commf} later in Section~\ref{sec:ParityMath.A}).
Before further discussion and clarification on these properties of products, 
it is convenient to introduce the fermionic canonical embedding.

\subsection{Fermionic canonical embedding}
\label{sec:Tensors.cEmb}

In the qubit case, we also have the \emph{canonical embedding} map
with respect to the tensor product,
for the subsystems $X\subseteq Y\subseteq M$, as
\begin{subequations}
\label{eq:cEmbs}
\begin{equation}
\label{eq:cEmb}
\begin{split}
\iota_{X,Y} :\quad 
\mathcal{A}_X \quad&\longrightarrow\quad \mathcal{A}_Y,\\
A_X \quad&\longmapsto\quad A_X \otimes \Id_{\bar{X}},
\end{split}
\end{equation}
using the notation $\bar{X} = Y\setminus X$.
We would like to have a similar tool in the fermionic case,
so let us define the \emph{fermionic canonical embedding} as
\begin{equation}
\label{eq:cEmbf}
\begin{split}
\iotaf_{X,Y} :\quad 
\f{\mathcal{A}}_X \quad&\longrightarrow\quad \f{\mathcal{A}}_Y,\\
\f{A}_X \quad&\longmapsto\quad \f{A}_X \otimesf \Idf_{\bar{X}}.
\end{split}
\end{equation}
\end{subequations}
Let us have the nested subsystems, or mode subsets, $X\subseteq Y\subseteq Z\subseteq M$,
then we have
\begin{subequations}
\begin{align}
\label{eq:cEmbCons}
\iota_{Y,Z}\circ\iota_{X,Y} &= \iota_{X,Z},\\
\label{eq:cEmbfCons}
\iotaf_{Y,Z}\circ\iotaf_{X,Y} &= \iotaf_{X,Z}.
\end{align}
\end{subequations}
(These follow from the associativity properties
\eqref{eq:TPAssoc} and \eqref{eq:TPfAssoc} in the two cases, respectively.)

It is important to see that the standard \eqref{eq:cEmb} and fermionic \eqref{eq:cEmbf} canonical embeddings
are unitarily equivalent, that is, there exists $U_{X\bar{X}}\in\LieGrp{U}(\mathcal{H}_Y)$, such that
\begin{equation}
\label{eq:iotaU}
\f{A}_X\otimesf\Idf_{\bar{X}} = U_{X\bar{X}} \bigl(\f{A}_X\otimes\Idf_{\bar{X}}\bigr) U_{X\bar{X}}^\dagger.
\end{equation}
(For the proof, by the construction of $U_{X\bar{X}}$, see Appendix~\ref{appsec:Tensors.cEmb}.
For the explicit form for few modes, see Appendix~\ref{appsec:Parity.Explu}.
Note that $U_{X\bar{X}}\neq U_{\bar{X}X}$.)
From this, it easily follows
that 
also the fermionic canonical embedding \eqref{eq:cEmbf} is a $*$-homomorphism,
\begin{subequations}
\begin{align}
\label{eq:cEmbHom}
\iota_{X,Y}(A_X)\iota_{X,Y}(B_X) &= \iota_{X,Y}(A_XB_X),&\qquad
\iota_{X,Y}(A_X^\dagger) &= \iota_{X,Y}(A_X)^\dagger,\\
\label{eq:cEmbfHom}
\iotaf_{X,Y}(\f{A}_X)\iotaf_{X,Y}(\f{B}_X) &= \iotaf_{X,Y}(\f{A}_X\f{B}_X),&\qquad
\iotaf_{X,Y}(\f{A}_X^\dagger) &= \iotaf_{X,Y}(\f{A}_X)^\dagger.
\end{align}
\end{subequations}
(The first one is obvious consequence of the tensor product;
the second one follows from this by using \eqref{eq:iotaU}.)

Note that, while $\iota_{X,Y}(A_X)$ is an \emph{identical extension} of $A_X$,
that is, it acts identically on the factor $\mathcal{H}_{\bar{X}}$ of $\mathcal{H}_Y$
(as $\iota_{X,Y}(A_X)\bigl(\cket{\psi_X}\otimes\cket{\psi_{\bar{X}}}\bigr)=(A_X\cket{\psi_X})\otimes\cket{\psi_{\bar{X}}}$),
this is not the case for $\iotaf_{X,Y}(\f{A}_X)$,
which acts identically on $\mathcal{H}_{\bar{X}}$ 
transformed by the \emph{global} unitary $U_{X\bar{X}}$, see in \eqref{eq:iotaU}.
It is important to see that the unitary transformations 
for the embeddings of $X$ and $\bar{X}$ are different,
and these two embeddings cannot be realized using a common unitary
(unless the parity superselection rule is imposed,
see \eqref{eq:TPSH} in Section~\ref{sec:ParityPhys.TPSA}).
(For explicit example, see Appendix~\ref{appsec:Parity.examples}.)

Turning to the special case of single-mode subsets,
we have that $\iotaf$ is the multi-mode generalization of $\Gammaf$,
\begin{equation}
\label{eq:GammafTPf}
\Gammaf_Y(A_i) = \f{A}_{\{i\}}\otimesf\Idf_{Y\setminus\{i\}}
\equiv \iotaf_{\{i\},Y}(\f{A}_{\{i\}}),
\end{equation}
see \eqref{eq:Gammaf}, with the notation $\f{A}_{\{i\}}=\Gammaf_{\{i\}}(A_i)= A_i$.
(This follows by linearity from
$\Gammaf_Y(E\indexddu{i}{\nu}{\nu'})
=\Phi_Y\bigl(\Gamma_Y(E\indexddu{i}{\nu}{\nu'})\bigr)
=\Phi_Y\bigl(E\indexddu{i}{\nu}{\nu'}\otimes\Id_{Y\setminus\{i\}}\bigr)
=\Phi_Y\bigl(\Phi_{\{i\}}^{-1}(\f{E}\indexddu{\{i\}}{\nu}{\nu'})\otimes \Phi_{Y\setminus\{i\}}^{-1}(\Idf_{Y\setminus\{i\}})\bigr)
=\f{E}\indexddu{\{i\}}{\nu}{\nu'}\otimesf\Idf_{Y\setminus\{i\}}$,
by using \eqref{eq:PhiGamma}, \eqref{eq:Gamma}, \eqref{eq:EYEYftraf} and \eqref{eq:TPf}.)
Moreover,
any ordered product of single-mode operators, $\prodord_{i\in X}\iotaf_{\{i\},X}\bigl(\f{A}_{\{i\}}\bigr)$, 
is embedded properly,
\begin{subequations}
\begin{equation}
\label{eq:cEmbfmotiv}
\iotaf_{X,Y} \Bigl(\prodord_{i\in X}\iotaf_{\{i\},X}(\f{A}_{\{i\}}) \Bigr) 
=\prodord_{i\in X}\iotaf_{X,Y}\Bigl(\iotaf_{\{i\},X}(\f{A}_{\{i\}}) \Bigr) 
=\prodord_{i\in X}\iotaf_{\{i\},Y}(\f{A}_{\{i\}}),
\end{equation}
using \eqref{eq:cEmbfHom} and \eqref{eq:cEmbfCons}.
In general,
let us have the \emph{ordered partition} $\ord{\upsilon}=\tuple{Y_1,Y_2,\dots,Y_{\abs{\ord{\upsilon}}}}\in \ord{\Pi}(Z)$,
for the mode subset $Z\subseteq M$,
and for all parts $Y\in\ord{\upsilon}$, let us have the ordered partitions $\ord{\xi}_Y\in\ord{\Pi}(Y)$ of them,
then combining \eqref{eq:cEmbfHom} and \eqref{eq:cEmbfCons} 
leads to the convenient property for the product of embeddings of products of embeddings
\begin{equation}
\label{eq:TPfpAssoc0}
\begin{split}
\prodord_{Y\in\ord{\upsilon}} \iotaf_{Y,Z}\Bigl(\prodord_{X_Y\in\ord{\xi}_Y}\iotaf_{X_Y,Y}(\f{A}_{X_Y})\Bigr)
&=\prodord_{Y\in\ord{\upsilon}}\prodord_{X_Y\in\ord{\xi}_Y} \iotaf_{Y,Z}\bigl(\iotaf_{X_Y,Y}(\f{A}_{X_Y})\bigr)\\
&=\prodord_{Y\in\ord{\upsilon}}\prodord_{X_Y\in\ord{\xi}_Y} \iotaf_{X_Y,Z}(\f{A}_{X_Y}),
\end{split}
\end{equation}
\end{subequations}
where the ordered product $\prodord$ is ordered as the ordered partition $\ord{\upsilon}$.

It is now convenient to introduce the following shorthand notation for
the ordered product of embeddings of operators of disjoint mode subsets as
\begin{equation}
\label{eq:TPfp}
\bigotimesfp_{X\in\ord{\xi}} \f{A}_X := \prodord_{X\in\ord{\xi}} \iotaf_{X,Y} (\f{A}_X)
\end{equation}
for the ordered partition $\ord{\xi}\in\ord{\Pi}(Y)$.
Note that, with a slight but convenient abuse of the notation, the ordering in a product 
$\f{A}_{X_1}\otimesfp\f{A}_{X_2}\otimesfp\f{A}_{X_3}\otimesfp\dots\equiv
\iotaf_{X_1,Y}(\f{A}_{X_1})
\iotaf_{X_2,Y}(\f{A}_{X_2})
\iotaf_{X_3,Y}(\f{A}_{X_3})\dots$
carries now the ordering of the usual product on the right-hand side of \eqref{eq:TPfp}.
We have that also this product is consistent with the refinement of the partitions.
Indeed, let us have $Z\subseteq M$, the ordered partition $\ord{\upsilon}\in\ord{\Pi}(Z)$,
and for all parts $Y\in\ord{\upsilon}$, let us have the ordered partitions $\ord{\xi}_Y\in\ord{\Pi}(Y)$ of them,
by which let $\ord{\xi}\in \ord{\Pi}(Z)$ be the ``ordered merging'' of the ordered partitions $\ord{\xi}_Y$,
$\ord{\xi}:= \ord{\bigcup}_{Y\in\ord{\upsilon}} \ord{\xi}_Y$;
then we have the \emph{associativity} as
\begin{equation}
\label{eq:TPfpAssoc}
\bigotimesfp_{Y\in\ord{\upsilon}} \Bigl( \bigotimesfp_{X_Y\in\ord{\xi}_Y} \f{A}_{X_Y} \Bigr)
 = \bigotimesfp_{X\in\ord{\xi}} \f{A}_X.
\end{equation}
(This follows from \eqref{eq:TPfpAssoc0} and \eqref{eq:TPfp}; it is just the associativity of the product of embeddings.)
This justifies the use of the ``associative'' tensor product sign.
This means that the ordering in which the products are calculated does not matter (associativity),
but the ordering of the factors does (noncommutativity):
the ordering of $\ord{\xi}$ here refers only to that 
the factors on the right-hand side of \eqref{eq:TPfp} are not commuting.

\subsection{Fermionic tensorial and algebraic product operators}
\label{sec:Tensors.Lambda}

After introducing the fermionic tensor product in Section~\ref{sec:Tensors.TP},
we briefly mentioned that it does not lead to a proper tensor product structure,
which manifested itself in several aspects, for example,
there exist operators $\f{A}_X\in\f{\mathcal{A}}_X$ and $\f{B}_{\bar{X}}\in\f{\mathcal{A}}_{\bar{X}}$ such that
$\f{A}_X\otimesf \f{B}_{\bar{X}} \neq 
\bigl(\f{A}_X\otimesf\Idf_{\bar{X}}\bigr)\bigl( \Idf_X\otimesf \f{B}_{\bar{X}}\bigr)
\equiv\f{A}_X\otimesfp\f{B}_{\bar{X}}$.
Therefore it is useful to introduce the map
\begin{subequations}
\label{eq:Lambdaboth}
\begin{equation}
\label{eq:Lambda2}
\begin{split}
\f{\Lambda}_{X\bar{X}}:\quad 
\f{\mathcal{A}}_Y \quad&\longrightarrow\quad \f{\mathcal{A}}_Y,\\
\f{A}_X\otimesf \f{B}_{\bar{X}} \quad&\longmapsto\quad 
\f{A}_X\otimesfp \f{B}_{\bar{X}}
\equiv \iotaf_{X,Y}(\f{A}_X)\iotaf_{\bar{X},Y}(\f{B}_{\bar{X}}),
\end{split}
\end{equation}
encoding these properties,
bringing back and forth between the fermionic tensorial and the algebraic points of view.
(Note that $\f{\Lambda}_{X\bar{X}}\neq\f{\Lambda}_{\bar{X}X}$.)
We also have its generalization for arbitrary number of mode subsets,
labeled by the ordered partition $\ord{\xi}=\tuple{X_1,X_2,\dots,X_{\abs{\ord{\xi}}}}\in \ord{\Pi}(Y)$, as
\begin{equation}
\label{eq:Lambda}
\begin{split}
\f{\Lambda}_{\ord{\xi}}:\quad 
\f{\mathcal{A}}_Y \quad&\longrightarrow\quad \f{\mathcal{A}}_Y,\\
\bigotimesf_{X\in\xi} \f{A}_X \quad&\longmapsto\quad 
\bigotimesfp_{X\in\ord{\xi}} \f{A}_X
\equiv \prodord_{X\in\ord{\xi}}\iotaf_{X,Y}(\f{A}_X).
\end{split}
\end{equation}
\end{subequations}
This map is also given by multiplication of the \eqref{eq:EYf} basis with $\pm1$ phase factors.
(For the phase factors, see Appendices \ref{appsec:Tensors.Lambda2} and \ref{appsec:Tensors.Lambda}.
For the explicit form for few modes, see Appendix~\ref{appsec:Tensors.Expll}.
Note that $\f{\Lambda}_{X\bar{X}}$ is just a special case of this for $\ord{\xi}=\tuple{X,\bar{X}}$.
We use a simplified notation for ordered partitions in lower index position, 
omitting the parentheses and colons in the writing of tuples, since this does not cause confusion.)
Because of that form,
it is clearly \emph{unitary},
\begin{equation}
\label{eq:LambdaU}
\bigskalpHS{\f{\Lambda}_{\ord{\xi}}(\f{A}_Y)}{\f{\Lambda}_{\ord{\xi}}(\f{B}_Y)}=\bigskalpHS{\f{A}_Y}{\f{B}_Y}.
\end{equation}
We also have the \emph{compatibility with the Hilbert--Schmidt inner product} \eqref{eq:HS}
\begin{equation}
\label{eq:TPfpHS}
\BigskalpHS{\bigotimesfp_{X\in\ord{\xi}}\f{A}_X}{\bigotimesfp_{X\in\ord{\xi}}\f{B}_X}
= \prod_{X\in\ord{\xi}}\bigskalpHS{\f{A}_X}{\f{B}_X}.
\end{equation}
(This follows from \eqref{eq:TPfHS}, by using \eqref{eq:Lambda} and \eqref{eq:LambdaU}.)
The identity
\begin{subequations}
\begin{equation}
\label{eq:Lambdaiotaf}
  \f{\Lambda}_{X\bar{X}}     \circ\iotaf_{X,Y} 
= \f{\Lambda}_{X\bar{X}}^{-1}\circ\iotaf_{X,Y}
= \f{\Lambda}_{\bar{X}X}     \circ\iotaf_{X,Y} 
= \f{\Lambda}_{\bar{X}X}^{-1}\circ\iotaf_{X,Y} 
= \iotaf_{X,Y}
\end{equation}
is also easy to check by definitions \eqref{eq:cEmbf} and \eqref{eq:Lambda2},
and by noting that the identity operator is $\Idf_Y=\Idf_X\otimesf\Idf_{\bar{X}}$, see in Section~\ref{sec:Tensors.TP}.
In particular, we have
\begin{equation}
\label{eq:ffpiotaf}
\iotaf_{X,Y}(\f{A}_X)
= \f{A}_X \otimesf \Idf_{\bar{X}}
= \f{A}_X \otimesfp \Idf_{\bar{X}}
= \Idf_{\bar{X}} \otimesfp \f{A}_X
\end{equation}
for the fermionic canonical embedding \eqref{eq:cEmbf}.
\end{subequations}

In the following, we will consider these two ``representations'' of the products of operators of disjoint mode subsets, 
the ``fermionic tensorial representation'' and the ``algebraic representation''.
We call these products on the left-hand side and right-hand side of \eqref{eq:Lambda}
\emph{fermionic $\xi$-elementary tensors,} and
\emph{fermionic $\ord{\xi}$-elementary products,} respectively.

To elaborate on the meaning of $\f{\Lambda}_{\ord{\xi}}$ in \eqref{eq:Lambda},
we have that \eqref{eq:algfstuff} for single-mode subsets ($\abs{X}=1$ for all $X\in\xi$) yields
\begin{subequations}
\begin{equation}
\label{eq:specialsingle}
\bigotimesfp_{i\in Y}\f{A}_{\{i\}} = 
\bigotimesf_{i\in Y} \f{A}_{\{i\}}, 
\end{equation}
by using \eqref{eq:GammafTPf} and \eqref{eq:TPfp}.
Here on the left-hand side, the increasing ordering is understood.
Moreover, 
if $\ord{\xi} = \tuple{X_1,X_2,\dots,X_{\abs{\ord{\xi}}}}$ is 
such that 
$X_s<X_r$ (elementwisely) for all $s<r$,
that is, the mode subsets $X\in\ord{\xi}$ contain modes neighboring with respect to $Y$, 
and ordered accordingly to the Jordan--Wigner ordering of the modes, 
we have
\begin{equation}
\label{eq:specialmulti}
\bigotimesfp_{X\in\ord{\xi}}\f{A}_X = \bigotimesf_{X\in\xi} \f{A}_X,
\end{equation}
\end{subequations}
where $\xi$ is the unordered partition consisting of the parts of the ordered partition $\ord{\xi}$.
(For the proof, it is enough to consider the case $\f{A}_X=\bigotimesf_{j\in X}\f{A}_{\{j\}}$,
since the elementary fermionic tensors span $\f{\mathcal{A}}_X$.
Then, applying \eqref{eq:specialsingle}, \eqref{eq:TPfpAssoc}, \eqref{eq:specialsingle} again and \eqref{eq:TPfAssoc}, we have
\begin{equation*}
\bigotimesfp_{X\in\ord{\xi}} \f{A}_X \equiv
\bigotimesfp_{X\in\ord{\xi}}  \bigotimesf_{j\in X}\f{A}_{\{j\}} 
=\bigotimesfp_{X\in\ord{\xi}}\bigotimesfp_{j\in X}\f{A}_{\{j\}}
=\bigotimesfp_{i\in Y}\f{A}_{\{i\}}
=\bigotimesf_{i\in Y} \f{A}_{\{i\}}
=\bigotimesf_{X\in\xi}\bigotimesf_{j\in X} \f{A}_{\{j\}}
\equiv \bigotimesf_{X\in\ord{\xi}} \f{A}_X,
\end{equation*}
where \eqref{eq:TPfpAssoc} could be applied because of the special form of the partition $\ord{\xi}$.)
That is, the fermionic tensorial and the algebraic products coincide,
$\f{\Lambda}_{\ord{\xi}}=\IIdf_Y$, 
\emph{for these very special partitions,}
which are usually considered in the literature.
For more general partitions, 
which contain mode subsets of non-neighboring modes or 
which are ordered differently than the Jordan--Wigner ordering,
the fermionic tensorial and the algebraic representations are different, and $\f{\Lambda}_{\ord{\xi}}$ is nontrivial.
 
Note that if the number of modes is larger than two,
then it is not possible to give an ordering
with respect to which all partitions have mode subsets containing consecutive modes.
(This holds also for qubits, however, it does not cause any difficulty there.)
This may happen also when only some particular partitions are considered, which are relevant in some sense.
For example, in the case of fermionic topological codes \cite{Gu-2014},
when bipartitions given for all plaquettes by the plaquette and its complement are considered;
or in the case of $2$-separability, when it is enough to consider all the bipartitions
(or in a coarsened case,
when six modes, three spatial positions with two spin directions,
only spatial partitionings are considered \cite{Johansson-2016b});
or in the case of the more general $k$-producibility, $k$-partitionability, $k$-stretchability,
or general Level~II notions of partial entanglement or partial correlation \cite{Szalay-2015b,Szalay-2017,Szalay-2018,Szalay-2019} 
of fermionc modes \cite{Szalay-2017,Brandejs-2019}.

The property \eqref{eq:algfstuff} explains how the fermionic tensor product 
``reorders'' the one-mode operators in a product.
In this way, the fermionic tensor product 
is in accordance with the \emph{fixed} Jordan--Wigner ordering of the modes $1,2,\dots,L$.
About this, we have
\begin{equation}
\label{eq:Lambdareorder}
\prodord_{i\in Y}\iotaf_{\{i\},Y}(\f{A}_{\{i\}}) 
= \f{\Lambda}_{\ord{\xi}}^{-1}\Bigl(\prodord_{X\in\ord{\xi}}\prodord_{j\in X} \iotaf_{\{j\},Y}(\f{A}_{\{j\}}) \Bigr),
\end{equation}
illustrating the operator reordering by $\f{\Lambda}_{\ord{\xi}}$
for general $\ord{\xi}$.
(The proof of this is
 that the left-hand side by \eqref{eq:algfstuff} and \eqref{eq:GammafTPf} is
\begin{equation*}
\bigotimesf_{X\in\ord{\xi}}\prodord_{j\in X} \iotaf_{\{j\},X}(\f{A}_{\{j\}})
=\f{\Lambda}_{\ord{\xi}}^{-1}\biggl(
\prodord_{X\in\ord{\xi}} \iotaf_{X,Y} \Bigl(\prodord_{j\in X} \iotaf_{\{j\},X}(\f{A}_{\{j\}}) \Bigr) \biggr),
\end{equation*}
by \eqref{eq:Lambda},
which equals to the right-hand-side by \eqref{eq:TPfpAssoc0}.)
Note that $\f{\Lambda}_{\ord{\xi}}$ is not a simple unitary conjugation, coming from a permutational mode transformation,
so it cannot be implemented as a mode transformation used in fermionic orbital optimization \cite{Krumnow-2016}.
This is because it permutes operators in $\f{\mathcal{A}}_Y\isom\mathcal{H}_Y\otimes\mathcal{H}_Y^*$,
and it is not factorized into independent operations on $\mathcal{H}_Y$ and $\mathcal{H}_Y^*$.
$\f{\Lambda}_{\ord{\xi}}$ cannot even be considered as a general Bogoliubov transformation, its way of functioning is completely different.
First, $\f{\Lambda}_{\ord{\xi}}$ leaves single-mode operators $\Gammaf_Y(\f{A}_i)$ invariant (see \eqref{eq:cEmbfHom} and \eqref{eq:Lambda}),
so, as Bogoliubov transformation, it would be trivial.
However, its action on multi-mode operators is given explicitly \eqref{eq:Lambda}, contrary to Bogoliubov transformations,
where the action on multi-mode operators is given through the action on single-mode creation and annihilation operators.

\subsection{Fermionic partial trace}
\label{sec:Tensors.PT}

In the qubit case, we also have the \emph{partial trace} map
with respect to the tensor product \cite{Petz-2008,Wilde-2013},
for the subsystems $X\subseteq Y\subseteq M$, as
\begin{subequations}
\label{eq:PT}
\begin{equation}
\label{eq:TPPT}
\begin{split}
\Tr_{Y,X} :\quad \mathcal{A}_Y \quad&\longrightarrow\quad \mathcal{A}_X,\\
A_X\otimes B_{\bar{X}} \quad&\longmapsto\quad  A_X \Tr (B_{\bar{X}}),
\end{split}
\end{equation}
using the notation $\bar{X} = Y\setminus X$.
(Note that only the discarded subsystem $\bar{X}$ is usually written in the lower index of the partial trace \cite{Petz-2008,Wilde-2013},
however, from our point of view, it is more expressive to denote
the source subsystem $Y$ and the target subsystem $X$.)
We would like to have a similar tool in the fermionic case,
so let us define the \emph{fermionic partial trace} with respect to the fermionic tensor product \eqref{eq:TPf} as
\begin{equation}
\label{eq:TPfPTf}
\begin{split}
\Trf_{Y,X} :\quad \f{\mathcal{A}}_Y \quad&\longrightarrow\quad \f{\mathcal{A}}_X,\\
\f{A}_X\otimesf \f{B}_{\bar{X}} \quad&\longmapsto\quad  \f{A}_X \Trf(\f{B}_{\bar{X}}).
\end{split}
\end{equation}
\end{subequations}
Since this is $\Trf_{Y,X}\bigl(\f{A}_X\otimesf \f{B}_{\bar{X}}\bigr)
= \Trf_{Y,X}\bigl(\Phi_Y\bigl(\Phi_X^{-1}(\f{A}_X)\otimes\Phi_{\bar{X}}^{-1}(\f{B}_{\bar{X}})\bigr)\bigr)
:=\f{A}_X \Trf( \f{B}_{\bar{X}})$ by using \eqref{eq:TPf},
which is equivalent to
$\bigl(\Phi_X^{-1}\circ\Trf_{Y,X}\circ\Phi_Y\bigr)\bigl(\Phi_X^{-1}(\f{A}_X)\otimes\Phi_{\bar{X}}^{-1}(\f{B}_{\bar{X}})\bigl)
:=\Phi_X^{-1}(\f{A}_X)\Tr\bigl(\Phi_{\bar{X}}^{-1}(\f{B}_{\bar{X}})\bigr)
= \Tr_{Y,X}\bigl(\Phi_X^{-1}(\f{A}_X)\otimes\Phi_{\bar{X}}^{-1}(\f{B}_{\bar{X}})\bigr)$
by using \eqref{eq:TPPT}, one can read off the form $\Phi_X^{-1}\circ\Trf_{Y,X}\circ\Phi_Y=\Tr_{Y,X}$, that is,
\begin{equation}
\label{eq:PTf}
\Trf_{Y,X} = \Phi_X \circ \Tr_{Y,X} \circ \Phi_Y^{-1}.
\end{equation}
The maps $\Tr$ and $\Trf$ in \eqref{eq:TPPT} and \eqref{eq:TPfPTf} are the trace maps,
see the end of Section~\ref{sec:JW.Phi}.

It is important to see that, 
similarly to the qubit case,
the fermionic partial trace is the adjoint map of the fermionic canonical embedding
with respect to the Hilbert--Schmidt inner product \eqref{eq:HS},
that is,
\begin{subequations}
\label{eq:cembPTadjgen}
\begin{align}
\label{eq:cembPTadj}
\Tr_{Y,X}=\iota_{X,Y}^\daggerHS:\quad
\bigskalpHS{\iota_{X,Y}(A_X)}{B_Y}
&= \bigskalpHS{A_X}{\Tr_{Y,X}(B_Y)} \quad 
\forall A_X\in\mathcal{A}_X, B_Y\in\mathcal{A}_Y,\\
\label{eq:cembfPTfadj}
\Trf_{Y,X}=\iotaf_{X,Y}^\daggerHS:\quad
\bigskalpHS{\iotaf_{X,Y}(\f{A}_X)}{\f{B}_Y}
&= \bigskalpHS{\f{A}_X}{\Trf_{Y,X}(\f{B}_Y)}\quad 
\forall \f{A}_X\in\f{\mathcal{A}}_X, \f{B}_Y\in\f{\mathcal{A}}_Y.
\end{align}
\end{subequations}
We will also meet this later in Section~\ref{sec:States.redf} from a more physical point of view.
(The first one is a property of the partial trace, maybe not so well-known,
the second one follows in the same way, 
i.e., by linearity, it is enough to prove for fermionic $\set{X,\bar{X}}$-elementary tensors,
$\bigskalpHS{\f{A}_X\otimesf\Idf_{\bar{X}}}{\f{B}_X\otimesf\f{B}_{\bar{X}}} =
\bigskalpHS{\f{A}_X}{\f{B}_X}\Trf(\f{B}_{\bar{X}}) =
\bigskalpHS{\f{A}_X}{\f{B}_X\Trf(\f{B}_{\bar{X}})} =
\bigskalpHS{\f{A}_X}{\Trf_{Y,X}\bigl(\f{B}_X\otimesf\f{B}_{\bar{X}}\bigr)}$,
by using \eqref{eq:TPfHS},
the linearity of the Hilbert--Schmidt inner product \eqref{eq:HS} in its second argument,
and the definition \eqref{eq:TPfPTf}. 
Using the \emph{adjoint action} (defined for unitaries as $\Ad_U(A)=UAU^\dagger$),
the adjoint relationship \eqref{eq:cembfPTfadj} leads to the writing of the fermionic partial trace
\begin{equation}
\label{eq:PTfU}
\Trf_{Y,X}=\Tr_{Y,X}\circ\Ad_{U_{X\bar{X}}^\dagger},
\end{equation}
which is, although less expressive, but slightly simpler than \eqref{eq:PTf}.
(This follows from that $\iotaf_{X,Y}=\Ad_{U_{X\bar{X}}}\circ\iota_{X,Y}$, with the adjoint action of $U_{X\bar{X}}$ given in \eqref{eq:iotaU},
by which we have $\Trf_{Y,X}=\iotaf_{X,Y}^\daggerHS=\iota_{X,Y}^\daggerHS\circ\Ad_{U_{X\bar{X}}}^\daggerHS=\Tr_{Y,X}\circ\Ad_{U_{X\bar{X}}^\dagger}$
by \eqref{eq:cembPTadj} and \eqref{eq:cembfPTfadj}.)
So $\Trf_{Y,X}(\f{A}_Y) = \Phi_X\bigl(\Tr_{Y,X}\bigl(\Phi_Y^{-1}(\f{A}_Y)\bigr)\bigr) = \Tr_{Y,X}\bigl(U_{X\bar{X}}^\dagger \f{A}_Y U_{X\bar{X}}\bigr)$.

Thanks to the unitarity \eqref{eq:LambdaU} of $\f{\Lambda}_{X\bar{X}}$,
the adjoint relationship \eqref{eq:cembfPTfadj} leads to the adjoint of the identity \eqref{eq:Lambdaiotaf},
\begin{subequations}
\begin{equation}
\label{eq:PTfLambda}
 \Trf_{Y,X}\circ\f{\Lambda}_{X\bar{X}}      
=\Trf_{Y,X}\circ\f{\Lambda}_{X\bar{X}}^{-1}
=\Trf_{Y,X}\circ\f{\Lambda}_{\bar{X}X}      
=\Trf_{Y,X}\circ\f{\Lambda}_{\bar{X}X}^{-1} 
=\Trf_{Y,X}.
\end{equation}
(This follows as 
the steps 
$\bigskalpHS{\f{A}_X}{\Trf_{Y,X}(\f{B}_Y)}
= \bigskalpHS{\iotaf_{X,Y}(\f{A}_X)}{\f{B}_Y}
= \bigskalpHS{\f{\Lambda}_{\bar{X}X}^{-1}\bigl(\iotaf_{X,Y}(\f{A}_X)\bigr)}{\f{B}_Y}
= \bigskalpHS{\iotaf_{X,Y}(\f{A}_X)}{\f{\Lambda}_{\bar{X}X}(\f{B}_Y)}
= \bigskalpHS{\f{A}_X}{\Trf_{Y,X}\bigl(\f{\Lambda}_{\bar{X}X}(\f{B}_Y)\bigr)}$,
by using \eqref{eq:cembfPTfadj}, \eqref{eq:Lambdaiotaf}, \eqref{eq:LambdaU} and \eqref{eq:cembfPTfadj} again.
This holds for all $\f{A}_X$ and $\f{B}_Y$,
which leads to that the first term equals to the last in \eqref{eq:PTfLambda}.
The other equalities can be seen similarly.)
In particular, by definition \eqref{eq:Lambda2}, we have 
\begin{equation}
\label{eq:PTffp}
\begin{split}
\Trf_{Y,X}\bigl(\f{A}_X\otimesf \f{B}_{\bar{X}}\bigr)
 = \Trf_{Y,X}\bigl(\f{A}_X\otimesfp\f{B}_{\bar{X}}\bigr)
 = \Trf_{Y,X}\bigl(\f{B}_{\bar{X}}\otimesfp\f{A}_X\bigr)
 = \f{A}_X \Trf(\f{B}_{\bar{X}}),
\end{split}
\end{equation}
\end{subequations}
that is, 
the same fermionic partial trace can be used 
in the fermionic tensorial and also in the (arbitrarily ordered) algebraic representation of products.

We also have 
that for the nested subsystems, or mode subsets, $X\subseteq Y\subseteq Z\subseteq M$,
\begin{subequations}
\begin{align}
\label{eq:PTCons}
\Tr_{Y,X} \circ \Tr_{Z,Y} &= \Tr_{Z,X},\\
\label{eq:PTfCons}
\Trf_{Y,X} \circ \Trf_{Z,Y} &= \Trf_{Z,X}.
\end{align}
\end{subequations}
(The first one is obvious property of the partial trace \eqref{eq:TPPT};
the second one follows from this, by using \eqref{eq:PTf}.)

The effects of the qubit and the fermionic partial trace maps
on the corresponding basis elements \eqref{eq:EY}-\eqref{eq:EYf} are again similar,
\begin{subequations}
\begin{align}
\label{eq:PTEY}
\Tr_{Y,X} (E\indexddu{Y}{\vs{\nu}}{\vs{\nu}'})
&= \delta\indexdu{\vs{\nu}_{\bar{X}}}{\vs{\nu}_{\bar{X}}'}
 E\indexddu{X}{\vs{\nu}_X}{\vs{\nu}_X'},\\
\label{eq:PTfEYf}
\Trf_{Y,X} (\f{E}\indexddu{Y}{\vs{\nu}}{\vs{\nu}'})
&= \delta\indexdu{\vs{\nu}_{\bar{X}}}{\vs{\nu}_{\bar{X}}'}
 \f{E}\indexddu{X}{\vs{\nu}_X}{\vs{\nu}_X'}.
\end{align}
\end{subequations}
(The first one is by \eqref{eq:EYXs} and \eqref{eq:TPPT} and the orthonormalization of the basis in $\mathcal{H}_{\bar{X}}$;
the second one follows from this, by using \eqref{eq:EYEYftraf} and \eqref{eq:PTf},
or by \eqref{eq:EYfXs} and \eqref{eq:TPfPTf}.)

Thanks to the unitarity \eqref{eq:LambdaU},
the relationship \eqref{eq:cembfPTfadj} leads to the adjoint of the second identities in \eqref{eq:cEmbHom} and \eqref{eq:cEmbfHom},
\begin{subequations}
\begin{align}
\Tr_{Y,X} (A_Y^\dagger) &=  \Tr_{Y,X} (A_Y)^\dagger,\\
\Trf_{Y,X} (\f{A}_Y^\dagger) &= \Trf_{Y,X} (\f{A}_Y)^\dagger.
\end{align}
\end{subequations}
(The first one is well-known, it is simply by definition \eqref{eq:TPPT},
exploiting also that
$\bigotimes_{X\in\xi} A_X^\dagger
= \bigl(\bigotimes_{X\in\xi} A_X\bigr)^\dagger$.
The analogue identity does not hold for fermionic tensor products,
there exist operators $\f{A}_X\in\f{\mathcal{A}}_X$ for which
$\bigotimesf_{X\in\xi} \f{A}_X^\dagger
\neq \bigl(\bigotimesf_{X\in\xi} \f{A}_X\bigr)^\dagger$,
so, for the second one, we need to go for a more general proof.
This follows as $\bigskalpHS{\f{B}_X}{\Trf_{Y,X}(\f{A}_Y^\dagger)}
=\bigskalpHS{\iotaf_{X,Y}(\f{B}_X)}{\f{A}_Y^\dagger}
=\bigskalpHS{\f{A}_Y}{\iotaf_{X,Y}(\f{B}_X)^\dagger}
=\bigskalpHS{\f{A}_Y}{\iotaf_{X,Y}(\f{B}_X^\dagger)}
=\bigskalpHS{\Trf_{Y,X}(\f{A}_Y)}{\f{B}_X^\dagger}
=\bigskalpHS{\f{B}_X}{\Trf_{Y,X}(\f{A}_Y)^\dagger}$,
by using 
\eqref{eq:cembfPTfadj},
the second equality in \eqref{eq:cEmbfHom},
 and that $\skalpHS{A}{B}=\skalpHS{B^\dagger}{A^\dagger}$, see \eqref{eq:HS}.
The same derivation without the tildes works also for the qubit case.
Another derivation can be given by the use of \eqref{eq:PTfU}.)

\subsection{Fermionic products of maps}
\label{sec:Tensors.mTP}

Let us have a partition $\xi\in\Pi(Y)$.
In the qubit case, we have
the \emph{tensor product of the linear maps} 
$\Omega_X \in \Lin \mathcal{A}_X$ for all subsystems $X\in\xi$,
defined by 
their joint action on $\xi$-elementary tensors $\bigotimes_{X\in\xi}A_X$ as 
\begin{subequations}
\label{eq:mTPTPs}
\begin{equation}
\label{eq:mTPTP}
\Bigl(\bigotimes_{X\in\xi} \Omega_X\Bigr)\Bigl(\bigotimes_{X\in\xi}A_X\Bigr) = \bigotimes_{X\in\xi}\Omega_X(A_X).
\end{equation}
We would like to have a similar tool for the fermionic case,
so let us define the \emph{fermionic tensor products} of 
the linear maps $\f{\Omega}_X \in \Lin \f{\mathcal{A}}_X$,
given on fermionic $\xi$-elementary tensors and fermionic $\ord{\xi}$-elementary products as
\begin{align}
\label{eq:mTPfTPf}
\Bigl(\bigotimesf_{X\in\xi} \f{\Omega}_X\Bigr)
\Bigl(\bigotimesf_{X\in\xi}\f{A}_X\Bigr) 
&:= \bigotimesf_{X\in\xi}\f{\Omega}_X(\f{A}_X),\\
\label{eq:mTPfpTPfp}
\Bigl(\bigotimesfp_{X\in\ord{\xi}} \f{\Omega}_X\Bigr)
\Bigl( \bigotimesfp_{X\in\ord{\xi}}\f{A}_X \Bigr)
&:= \bigotimesfp_{X\in\ord{\xi}}\f{\Omega}_X(\f{A}_X),
\end{align}
\end{subequations}
acting naturally on the fermionic and algebraic representations of products,
respectively.
Again, with a slight but convenient abuse of the notation, the $\ord{\xi}=\tuple{X_1,X_2,\dots,X_{\abs{\ord{\xi}}}}$ ordering in a product
$\f{\Omega}_{X_1}\otimesfp\f{\Omega}_{X_2}\otimesfp\f{\Omega}_{X_3}\otimesfp\dots$ 
carries now the ordering of the $\ord{\xi}$-elementary product operator on which it is defined to act naturally.
From these, using \eqref{eq:mTPTP}, \eqref{eq:TPf} and \eqref{eq:Lambda}, one can read off the forms
\begin{subequations}
\begin{align}
\label{eq:mTPf}
\bigotimesf_{X\in\xi} \f{\Omega}_X 
&= \f{\Psi}_\xi\circ
\Bigl(\bigotimes_{X\in\xi} \f{\Omega}_X\Bigr)\circ
\f{\Psi}_\xi^{-1}
\equiv \Phi_Y \circ\Bigl(\bigotimes_{X\in\xi} \Phi_X^{-1}\circ\f{\Omega}_X\circ\Phi_X\Bigr)\circ\Phi_Y^{-1},\\
\label{eq:mTPfp}
\bigotimesfp_{X\in\ord{\xi}} \f{\Omega}_X 
&= \f{\Lambda}_{\ord{\xi}} \circ 
\Bigl(\bigotimesf_{X\in\xi} \f{\Omega}_X\Bigr) \circ
\f{\Lambda}_{\ord{\xi}}^{-1}
\equiv \f{\Lambda}_{\ord{\xi}} \circ \f{\Psi}_\xi\circ
\Bigl(\bigotimes_{X\in\xi} \f{\Omega}_X\Bigr)
\circ \f{\Psi}_\xi^{-1} \circ \f{\Lambda}_{\ord{\xi}}^{-1},
\end{align}
see also in the diagram
\begin{equation}
\vcenter{\xymatrix@M+=8bp@C+=32bp{
  \bigotimes_{X\in\xi}\f{A}_X
    \ar@{|->}[r]^{\f{\Psi}_{\xi}}
    \ar@{|->}[d]_{\bigotimes_{X\in\xi} \f{\Omega}_X}         &
 \bigotimesf_{X\in\xi}\f{A}_X 
    \ar@{|->}[r]^{\f{\Lambda}_{\ord{\xi}}}
    \ar@{|->}[d]_{\bigotimesf_{X\in\xi} \f{\Omega}_X}        &
 \bigotimesfp_{X\in\ord{\xi}}\f{A}_X 
    \ar@{|->}[d]_{\bigotimesfp_{X\in\ord{\xi}} \f{\Omega}_X} \\
 \bigotimes_{X\in\xi}\f{\Omega}_X(\f{A}_X) 
    \ar@{|->}[r]^{\f{\Psi}_{\xi}}                            &
 \bigotimesf_{X\in\xi}\f{\Omega}_X(\f{A}_X) 
    \ar@{|->}[r]^{\f{\Lambda}_{\ord{\xi}}}                   &
 \bigotimesfp_{X\in\ord{\xi}}\f{\Omega}_X(\f{A}_X).
}}
\end{equation}
\end{subequations}
In particular, we have that $\bigotimes_{X\in\xi} \Omega_X$ is a \emph{joint extension} of the \emph{unital} (identity preserving) maps $\Omega_X$,
that is,
\begin{subequations}
\begin{equation}
\Bigl(\bigotimes_{X'\in\xi} \Omega_{X'}\Bigr)\bigl(\iota_{X,Y}(A_X)\bigr)=\iota_{X,Y}\bigl(\Omega_X(A_X)\bigr)
\end{equation}
for all $X\in\xi$.
(Note that this does not hold for general non-unital maps,
since the left-hand side is $\Omega_X(A_X)\otimes \bigl(\bigotimes_{X'\in\xi,X'\neq X} \Omega_{X'}(\Id_{X'})\bigr)$,
see \eqref{eq:cEmb} and \eqref{eq:mTPTP}, 
which has to be equal to $\Omega_X(A_X)\otimes\Id_{Y\setminus X}$,
which can always be violated by non-unital maps for which $\Omega_X(\Id_X)\notin \field{C}\Id_X$.)
Analogously, $\bigotimesf_{X\in\ord{\xi}} \f{\Omega}_X$ and
$\bigotimesfp_{X\in\ord{\xi}} \f{\Omega}_X$ are \emph{joint extensions} of the \emph{unital} maps $\f{\Omega}_X$,
\begin{align}
\label{eq:umJExtffp}
 \Bigl( \bigotimesf_{X'\in    {\xi}} \f{\Omega}_{X'}\Bigr)\bigl(\iotaf_{X,Y}(\f{A}_X)\bigr)
=\iotaf_{X,Y}\bigl(\f{\Omega}_X(\f{A}_X)\bigr),\\
\Bigl(\bigotimesfp_{X'\in\ord{\xi}} \f{\Omega}_{X'}\Bigr)\bigl(\iotaf_{X,Y}(\f{A}_X)\bigr)
=\iotaf_{X,Y}\bigl(\f{\Omega}_X(\f{A}_X)\bigr),
\end{align}
\end{subequations}
for all $X\in\xi$.
(The first one can be seen by \eqref{eq:cEmbf} and \eqref{eq:mTPfTPf},
the second one follows from this, by using \eqref{eq:mTPfp} and \eqref{eq:ffpiotaf}.)

Both the usual and the fermionic tensor products of maps are consistent with the refinement of the partitions.
Indeed, let us have $Z\subseteq M$, the partition $\upsilon\in\Pi(Z)$,
and for all parts $Y\in\upsilon$, let us have the partitions $\xi_Y\in\Pi(Y)$ of them,
by which let $\xi\in\Pi(Z)$ be the ``merging'' of the partitions $\xi_Y$,
$\xi:=\bigcup_{Y\in\upsilon}\xi_Y$;
and
let us have 
the ordered partition $\ord{\upsilon}\in\ord{\Pi}(Z)$,
and for all parts $Y\in\ord{\upsilon}$, let us have the ordered partitions $\ord{\xi}_Y\in\ord{\Pi}(Y)$ of them,
by which let $\ord{\xi}\in \ord{\Pi}(Z)$ be the ``ordered merging'' of the ordered partitions $\ord{\xi}_Y$,
$\ord{\xi}:= \ord{\bigcup}_{Y\in\ord{\upsilon}} \ord{\xi}_Y$;
then we have the \emph{associativity} for the products of maps as
\begin{subequations}
\label{eq:mTPGenAssoc}
\begin{align}
\label{eq:mTPAssoc}
\bigotimes_{Y\in\upsilon} \Bigl( \bigotimes_{X_Y\in\xi_Y} \Omega_{X_Y} \Bigr) 
 &= \bigotimes_{X\in\xi} \Omega_X,\\
\label{eq:mTPfAssoc}
\bigotimesf_{Y\in\upsilon}\Bigl( \bigotimesf_{X_Y\in\xi_Y} \f{\Omega}_{X_Y}\Bigr)
 &= \bigotimesf_{X\in\xi} \f{\Omega}_X,\\
\label{eq:mTPfpAssoc}
\bigotimesfp_{Y\in\ord{\upsilon}} \Bigl( \bigotimesfp_{X_Y\in\ord{\xi}_Y} \f{\Omega}_{X_Y}\Bigr)
 &= \bigotimesfp_{X\in\ord{\xi}} \f{\Omega}_X.
\end{align}
\end{subequations}
(The first one is a property of the tensor product of maps,
following from the associativity \eqref{eq:TPAssoc} and the definition \eqref{eq:mTPTP};
the second one follows from \eqref{eq:TPfAssoc} and \eqref{eq:mTPfTPf},
the third one follows from \eqref{eq:TPfpAssoc} and \eqref{eq:mTPfpTPfp} in exactly the same way,
i.e., it is easy to check by acting on fermionic $\xi$-elementary tensors $\bigotimesf_{X\in\xi}\f{A}_X$
and fermionic $\ord{\xi}$-elementary products $\bigotimesfp_{X\in\ord{\xi}}\f{A}_X$.)
This justifies the use of the ``associative'' tensor product signs, that is, the ordering in which the product is performed is arbitrary.
(Note that the ordering in $\ord{\xi}$ is the ordering of the product of operators it acts on.)
We also have that the identity maps are
$\bigotimes_{X\in\xi}\IId_X = \IId_Y$,
$\bigotimesf_{X\in\xi}\IIdf_X = \IIdf_Y$
 and
$\bigotimesfp_{X\in\ord{\xi}}\IIdf_X = \IIdf_Y$.

Contrary to the level of the algebras,
on the level of the maps, the tensor product obeys
not only the linear 
but also the $*$-algebraic structure also in the fermionic cases, 
that is,
\begin{subequations}
\label{eq:mTPGenHom}
\begin{align}
\label{eq:mTPHom}
\Bigl(\bigotimes_{X\in\xi} \Omega_X\Bigr)\circ\Bigl(\bigotimes_{X\in\xi}\Xi_X\Bigr)
 &= \bigotimes_{X\in\xi}\bigl(\Omega_X\circ\Xi_X\bigr),&
 \bigotimes_{X\in\xi}     \Omega_X^\daggerHS &= \Bigl( \bigotimes_{X\in\xi}     \Omega_X\Bigr)^\daggerHS,\\
\label{eq:mTPfHom}
\Bigl(\bigotimesf_{X\in\xi} \f{\Omega}_X\Bigr)\circ\Bigl(\bigotimesf_{X\in\xi}\f{\Xi}_X\Bigr)
 &= \bigotimesf_{X\in\xi}\bigl(\f{\Omega}_X\circ\f{\Xi}_X\bigr),&
\bigotimesf_{X\in\xi} \f{\Omega}_X^\daggerHS &= \Bigl(\bigotimesf_{X\in\xi} \f{\Omega}_X\Bigr)^\daggerHS,\\
\label{eq:mTPfpHom}
\Bigl(\bigotimesfp_{X\in\ord{\xi}} \f{\Omega}_X\Bigr)\circ\Bigl(\bigotimesfp_{X\in\ord{\xi}}\f{\Xi}_X\Bigr)
 &= \bigotimesfp_{X\in\ord{\xi}}\bigl(\f{\Omega}_X\circ\f{\Xi}_X\bigr),&
\bigotimesfp_{X\in\ord{\xi}} \f{\Omega}_X^\daggerHS &= \Bigl(\bigotimesfp_{X\in\ord{\xi}} \f{\Omega}_X\Bigr)^\daggerHS.
\end{align}
\end{subequations}
(The first composition and adjoint identities
are properties of the tensor product, following from \eqref{eq:mTPTP} and \eqref{eq:TPHS};
the second ones follow from \eqref{eq:mTPfTPf} and \eqref{eq:TPfHS}
the third ones follow from \eqref{eq:mTPfpTPfp} and \eqref{eq:TPfpHS} in exactly the same way,
i.e., it is easy to check by acting on fermionic $\xi$-elementary tensors $\bigotimesf_{X\in\xi}\f{A}_X$
and $\ord{\xi}$-elementary products $\bigotimesfp_{X\in\ord{\xi}}\f{A}_X$.)
In particular, by \eqref{eq:mTPfHom} and \eqref{eq:mTPfpHom} we have,
similarly to 
\begin{subequations}
\begin{align}
\label{eq:noMapLambda}
\prod_{X\in\xi}\bigl(\Omega_X\otimes\IId_{\bar{X}}\bigr)
&= \bigotimes_{X\in\xi}\Omega_X 
\intertext{in the qubit case, that}
\label{eq:noMapLambdaf}
\prod_{X\in\xi}\bigl(\f{\Omega}_X\otimesf\IIdf_{\bar{X}}\bigr)
&= \bigotimesf_{X\in\xi}\f{\Omega}_X,\\
\label{eq:noMapLambdafp}
\prod_{X\in\xi}\bigl(\f{\Omega}_X\otimesfp_{\ord{\xi}}\IIdf_{\bar{X}}\bigr)
&= \bigotimesfp_{X\in\ord{\xi}}\f{\Omega}_X,
\end{align}
\end{subequations}
(where $\prod$ stands for composition),
therefore compositions of extensions from disjoint mode subsets are commutative,
so there is no need for a \eqref{eq:Lambda}-kind of map on this level.
(Again, the notation $\ord{\xi}$ is only about
the ordering of the product it acts on.)

Note again that 
the ordered partition in \eqref{eq:mTPfpTPfp} is a property of $\otimesfp$,
it is about the ordering of the product it acts on,
not about the ordering of the composition of the extended maps, which does not matter, as we have seen in \eqref{eq:noMapLambdafp}.
Also, it is important to see that,
although it was easy to construct the fermionic product \eqref{eq:mTPfp} so that
it holds for a fixed ordering, see \eqref{eq:mTPfpTPfp},
this could not be done for arbitrary ordering at the same time,
as there exist operators $\f{A}_X\in\f{\mathcal{A}}_X$ and maps $\f{\Omega}_X\in\f{\mathcal{B}}_X$ such that
$\bigl(\bigotimesfp_{X\in\ord{\xi}} \f{\Omega}_X\bigr)
\bigl( \bigotimesfp_{X\in\ord{\xi}'}\f{A}_X \bigr)
\neq \bigotimesfp_{X\in\ord{\xi}'}\f{\Omega}_X(\f{A}_X)$ 
for $\ord{\xi}'$ ordered differently than $\ord{\xi}$
(unless the parity superselection rule is imposed, see \eqref{eq:TPSA} in Section~\ref{sec:ParityPhys.TPSA},
after which this is illustrated in Section~\ref{sec:ParityPhys.Loc}).

The constructions presented in this section
also illustrate that the definition of the map product in the fermionic case is somewhat arbitrary,
since there is no tensor product structure on $\f{\mathcal{A}}_Y$, which would provide a natural, canonical one,
which would be working in the same way for arbitrarily ordered products.
We already have two products, \eqref{eq:mTPfTPf} and \eqref{eq:mTPfpTPfp},
and the latter is different for the different orderings by which it is defined.
Properties necessary to have something useful are given in \eqref{eq:mTPGenAssoc} and \eqref{eq:mTPGenHom}.

\subsection{Fermionic embeddings of maps}
\label{sec:Tensors.mcEmb}

In the qubit case, we also have the
\emph{canonical embedding of maps} with respect to the tensor product of maps,
for the subsystems $X\subseteq Y$, as
\begin{subequations}
\label{eq:mapInjs}
\begin{equation}
\label{eq:mapInj}
\begin{split}
\iotabb_{X,Y} :\quad 
\Lin\mathcal{A}_X \quad&\longrightarrow\quad \Lin\mathcal{A}_Y,\\
\Omega_X \quad&\longmapsto\quad \Omega_X \otimes \IId_{\bar{X}},
\end{split}
\end{equation}
using the notation $\bar{X} = Y\setminus X$.
We would like to have a similar tool in the fermionic case,
so let us define the \emph{fermionic embeddings of maps} as
\begin{align}
\label{eq:mapInjf}
\begin{split}
\iotabbf_{X,Y} :\quad 
\Lin\f{\mathcal{A}}_X \quad&\longrightarrow\quad \Lin\f{\mathcal{A}}_Y,\\
\f{\Omega}_X \quad&\longmapsto\quad \f{\Omega}_X \otimesf \IIdf_{\bar{X}},
\end{split}\\
\label{eq:mapInjfp}
\begin{split}
\iotabbfp_{X,Y} :\quad 
\Lin\f{\mathcal{A}}_X \quad&\longrightarrow\quad \Lin\f{\mathcal{A}}_Y,\\
\f{\Omega}_X \quad&\longmapsto\quad \f{\Omega}_X \otimesfp \IIdf_{\bar{X}} = 
\f{\Lambda}_{X\bar{X}}\circ\bigl(\f{\Omega}_X \otimesf \IIdf_{\bar{X}}\bigr)\circ\f{\Lambda}_{X\bar{X}}^{-1}.
\end{split}
\end{align}
\end{subequations}
Then we have that 
\begin{subequations}
\begin{align}
\label{eq:cmEmb}
\iotabb_{X,Y}(\Omega_X)\bigl(A_X\otimes B_{\bar{X}}\bigr) &= \Omega_X(A_X)\otimes B_{\bar{X}},\\
\label{eq:cmEmbf}
\iotabbf_{X,Y}(\f{\Omega}_X)\bigr(\f{A}_X\otimesf \f{B}_{\bar{X}}\bigr) &= \f{\Omega}_X(\f{A}_X)\otimesf \f{B}_{\bar{X}},\\
\label{eq:cmEmbfp}
\iotabbfp_{X,Y}(\f{\Omega}_X)\bigl(\f{A}_X\otimesfp\f{B}_{\bar{X}}\bigr)
&= \f{\Omega}_X(\f{A}_X)\otimesfp\f{B}_{\bar{X}}.
\end{align}
(These follow from \eqref{eq:mTPTP}, \eqref{eq:mTPfTPf} and \eqref{eq:mTPfpTPfp}, respectively.)
One can see this also in the diagram
\begin{equation}
\vcenter{\xymatrix@M+=8bp@C+=32bp{
  \f{A}_X\otimes\f{B}_{\bar{X}} 
    \ar@{|->}[r]^{\f{\Psi}_{X\bar{X}}}
    \ar@{|->}[d]_{\iotabb_{X,Y}(\f{\Omega}_X)}      &
  \f{A}_X\otimesf\f{B}_{\bar{X}} 
    \ar@{|->}[r]^{\f{\Lambda}_{X\bar{X}}}
    \ar@{|->}[d]_{\iotabbf_{X,Y}(\f{\Omega}_X)}     &
  \f{A}_X\otimesfp\f{B}_{\bar{X}}
    \ar@{|->}[d]_{\iotabbfp_{X,Y}(\f{\Omega}_X)}    \\
  \f{\Omega}_X(\f{A}_X)\otimes \f{B}_{\bar{X}} 
    \ar@{|->}[r]^{\f{\Psi}_{X\bar{X}}}              &
  \f{\Omega}_X(\f{A}_X)\otimesf \f{B}_{\bar{X}} 
    \ar@{|->}[r]^{\f{\Lambda}_{X\bar{X}}}           &
  \f{\Omega}_X(\f{A}_X)\otimesfp \f{B}_{\bar{X}}.
}}
\end{equation}
\end{subequations}
In particular, we have that these are \emph{extensions} of maps, that is,
\begin{subequations}
\begin{align}
\label{eq:cmEmbcEmb}
\iotabb_{X,Y}(\f{\Omega}_X)  \bigl(\iota_{X,Y}(A_X)\bigr)  &= \iota_{X,Y} \bigl(\f{\Omega}_X(A_X)\bigr),\\
\label{eq:cmEmbfcEmbf}
\iotabbf_{X,Y}(\f{\Omega}_X) \bigl(\iotaf_{X,Y}(\f{A}_X)\bigr) &= \iotaf_{X,Y}\bigl(\f{\Omega}_X(\f{A}_X)\bigr),\\
\label{eq:cmEmbfpcEmbf}
\iotabbfp_{X,Y}(\f{\Omega}_X)\bigl(\iotaf_{X,Y}(\f{A}_X)\bigr) &= \iotaf_{X,Y}\bigl(\f{\Omega}_X(\f{A}_X)\bigr).
\end{align}
\end{subequations}
(The first two follow from \eqref{eq:cmEmb} and \eqref{eq:cmEmbf},
 with the definitions \eqref{eq:cEmb} and \eqref{eq:cEmbf}, respectively,
the third one follows from the second, by \eqref{eq:Lambdaiotaf} or \eqref{eq:ffpiotaf}.)
These are also \emph{identical extensions}, if the maps are unital, that is,
\begin{subequations}
\begin{align}
\label{eq:cmEmbcEmbi}
\iotabb_{X,Y}(\f{\Omega}_X)  \bigl(\iota_{\bar{X},Y}(B_{\bar{X}})\bigr)  &= \iota_{\bar{X},Y} (B_{\bar{X}}),\\
\label{eq:cmEmbfcEmbif}
\iotabbf_{X,Y}(\f{\Omega}_X) \bigl(\iotaf_{\bar{X},Y}(\f{B}_{\bar{X}})\bigr) &= \iotaf_{\bar{X},Y}(\f{B}_{\bar{X}}),\\
\label{eq:cmEmbfpcEmbif}
\iotabbfp_{X,Y}(\f{\Omega}_X)\bigl(\iotaf_{\bar{X},Y}(\f{B}_{\bar{X}})\bigr) &= \iotaf_{\bar{X},Y}(\f{B}_{\bar{X}}).
\end{align}
\end{subequations}
(These follow similarly.)

Also, we have that
for the nested subsystems, or mode subsets $X\subseteq Y\subseteq Z\subseteq M$,
\begin{subequations}
\label{eq:cmEmbGenCons}
\begin{align}
\label{eq:cmEmbCons}
\iotabb_{Y,Z}   \circ \iotabb_{X,Y}   &= \iotabb_{X,Z},  \\
\label{eq:cmEmbfCons}
\iotabbf_{Y,Z}  \circ \iotabbf_{X,Y}  &= \iotabbf_{X,Z}, \\
\label{eq:cmEmbfpCons}
\iotabbfp_{Y,Z} \circ \iotabbfp_{X,Y} &= \iotabbfp_{X,Z}.
\end{align}
\end{subequations}
(These follow from the associativity properties \eqref{eq:mTPAssoc}, \eqref{eq:mTPfAssoc}  and \eqref{eq:mTPfpAssoc}
in the three cases, respectively.)

It easily follows that
the map embeddings are $*$-homomorphisms,
\begin{subequations}
\label{eq:cmEmbGenHom}
\begin{align}
\label{eq:cmEmbHom}
\iotabb_{X,Y}(\Omega_X)\circ\iotabb_{X,Y}(\Xi_X) &= \iotabb_{X,Y}\bigl(\Omega_X\circ\Xi_X\bigr),&\qquad
\iotabb_{X,Y}(\Omega_X^\daggerHS) &= \iotabb_{X,Y}(\Omega_X)^\daggerHS,\\
\label{eq:cmEmbfHom}
\iotabbf_{X,Y}(\f{\Omega}_X)\circ\iotabbf_{X,Y}(\f{\Xi}_X) &= \iotabbf_{X,Y}\bigl(\f{\Omega}_X\circ\f{\Xi}_X\bigr),&\qquad
\iotabbf_{X,Y}(\f{\Omega}_X^\daggerHS) &= \iotabbf_{X,Y}(\f{\Omega}_X)^\daggerHS,\\
\label{eq:cmEmbfpHom}
\iotabbfp_{X,Y}(\f{\Omega}_X)\circ\iotabbfp_{X,Y}(\f{\Xi}_X) &= \iotabbfp_{X,Y}\bigl(\f{\Omega}_X\circ\f{\Xi}_X\bigr),&\qquad
\iotabbfp_{X,Y}(\f{\Omega}_X^\daggerHS) &= \iotabbfp_{X,Y}(\f{\Omega}_X)^\daggerHS.
\end{align}
\end{subequations}
(The composition and adjoint identities follow from the corresponding identities in
 \eqref{eq:mTPHom}, \eqref{eq:mTPfHom} and \eqref{eq:mTPfpHom} in the three cases, respectively.)

A quick exercise is to show that the partial traces are the embeddings of the traces,
\begin{subequations}
\begin{align}
\label{eq:PTemb}
\Tr_{Y,X}   &= \Tr \otimes\IId_X 
 \equiv \iotabb_{\bar{X},Y}(\Tr),\\
\label{eq:PTfemb}
\Trf_{Y,X}  &= \Trf \otimesf\IIdf_X
 \equiv \iotabbf_{\bar{X},Y}(\Trf),\\
\label{eq:PTfpemb}
\Trf_{Y,X}  &= \Trf \otimesfp\IIdf_X
 \equiv \iotabbfp_{\bar{X},Y}(\Trf).
\end{align}
\end{subequations}
(Indeed, 
the first one is obvious;
the second one follows from this, by using \eqref{eq:mTPf} as
$\Trf \otimesf\IIdf_X 
= \Phi_{X}
  \circ\bigl( \Phi_\emptyset^{-1}\circ\Trf\circ\Phi_{\bar{X}} \otimes \Phi_X^{-1}\circ\IIdf_X\circ\Phi_X\bigr)
  \circ\Phi_Y^{-1}
= \Phi_{X}\circ\bigl(\Tr\otimes\IId_X\bigr)\circ\Phi_Y^{-1} 
= \Phi_{X}\circ\Tr_{Y,X}\circ\Phi_Y^{-1}=\Trf_{Y,X}$, with the definition \eqref{eq:PTf};
the third one follows from this by \eqref{eq:mTPfp}.)

Note that while $\iotabb_{X,Y}(\Omega_X)$ is a \emph{strong extension} of $\Omega_X$,
that is, 
\begin{subequations}
\begin{align}
\label{eq:mIdExt1}
\iotabb_{X,Y}(\Omega_X)\bigl(\iota_{X,Y}(A_X)\iota_{\bar{X},Y}(B_{\bar{X}})\bigr)
&= \Bigl(\iotabb_{X,Y}(\Omega_X)\bigl(\iota_{X,Y}(A_X)\bigr)\Bigr)\iota_{\bar{X},Y}(B_{\bar{X}}), \\
\label{eq:mIdExt2}
\iotabb_{X,Y}(\Omega_X)\bigl(\iota_{\bar{X},Y}(B_{\bar{X}})\iota_{X,Y}(A_X)\bigr)
&= \iota_{\bar{X},Y}(B_{\bar{X}})\Bigl(\iotabb_{X,Y}(\Omega_X)\bigl(\iota_{X,Y}(A_X)\bigr)\Bigr),
\end{align}
\end{subequations}
because of
$\iota_{X,Y}(A_X)\iota_{\bar{X},Y}(B_{\bar{X}})
=\iota_{\bar{X},Y}(B_{\bar{X}})\iota_{X,Y}(A_X)
=A_X\otimes B_{\bar{X}}$;
this is not the case for 
$\iotabbfp_{X,Y}(\f{\Omega}_X)$,
we have only
\begin{subequations}
\label{eq:mIdExtfp}
\begin{align}
\label{eq:mIdExtfp1}
\iotabbfp_{X,Y}(\f{\Omega}_X)\bigl(\iotaf_{X,Y}(\f{A}_X)\iotaf_{\bar{X},Y}(\f{B}_{\bar{X}})\bigr)
&=    \Bigl(\iotabbfp_{X,Y}(\f{\Omega}_X)\bigl(\iota_{X,Y}(\f{A}_X)\bigr)\Bigr)\iotaf_{\bar{X},Y}(\f{B}_{\bar{X}}), \\
\intertext{but there exist operators $\f{A}_X\in\f{\mathcal{A}}_X$ and $\f{B}_{\bar{X}}\in\f{\mathcal{A}}_{\bar{X}}$ such that}
\label{eq:mIdExtfp2}
\iotabbfp_{X,Y}(\f{\Omega}_X)\bigl(\iotaf_{\bar{X},Y}(\f{B}_{\bar{X}})\iotaf_{X,Y}(\f{A}_X)\bigr)
&\neq \iotaf_{\bar{X},Y}(\f{B}_{\bar{X}})\Bigl(\iotabbfp_{X,Y}(\f{\Omega}_X)\bigl(\iotaf_{X,Y}(\f{A}_X)\bigr)\Bigr).
\end{align}
\end{subequations}
It is important to see that 
although it was easy to construct the fermionic map embedding \eqref{eq:cmEmbfp} so that
 the first part \eqref{eq:mIdExtfp1} of the definition of strong extensions holds \eqref{eq:cmEmbfp},
this could not be done for the second part \eqref{eq:mIdExtfp2} at the same time
(unless the parity superselection rule is imposed, see \eqref{eq:TPSA} in Section~\ref{sec:ParityPhys.TPSA},
after which this is illustrated in Section~\ref{sec:ParityPhys.Loc}).
This is an important point, 
because the notion of \emph{locality} is given by \emph{strong extensions of maps}.
We will turn back to locality in Section~\ref{sec:ParityPhys.Loc}.
Note that for $\iotabbf_{X,Y}(\f{\Omega}_X)$, acting on fermionic tensor products, we have
$\iotabbf_{X,Y}(\f{\Omega}_X)\bigl(\f{A}_X\otimesf \f{B}_{\bar{X}}\bigr)= \bigl(\f{\Omega}_X(\f{A}_X)\bigr)\otimesf \f{B}_{\bar{X}}$,
from which no \eqref{eq:mIdExtfp}-like property follows.

The constructions presented in this section
also illustrate that the definition of the map embedding in the fermionic case is somewhat arbitrary,
since there is no tensor product structure on $\f{\mathcal{A}}_Y$, which would provide a natural, canonical embedding.
We already have two embeddings, \eqref{eq:mapInjf} and \eqref{eq:mapInjfp},
and the latter could even be defined in a reverse way, for which \eqref{eq:mIdExtfp2} holds instead of \eqref{eq:mIdExtfp1}.
Properties necessary to have something useful are given in \eqref{eq:cmEmbGenCons} and \eqref{eq:cmEmbGenHom}.

\section{States and reduced states}
\label{sec:States}

In the previous sections, 
we gave some general tools for calculations of second quantized fermionic systems.
In this section, we turn to quantum states,
and the central notion we have to consider is \emph{positivity}, being so important in life,
\begin{subequations}
\begin{align}
\label{eq:posdef}
R\geq0
&\arriff \forall\cket{\psi}\in\mathcal{H}: \bra{\psi}R\cket{\psi}\geq0\\
\label{eq:posabssquare}
&\arriff \exists A : R = A^\dagger A\\
\label{eq:posHS}
&\arriff \forall A\geq0: \skalpHS{R}{A}\geq0\\
\label{eq:possquare}
&\arriff \exists A=A^\dagger : R = A^2.
\end{align}
\end{subequations}
(The first one is the standard definition of positivity, 
the other ones are easy to prove.)
From \eqref{eq:posabssquare}, it easily follows that positivity is preserved by $*$-homomorphisms.

\subsection{States}
\label{sec:States.states}

The \emph{state} \cite{Bratteli-s} of a quantum system is given 
by a linear functional $\rho\in\mathcal{A}^*$ on the algebra $\mathcal{A}$
containing the observables of the system (as normal elements),
which is positive $\rho(A^\dagger A) \geq 0$ and normalized $\rho(\Id) = 1$,
expressing the \emph{expectation value} of the operators $A\in\mathcal{A}$ as
$\bracket{A} = \rho(A)$.
For the \emph{finite dimensional} Hilbert space $\mathcal{A}$,
this functional can be given by the inner product \eqref{eq:HS} with an operator, the \emph{density operator} $\rho\in\mathcal{A}$,
which is positive $\rho^\dagger = \rho \geq 0$ and normalized $\Tr(\rho) = 1$,
by which 
\begin{equation}
\label{eq:expectation}
\bracket{A} = \rho(A) = \skalpHS{\rho}{A} = \Tr(\rho^\dagger A).
\end{equation}
For simplicity, we always think of states in this latter sense,
and we do not distinguish states and density operators in writing.
Let us denote the \emph{set of states} as
\begin{equation}
\label{eq:state}
\mathcal{D} := \bigsset{\rho\in\mathcal{A}}{\rho^\dagger = \rho \geq 0, \Tr(\rho) = 1}
\isom \bigsset{\rho\in\mathcal{A}^*}{\rho(A^\dagger A) \geq 0, \rho(\Id) = 1}.
\end{equation}
It is a convex set.
A state is \emph{pure}, if and only if $\rho^2=\rho$ holds for the density operator,
which leads to the form $\rho=\cket{\psi}\bra{\psi}$,
with state vectors, $\norm{\psi}=1$.

\subsection{Qubit reduction: operators}
\label{sec:States.red}

We expect that 
for all nested subsystems $X\subseteq Y\subseteq M$,
for all operators $A_X \in \mathcal{A}_X$ of subsystem $X$,
we have
$\bracket{A_X} = \bracket{\iota_{X,Y}(A_X)}$.
That is, the reduced state of a subsystem is the one
which gives the same expectation for the operators of the subsystem
as the state of the larger system
for the operators of the subsystem considered as a part of the larger system \cite{Petz-2008}.
Expressing this with the Hilbert--Schmidt inner product \eqref{eq:HS}
and the state $\rho_Y\in\mathcal{D}_Y := \sset{\rho_Y\in\mathcal{A}_Y}{\rho_Y^\dagger = \rho_Y \geq 0, \Tr(\rho_Y) = 1}$, we have
that the reduced state $\rho_X\in\mathcal{D}_X$ is given by the partial trace \eqref{eq:TPPT},
as is well-known,
\begin{equation}
\label{eq:defreduced}
\Bigl(\forall A_X\in\mathcal{A}_X:\quad
     \skalpHS{\rho_X}{A_X}
= \bigskalpHS{\rho_Y}{\iota_{X,Y}(A_X)} \Bigr)
\quad\Longleftrightarrow\quad \rho_X = \Tr_{Y,X}(\rho_Y).
\end{equation}

The usual textbook derivation of this, 
by writing $\rho_Y = \sum_r B_{X,r}\otimes C_{\bar{X},r}$,
with the notation $\bar{X} = Y\setminus X$,
and using the linearity of the Hilbert--Schmidt inner product \eqref{eq:HS}, is
\begin{equation*}
\begin{split}
\bigskalpHS{\rho_Y}{A_X\otimes \Id_{\bar{X}}}
&\equals \sum_r\Tr\bigl( (B_{X,r}\otimes C_{\bar{X},r})^\dagger(A_X\otimes \Id_{\bar{X}}) \bigr) \\
&\equals \sum_r\Tr\bigl( (B_{X,r}^\dagger A_X) \otimes C_{\bar{X},r}^\dagger\bigr)\\
&\equalsref{eq:TPHS} \sum_r\Tr(B_{X,r}^\dagger A_X)\Tr(C_{\bar{X},r}^\dagger)\\
&\equals \Tr \Bigl(\Bigl(\sum_r B_{X,r}^\dagger\Tr(C_{\bar{X},r}^\dagger)\Bigr)A_X\Bigr),
\end{split}
\end{equation*}
which leads to
\begin{equation}
\forall A_X\in\mathcal{A}_X: \quad
\Tr(\rho_X^\dagger A_X) = \Tr \Bigl(\Bigl(\sum_r B_{X,r}\Tr(C_{\bar{X},r})\Bigr)^\dagger A_X\Bigr),
\end{equation}
which, because the Hilbert--Schmidt inner product is nondegenerate, leads to
\begin{equation}
\rho_X = \sum_r B_{X,r}\Tr(C_{\bar{X},r}) 
= \Tr_{Y,X} \Bigl(\sum_r B_{X,r}\otimes C_{\bar{X},r}\Bigr) = \Tr_{Y,X}(\rho_Y).
\end{equation}

A more elevated derivation of this can be presented by noting that
we started from the canonical embedding \eqref{eq:cEmb} of the operators of subsystems into larger subsystems,
then the left-hand side of \eqref{eq:defreduced} expresses that the
state reduction is the adjoint map of the canonical embedding,
which is known to be the partial trace \eqref{eq:cembPTadj},
$\bigskalpHS{\Tr_{Y,X}(\rho_Y)}{A_X}= \bigskalpHS{\rho_Y}{\iota_{X,Y}(A_X)}$
for all $A_X\in\mathcal{A}_X$ and $\rho_Y\in\mathcal{A}_Y$ (since $\mathcal{A}_Y$ is spanned by $\mathcal{D}_Y$).
On the other hand, the canonical embedding \eqref{eq:cEmb} is natural, when there exists a tensor product structure.

\subsection{Fermionic reduction: operators}
\label{sec:States.redf}

The states are linear functionals on the algebra containing the observables of the system (as normal elements).
In the fermionic case, not all the (normal) elements of the operator algebra $\f{\mathcal{A}}_Y$ are observable ones,
and, strictly speaking, the states act only on the (sub)algebra containing the observables, also called physical subalgebra
(see in Sections~\ref{sec:ParityMath} and \ref{sec:ParityPhys}).
In what follows, first we formulate the construction for the whole operator algebra $\f{\mathcal{A}}_Y$,
and we make the restriction to the physical subalgebra later (see in Section~\ref{sec:ParityPhys}).

Similarly to the qubit case,
we expect that for all nested mode subsets $X\subseteq Y\subseteq M$,
for all operators $\f{A}_X\in\f{\mathcal{A}}_X$ of mode subset $X$,
we have $\bracket{\f{A}_X} = \bracket{\iotaf_{X,Y}(\f{A}_X)}$.
Expressing this with the Hilbert--Schmidt inner product \eqref{eq:HS}
and the state $\f{\rho}_Y\in\f{\mathcal{D}}_Y := \sset{\f{\rho}_Y\in\f{\mathcal{A}}_Y}{\f{\rho}_Y^\dagger = \f{\rho}_Y \geq 0, \Tr(\f{\rho}_Y) = 1}$, we have
the definition for the reduced state $\f{\rho}_X\in\f{\mathcal{D}}_X$ as
\begin{equation}
\label{eq:defreducedf}
\forall \f{A}_X\in\f{\mathcal{A}}_X: \quad
  \bigskalpHS{\f{\rho}_X}{ \f{A}_X }
= \bigskalpHS{\f{\rho}_Y}{ \iotaf_{X,Y}(\f{A}_X) }.
\end{equation}

Using the tools we have constructed in Section~\ref{sec:Tensors},
now the simple steps
\begin{equation*}
\begin{split}
\bigskalpHS{\f{\rho}_Y}{ \iotaf_{X,Y}(\f{A}_X) }
&\equalsref{eq:cEmbf}   \bigskalpHS{\f{\rho}_Y}{ \f{A}_X \otimesf \Idf_{\bar{X}} }\\
&\equalsref{eq:TPf}    \bigskalpHS{\f{\rho}_Y}{ \Phi_Y\bigl(\Phi_X^{-1}(\f{A}_X) \otimes \Phi_{\bar{X}}^{-1}(\Idf_{\bar{X}})\bigr) }\\
&\equalsref{eq:PhiU}      \bigskalpHS{\Phi_Y^{-1}(\f{\rho}_Y)}{ \Phi_X^{-1}(\f{A}_X) \otimes \Id_{\bar{X}} }\\
&\equalsref{eq:defreduced} \bigskalpHS{\Tr_{Y,X}\bigl(\Phi_Y^{-1}(\f{\rho}_Y)\bigr)}{ \Phi_X^{-1}(\f{A}_X) }\\
&\equalsref{eq:PhiU} \bigskalpHS{\Phi_X\bigr(\Tr_{Y,X}\bigl(\Phi_Y^{-1}(\f{\rho}_Y)\bigr)\bigr)}{ \f{A}_X }\\
&\equalsref{eq:PTf}  \bigskalpHS{\Trf_{Y,X}(\f{\rho}_Y)}{ \f{A}_X }
\end{split}
\end{equation*}
lead to
\begin{equation}
\forall \f{A}_X\in\f{\mathcal{A}}_X: \quad
  \bigskalpHS{\f{\rho}_X}{ \f{A}_X }
= \bigskalpHS{\Trf_{Y,X}(\f{\rho}_Y)}{ \f{A}_X },
\end{equation}
which, because the Hilbert--Schmidt inner product is nondegenerate, leads to
that the reduced state can be obtained by the use of the fermionic partial trace \eqref{eq:PTf},
\begin{equation}
\label{eq:reducedfPTf}
\f{\rho}_X  = \Trf_{Y,X}(\f{\rho}_Y).
\end{equation}
Note that the same derivation as in the qubit case would not work,
since for the fermionic tensor product \eqref{eq:TPf},
there exist operators $\f{B}_{X,r},\f{A}_X\in\f{\mathcal{A}}_X$ and $\f{C}_{\bar{X},r}\in\f{\mathcal{A}}_{\bar{X}}$ for the mode subsets
such that
 $(\f{B}_{X,r}\otimesf \f{C}_{\bar{X},r})(\f{A}_X\otimesf \Idf_{\bar{X}})\neq (\f{B}_{X,r}\f{A}_X) \otimesf \f{C}_{\bar{X},r}$
(see in Section~\ref{sec:Tensors.Lambda}).
Quite the contrary, the result \eqref{eq:defreduced} of the qubit case was used here.

A more elevated derivation of this can be presented by noting that
we started from the fermionic canonical embedding \eqref{eq:cEmbf} of the operators of modes into larger sets of modes,
then equation \eqref{eq:defreducedf} expresses that the
fermionic state reduction is the adjoint map of the fermionic canonical embedding,
which is known to be the fermionic partial trace \eqref{eq:cembfPTfadj},
$\bigskalpHS{\Trf_{Y,X}(\f{\rho}_Y)}{\f{A}_X}= \bigskalpHS{\f{\rho}_Y}{\iotaf_{X,Y}(\f{A}_X)}$
for all $\f{A}_X\in\f{\mathcal{A}}_X$ and $\f{\rho}_Y\in\f{\mathcal{A}}_Y$ (since $\f{\mathcal{A}}_Y$ is spanned by $\f{\mathcal{D}}_Y$).
Although choosing an embedding is rather arbitrary when there is no tensor product structure,
the fermionic canonical embedding \eqref{eq:cEmbf} is natural in the sense of \eqref{eq:GammafTPf} and \eqref{eq:cEmbfmotiv}.

\subsection{Fermionic reduction: matrices}
\label{sec:States.redfmx}

For illustrational, as well as practical (symbolical or numerical) purposes, 
one may want to expand the \emph{density operators} $\f{\rho}_Y\in\f{\mathcal{D}}_Y$
in the standard and the fermionic bases \eqref{eq:EY}-\eqref{eq:EYf},
using the Hilbert--Schmidt inner product \eqref{eq:HS}, as
\begin{equation}
\label{eq:rhoYexp}
\f{\rho}_Y
 = \sum_{\vs{\nu},\vs{\nu}'}
  \bigskalpHS{E\indexddu{Y}{\vs{\nu}}{\vs{\nu}'}}{\f{\rho}_Y}
   E\indexddu{Y}{\vs{\nu}}{\vs{\nu}'}
 = \sum_{\vs{\nu},\vs{\nu}'}
  \bigskalpHS{\f{E}\indexddu{Y}{\vs{\nu}}{\vs{\nu}'}}{\f{\rho}_Y}
   \f{E}\indexddu{Y}{\vs{\nu}}{\vs{\nu}'},
\end{equation}
leading to the standard and the fermionic \emph{density matrices} of the fermionic state
\begin{subequations}
\begin{align}
\label{eq:rhoYstd}
R\indexdud{Y}{\vs{\nu}}{\vs{\nu}'}
&:=
   \bigskalpHS{E\indexddu{Y}{\vs{\nu}}{\vs{\nu}'}}{\f{\rho}_Y},\\
\label{eq:rhoYferm}
\f{R}\indexdud{Y}{\vs{\nu}}{\vs{\nu}'}
&:= 
   \bigskalpHS{\f{E}\indexddu{Y}{\vs{\nu}}{\vs{\nu}'}}{\f{\rho}_Y}.
\end{align}
\end{subequations}
Using \eqref{eq:PhiYEY}, the two matrices are related as
\begin{equation}
\label{eq:Rtraf}
\f{R}\indexdud{Y}{\vs{\nu}}{\vs{\nu}'} 
=   f\indexdud{Y}{\vs{\nu}}{\vs{\nu}'} 
    R\indexdud{Y}{\vs{\nu}}{\vs{\nu}'},
\end{equation}
by the elementwise product with the phase factors \eqref{eq:phasefY}.
For example, using numerical algorithms,
the expectation values of \emph{fermionic transition operators} 
$T\indexddu{Y}{\vs{\nu}}{\vs{\nu}'} = \f{E}\indexddu{Y}{\vs{\nu}'}{\vs{\nu}}$
can be evaluated easily, leading to the fermionic matrix elements \cite{Rissler-2006,Barcza-2015}.

Note that the \emph{density operator} $\f{\rho}_Y$ is self-adjoint, 
$\f{\rho}_Y^\dagger = \f{\rho}_Y$,
however, its \emph{density matrix} is the same as its matrix-adjoint,
$(R\indexddu{Y}{\vs{\nu}'}{\vs{\nu}})^* = R\indexdud{Y}{\vs{\nu}}{\vs{\nu}'}$,
\emph{in the standard basis only},
but not in the fermionic basis,
$(\f{R}\indexdud{Y}{\vs{\nu}'}{\vs{\nu}})^* \neq \f{R}\indexdud{Y}{\vs{\nu}}{\vs{\nu}'}$,
we have 
$f\indexddu{Y}{\vs{\nu}'}{\vs{\nu}}
(\f{R}\indexdud{Y}{\vs{\nu}'}{\vs{\nu}})^*  = 
f\indexddu{Y}{\vs{\nu}}{\vs{\nu}'}
\f{R}\indexdud{Y}{\vs{\nu}}{\vs{\nu}'}$
with the fermionic phase factors \eqref{eq:phasefY} instead.
This is because the adjoint is defined by the inner product of the Hilbert space $\mathcal{H}_Y$ 
(see at the beginning of Section~\ref{sec:JW}),
and $\Phi_Y$ is not a $*$-homomorphism (see in Section~\ref{sec:JW.fermions}).
Following these lines, note that the spectrum of a fermionic density operator $\f{\rho}_Y$
is the same as the eigenvalues of its density matrix \emph{in the standard basis}, $R\indexdud{Y}{\vs{\nu}}{\vs{\nu}'}$.
Consequently, 
the entropy of the state $\f{\rho}_Y$ can be obtained as
the entropy of the eigenvalues of its density matrix in the standard basis $R\indexdud{Y}{\vs{\nu}}{\vs{\nu}'}$,
which is used also in numerical calculations \cite{Murg-2010,Veis-2016,Szalay-2017,Legeza-2018,Brandejs-2019}.

Using the fermionic reduction \eqref{eq:reducedfPTf},
we have that,
for the nested mode subsets $X\subseteq Y\subseteq M$,
the matrix of the reduced density operator 
$\f{\rho}_X = \Trf_{Y,X}(\f{\rho}_Y)$
can be given by the \emph{usual partial index contraction in the fermionic basis \eqref{eq:EYf}},
while \emph{phase factors \eqref{eq:phasefY} arise in the standard basis \eqref{eq:EY}},
\begin{subequations}
\begin{align}
\label{eq:rhoXstd}
R\indexdud{X}{\vs{\nu}_X}{\vs{\nu}_X'}
&= f\indexddu{X}{\vs{\nu}_X}{\vs{\nu}_X'}
\sum_{\vs{\nu}_{\bar{X}},\vs{\nu}_{\bar{X}}'} \delta^{\vs{\nu}_{\bar{X}},\vs{\nu}_{\bar{X}}'}
f\indexddu{Y}{\vs{\nu}}{\vs{\nu}'}
R\indexdud{Y}{\vs{\nu}}{\vs{\nu}'},\\
\label{eq:rhoXferm}
\f{R}\indexdud{X}{\vs{\nu}_X}{\vs{\nu}_X'}
&= \sum_{\vs{\nu}_{\bar{X}},\vs{\nu}_{\bar{X}}'} \delta^{\vs{\nu}_{\bar{X}},\vs{\nu}_{\bar{X}}'}
\f{R}\indexdud{Y}{\vs{\nu}}{\vs{\nu}'}.
\end{align}
\end{subequations}
(The second one is by \eqref{eq:PTfEYf},
the first one follows from this, by using \eqref{eq:Rtraf}.)
\eqref{eq:rhoXstd} is the general, easy-to-implement way of obtaining reduced density matrices
in the standard, computation basis, 
in complete agreement with earlier, ``Fock space formalism based'' \cite{Friis-2013}, 
or ``creation-annihilation operator formalism based'' \cite{Amosov-2016} ways.
(For the explicit form of the reduced density matrices $R\indexdud{X}{\vs{\nu}_X}{\vs{\nu}_X'}$
for few modes, see Appendix~\ref{appsec:States.ExplR}.)
Note that in the special case when the reduction is taken to the first consecutive modes
with respect to the \emph{fixed} Jordan-Wigner ordering,
then the phase factors are trivial,
and the effect of the fermionic partial trace coincides with that of the standard partial trace.
(For the proof, see Appendix~\ref{appsec:Tensors.ptr}.)
So fermionic reduced density matrices
(and then entropic correlation and entanglement measures)
can also be calculated by the use of the usual partial trace 
after applying particular mode reorderings,
by permutation mode transformation \cite{Krumnow-2016}.

\subsection{Properties of the fermionic state reduction}
\label{sec:States.redfprop}

Taking a look at the formula \eqref{eq:rhoXstd},
containing the rather involved phase factors \eqref{eq:phasefY} (see in Appendices~\ref{appsec:JW.Explf} and \ref{appsec:States.ExplR})
coming from a nonpositive map $\Phi_Y$,
it might seem to be unexpected that the fermionic partial trace map \eqref{eq:TPfPTf}
is a \emph{quantum channel,} that is, a trace preserving completely positive map \cite{Nielsen-2000,Petz-2008,Wilde-2013},
mapping density operators of a larger mode subset to density operators of a smaller one,
$\Trf_{Y,X}: \f{\mathcal{D}}_Y \to \f{\mathcal{D}}_X$, also when $Y$ is a part of an even larger system $M$.
First of all, the complete positivity can clearly be seen from the alternative formula \eqref{eq:PTfU}.
However, we show here also that
complete positivity follows directly 
from the definition \eqref{eq:defreducedf} of the state reduction,
there is no need for derivations with concrete maps \eqref{eq:PT}, \eqref{eq:PTf}, \eqref{eq:PTfU}, realizing state reductions in different scenarios,
especially, no need for explicit calculations with fermionic partial trace \eqref{eq:rhoXstd} containing the highly nontrivial phase factors \eqref{eq:phasefY}.

First, it is easy to see that the fermionic state reduction \eqref{eq:defreducedf} \emph{preserves the trace},
that is, if 
$\bigskalpHS{\f{\rho}_X}{ \f{A}_X }
= \bigskalpHS{\f{\rho}_Y}{ \iotaf_{X,Y}(\f{A}_X) }$ holds for all $\f{A}_X\in\f{\mathcal{A}}_X$,
then $\Trf(\f{\rho}_X)=\Trf(\f{\rho}_Y)$.
(This follows from the choice $\f{A}_X = \f{I}_X$ in \eqref{eq:defreducedf},
and the definition \eqref{eq:HS} of the Hilbert--Schmidt inner product.)
Since the fermionic state reduction \eqref{eq:defreducedf} can be obtained by the use of the fermionic partial trace \eqref{eq:reducedfPTf},
the latter also \emph{preserves the trace,}
\begin{equation}
\label{eq:PTfTP}
\Trf\bigl(\Trf_{Y,X}(\f{\rho}_Y)\bigr) = \Trf(\f{\rho}_Y).
\end{equation}
(Another proof can be given by the properties of the fermionic partial trace,
since $\Trf_{Y,\emptyset} = \Trf$,
the trace preservation is just setting $X=\emptyset$ in \eqref{eq:PTfCons}.)

It is also easy to see that the fermionic state reduction \eqref{eq:defreducedf} is \emph{positive},
that is, if 
$\bigskalpHS{\f{\rho}_X}{ \f{A}_X }
= \bigskalpHS{\f{\rho}_Y}{ \iotaf_{X,Y}(\f{A}_X) }$ holds for all $\f{A}_X\in\f{\mathcal{A}}_X$,
then
for all $\f{\rho}_Y\geq0$, we have $\f{\rho}_X\geq0$.
Indeed, 
\begin{equation*}
\begin{split}
0\leq\f{\rho}_Y
&\arrthenref{eq:posHS} 
\forall \f{A}_Y\in \f{\mathcal{A}}_Y, 0\leq\f{A}_Y: 
0\leq \bigskalpHS{\f{\rho}_Y}{ \f{A}_Y } \\
&\arrthen 
\forall \f{A}_X\in \f{\mathcal{A}}_X, 0\leq\iotaf_{X,Y}(\f{A}_X)\in \f{\mathcal{A}}_Y:
0\leq \bigskalpHS{\f{\rho}_Y}{ \iotaf_{X,Y}(\f{A}_X) }\\
&\arrthenref{eq:iotaU}
\forall \f{A}_X\in \f{\mathcal{A}}_X, 0\leq\f{A}_X:
0\leq \bigskalpHS{\f{\rho}_Y}{ \iotaf_{X,Y}(\f{A}_X) }\\
&\arrthenref{eq:defreducedf} 
\forall \f{A}_X\in \f{\mathcal{A}}_X, 0\leq\f{A}_X:
0\leq \bigskalpHS{\f{\rho}_X}{ \f{A}_X }\\
&\arrthenref{eq:posHS}
0\leq\f{\rho}_X.
\end{split}
\end{equation*}
Since the fermionic state reduction \eqref{eq:defreducedf} can be obtained by the use of the fermionic partial trace \eqref{eq:reducedfPTf},
the latter is also \emph{positive},
\begin{equation}
\label{eq:PTfpos}
\f{\rho}_Y \geq 0 \quad\Longrightarrow\quad \Trf_{Y,X}(\f{\rho}_Y) \geq 0.
\end{equation}

It is also easy to see that the fermionic state reduction \eqref{eq:defreducedf} is \emph{completely positive},
that is, it is still positive if it acts on a part of an arbitrarily larger system.
Indeed, the state reduction $\f{\rho}_Y\mapsto\f{\rho}_X$ is defined as 
$\bigskalpHS{\f{\rho}_X}{ \f{A}_X }
= \bigskalpHS{\f{\rho}_Y}{ \iotaf_{X,Y}(\f{A}_X) }$ for all $\f{A}_X\in\f{\mathcal{A}}_X$,
and we already have this definition for all $Y$, and $X\subseteq Y$.
Now, if we consider mode subset $Y$ as a part of a larger mode subset $Y\cup W$ (with $Y\cap W =\emptyset$),
then the definition of state reduction is written as
$\f{\rho}_{Y\cup W}\mapsto\f{\rho}_{X\cup W}$ by
$\bigskalpHS{\f{\rho}_{X\cup W}}{ \f{A}_{X\cup W} }
= \bigskalpHS{\f{\rho}_{Y\cup W}}{ \iotaf_{X\cup W,Y\cup W}(\f{A}_{X\cup W}) }$ for all $\f{A}_{X\cup W}\in\f{\mathcal{A}}_{X\cup W}$,
which is the state reduction from mode subset $Y\cup W$ to mode subset $X\cup W$,
which is positive, as we have already seen.
Since the fermionic state reduction \eqref{eq:defreducedf} can be obtained by the use of the fermionic partial trace \eqref{eq:reducedfPTf},
the latter is also \emph{completely positive},

A more elevated derivation of this
can be presented in the framework of algebraic quantum mechanics,
where it follows from the properties of conditional expectations and state extension \cite{Petz-2008,Araki-2003a},
in a kind of dual treatment.
A less elevated derivation of the complete positivity of the fermionic partial trace \eqref{eq:PTf}
can also be given by \eqref{eq:PTfemb} and \eqref{eq:cmEmbfCons},
and an even less elevated, direct proof by the positivity of the Choi map \cite{Wilde-2013},
which is a good finger-exercise for playing with phase factors.

\section{Correlation and entanglement}
\label{sec:CorrEnt}

In the theory of quantum correlation and entanglement, 
the central notions are those of \emph{product states} and \emph{local maps.}
In this section, we recall the concepts for the qubit case,
then attempt to write them for the fermionic case.
This cannot be done satisfactorily,
since the notion of locality is not established yet.
This can be resolved by the imposition of the parity superselection rule,
which will be done in the subsequent sections.

\subsection{Qubit correlation and entanglement}
\label{sec:CorrEnt.qubits}

A composite quantum subsystem $Y\subseteq M$ is \emph{$\xi$-uncorrelated},
that is, uncorrelated with respect to the partition $\xi=\{X_1,X_2,\dots,X_{\abs{\xi}}\}\in\Pi(Y)$,
if for all subsystems $X\in\xi$, for all operators $A_X\in\mathcal{A}_X$,
we have 
$\bigbracket{\prod_{X\in\xi} \iota_{X,Y}(A_X)}
\equiv \bigbracket{\bigotimes_{X\in\xi} A_X}
= \prod_{X\in\xi}\bigbracket{A_X}$.
That is, the expectation value of $\xi$-product operators factorizes.
Expressing this with the Hilbert--Schmidt inner product \eqref{eq:HS}
and the state $\rho_Y\in\mathcal{D}_Y$, we have
that the $\xi$-uncorrelated states of subsystem $Y$ 
(states of the $\xi$-uncorrelated subsystem $Y$) are the $\xi$-product states,
\begin{equation}
\label{eq:defunc}
\biggl(\forall X\in\xi,
\forall A_X\in\mathcal{A}_X: \quad
 \BigskalpHS{\rho_Y}{\bigotimes_{X\in\xi} A_X} 
= \prod_{X\in\xi}\bigskalpHS{\rho_X}{A_X}\biggr)
\quad\Longleftrightarrow\quad 
\rho_Y = \bigotimes_{X\in\xi}\rho_X,
\end{equation}
where $\rho_X=\Tr_{Y,X}(\rho_Y)$ is the reduced state \eqref{eq:defreduced}.

The textbook derivation of this is by the use of \eqref{eq:TPHS}, leading to
\begin{equation}
\label{eq:derivunc}
\forall X\in\xi,
\forall A_X\in\mathcal{A}_X: \quad
\BigskalpHS{\rho_Y}{\bigotimes_{X\in\xi} A_X}
=\BigskalpHS{\bigotimes_{X\in\xi}\rho_X}{\bigotimes_{X\in\xi}A_X}, 
\end{equation}
which, because the Hilbert--Schmidt inner product is nondegenerate,
and the $\xi$-elementary tensors $\bigotimes_{X\in\xi}A_X$ span $\mathcal{A}_Y$, leads to
\begin{equation}
\label{eq:uncprod}
\rho_Y = \bigotimes_{X\in\xi}\rho_X.
\end{equation}

Let us denote 
the set of \emph{$\xi$-uncorrelated states of subsystem $Y$} as
\begin{equation}
\label{eq:Dunc}
\mathcal{D}_{\text{$\xi$-unc}}
:= \Bigsset{\rho_Y\in\mathcal{D}_Y}{\rho_Y 
= \bigotimes_{X\in\xi}\Tr_{Y,X}(\rho_Y)}.
\end{equation}
These states can be prepared from pure $\xi$-product states by $\xi$-Local Operations ($\xi$-LO),
that is, maps of the form $\bigotimes_{X\in\xi}\Omega_X$, where $\Omega_X$ are trace preserving completely positive maps (TPCP, \cite{Petz-2008,Wilde-2013}).
The other states are \emph{$\xi$-correlated}, contained in $\mathcal{D}_Y\setminus\mathcal{D}_{\text{$\xi$-unc}}$;
to prepare them, some communication (interaction) is needed among the subsystems.

A composite quantum subsystem $Y\subseteq M$ is \emph{$\xi$-separable}, that is, separable
with respect to the partition $\xi= \{X_1,X_2,\dots,X_{\abs{\xi}}\}\in\Pi(Y)$,
if it can be prepared from $\xi$-uncorrelated subsystems 
by the use of 
$\xi$-Local Operations and Classical Communications ($\xi$-LOCC, \cite{Bennett-1996a,Bennett-1996b,Werner-1989,Chitambar-2014})
that is, maps generated by $\xi$-LOs,
the application of which
may depend on outcomes of measurements performed locally,
that is, described by TPCP maps $\Omega_X=\sum_m\Omega_{X,m}$,
where the measurement outcomes
are given by trace nonincreasing CP maps $\Omega_{X,m}$.
It turns out that
the set of \emph{$\xi$-separable states of subsystem $Y$} (states of the $\xi$-separable subsystem $Y$) is 
the convex hull \cite{Werner-1989} of that of $\xi$-uncorrelated ones \cite{Dur-1999,Dur-2000,Seevinck-2008,Szalay-2015b,Szalay-2017,Szalay-2018},
\begin{equation}
\label{eq:Dsep}
\mathcal{D}_{\text{$\xi$-sep}} := \Conv \mathcal{D}_{\text{$\xi$-unc}}, 
\end{equation}
expressing statistical mixtures.
(The \emph{convex hull} of a set $V$ in a real vector space is the set of all the possible convex combinations of its elements,
$\Conv V = \sset{\sum_i p_iv_i}{v_i\in V, p_i>0, \sum_ip_i=1}$.)
Note that the states of classical systems can always be expressed as mixtures of uncorrelated states.
The other states are \emph{$\xi$-entangled}, contained in $\mathcal{D}_Y\setminus\mathcal{D}_{\text{$\xi$-sep}}$;
to prepare them, some quantum communication (quantum interaction) is needed among the subsystems,
expressing the quantum nature of this kind of correlation.

\subsection{Fermionic correlation and entanglement}
\label{sec:CorrEnt.fermion}

For qubits we naturally had the identity
$\prod_{X\in\xi} \iota_{X,Y}(A_X)=\bigotimes_{X\in\xi} A_X$
for products of operators of disjoint subsystems.
For fermions, this is not the case,
and we have different ways for the representation of such kind of product operators.
(Recall that the map $\f{\Lambda}_{\ord{\xi}}$, given in \eqref{eq:Lambda}, connects the two points of view.)
From now, we use the usual algebraic representation, 
and 
we will have some remarks on the fermionic tensorial representation
later in Section~\ref{sec:Summ}. 

Similarly to the qubit case, 
for the definition of uncorrelated states, 
we expect that the expectation value of fermionic $\ord{\xi}$-elementary product operators factorizes.
However, since the operators $\iotaf_{X,Y}(\f{A}_X)$ for different $X\in\xi$ are not commuting 
(we have not imposed parity superselection yet, see in Section~\ref{sec:ParityMath}),
such definition of uncorrelated states is given with respect to the ordering of the operators.
A state of a set of modes $Y\subseteq M$ is \emph{$\ord{\xi}$-uncorrelated}, that is, uncorrelated
with respect to the \emph{ordered} partition $\ord{\xi}=\tuple{X_1,X_2,\dots,X_{\abs{\ord{\xi}}}}\in\ord{\Pi}(Y)$,
if for all mode subsets $X\in\ord{\xi}$, for all operators $\f{A}_X\in\f{\mathcal{A}}_X$,
we have 
$\bigbracket{{\prodord}_{X\in\ord{\xi}} \iotaf_{X,Y}(\f{A}_X)}
\equiv\bigbracket{{\bigotimesfp}_{X\in\ord{\xi}}\f{A}_X}
= \prod_{X\in\ord{\xi}}\bigbracket{\f{A}_X}$.
Expressing these with the Hilbert--Schmidt inner product \eqref{eq:HS}
and the state $\f{\rho}_Y\in\f{\mathcal{D}}_Y$,
we have the following definition for the \emph{$\ord{\xi}$-uncorrelated} states
\begin{equation}
\label{eq:defuncfp}
\forall X\in\ord{\xi},
\forall \f{A}_X\in\f{\mathcal{A}}_X: \quad
\BigskalpHS{\f{\rho}_Y}{\bigotimesfp_{X\in\ord{\xi}}\f{A}_X}
= \prod_{X\in\xi}\bigskalpHS{\f{\rho}_X}{\f{A}_X}, 
\end{equation}
where $\f{\rho}_X=\f{\Tr}_{Y,X}(\f{\rho}_Y)$ is the fermionic reduced state \eqref{eq:defreducedf}, \eqref{eq:reducedfPTf}.

Using the tools we have constructed in Section~\ref{sec:Tensors}, now the simple steps
of applying \eqref{eq:TPfHS} and \eqref{eq:LambdaU}
lead to
\begin{equation}
\label{eq:derivuncfp}
\forall X\in\ord{\xi}, \forall \f{A}_X\in\f{\mathcal{A}}_X: \quad
\BigskalpHS{\f{\rho}_Y}{\bigotimesfp_{X\in\ord{\xi}} \f{A}_X}
= \BigskalpHS{\bigotimesfp_{X\in\ord{\xi}} \f{\rho}_X}{\bigotimesfp_{X\in\ord{\xi}}\f{A}_X}, 
\end{equation}
which, because the Hilbert--Schmidt inner product is nondegenerate,
and the fermionic $\ord{\xi}$-ele\-men\-tary product operators $\bigotimesfp_{X\in\ord{\xi}}\f{A}_X$ span $\f{\mathcal{A}}_Y$,
leads to that
the $\ord{\xi}$-uncorrelated states are the $\ord{\xi}$-product states
\begin{equation}
\label{eq:uncprodfp}
\f{\rho}_Y = \bigotimesfp_{X\in\ord{\xi}} \f{\rho}_X.
\end{equation}

Note that, at this point, 
if a state is uncorrelated with respect to an ordered partition,
it seems not to be necessarily uncorrelated with respect to a different ordering of the same parts.
Another related point is that, although we have $\f{\rho}_X=\Trf_{Y,X}(\f{\rho}_Y)\geq0$, 
their product in \eqref{eq:uncprodfp} 
is not positive semidefinite, not even self-adjoint in general.
These are because 
the terms in the product in \eqref{eq:uncprodfp} are not commuting 
in general.
We will come back to these issues in Section~\ref{sec:ParityPhys},
after recalling the parity superselection rule in Section~\ref{sec:ParityMath},
here we just mention the tricky part:
it turns out that if \eqref{eq:defuncfp} holds, then these terms are actually commuting \cite{Araki-2003b},
see also in \eqref{eq:prod.pos},
leading to an ordering-free definition of uncorrelated states also in the fermionic case.
Let us denote the set of \emph{$\xi$-uncorrelated states of mode subset $Y$} as
\begin{equation}
\label{eq:DfpupsuncwoSSR}
\f{\mathcal{D}}_{\text{$\xi$-unc}}
:= \Bigsset{\f{\rho}_Y\in\f{\mathcal{D}}_Y}{\f{\rho}_Y = \bigotimesfp_{X\in\xi} \Trf_{Y,X}(\f{\rho}_Y) }.
\end{equation}
The other states are \emph{$\xi$-correlated}, contained in
$\f{\mathcal{D}}_Y\setminus\f{\mathcal{D}}_{\text{$\xi$-unc}}$.
(This is the multipartite generalization of the state set $\mathcal{P}3$ in \cite{Banuls-2007}.
We do not consider $\mathcal{P}0\equiv\mathcal{P}1$ in \cite{Banuls-2007},
 since we do not impose parity superselection for operators if it is not imposed for states;
neither $\mathcal{P}2$, since the usual tensor product does not make any sense here.)

Following these lines, 
neither the LO-based definition of correlation 
nor the LOCC-based definition of separability or entanglement
can be formulated
for the whole algebra, since we have no locality yet.
We will elaborate on this in Section~\ref{sec:ParityPhys.Loc};
at this point we can only \emph{define} the set of \emph{$\xi$-separable states of mode subset $Y$} 
without any LOCC-based operational meaning by convexity as
\begin{equation}
\label{eq:DfpupssepwoSSR}
\f{\mathcal{D}}_{\text{$\xi$-sep}}:= \Conv \f{\mathcal{D}}_{\text{$\xi$-unc}}.  
\end{equation}
The other states are \emph{$\xi$-entangled}, contained in
$\f{\mathcal{D}}_Y\setminus\f{\mathcal{D}}_{\text{$\xi$-sep}}$.
(This is the multipartite generalization of the state set $\mathcal{S}3$ in \cite{Banuls-2007}.)

\section{Mathematics of the parity superselection for fermions}
\label{sec:ParityMath}

In Sections~\ref{sec:JW}, \ref{sec:Tensors} and \ref{sec:States},
we presented a mathematical toolbox for the efficient handling of fermionic systems in concrete representation.
These definitions and properties are given for the whole algebra $\f{\mathcal{A}}_Y$,
and are built up analogously to the standard ``qubit'' case.
However, the definition of correlation and entanglement in Section~\ref{sec:CorrEnt.fermion}
could not be endowed with their usual, local operation based motivation,
due to the lack of the notions of locality of maps,
rooted in the lack of independence of mode subsets, 
following mainly (but not completely) from the lack of \emph{commutativity} of the operator algebras of disjoint mode subsets;
that is,
there exist operators $\f{A}_X\in\f{\mathcal{A}}_X$ and $\f{B}_{\bar{X}}\in\f{\mathcal{A}}_{\bar{X}}$ for which
\begin{equation}
\label{eq:commutLambda}
\bigl[\f{A}_X\otimesf\Idf_{\bar{X}}, \Idf_X\otimesf \f{B}_{\bar{X}} \bigr] \neq 0.
\end{equation}
Re-establishing the commutation
by imposing the \emph{parity superselection rule} \cite{Wick-1952,Hegerfeldt-1968}
resolves these issues,
giving a mathematical reason for the superselection.
There are also important physical reasons for the imposing of the parity superselection.
We will turn to them in the next section.

In this section, we give the necessary definitions for handling the restrictions to the \emph{physical subspaces.}
First, in the \emph{Hilbert space,}
we consider the subspaces of state vectors of well-defined \emph{fermion-number parity.}
Second, in the \emph{operator algebra,}
we consider the subspaces of operators of well-defined \emph{fermion-operator parity,}
that is, those which preserve (even) or alter (odd) the fermion-number parity.
Third, in the \emph{map algebra,} 
we consider the subspaces of maps of well-defined \emph{fermion-map parity,}
that is, those which preserve (even) or alter (odd) the fermion-operator parity.
A short summary on these is given in Table \ref{tab:spaces}.

\begin{table}
\caption{Summary of the different linear spaces and elements}
\label{tab:spaces}
\setlength{\tabcolsep}{12pt}
\begin{tabular}{c c c c c}
\hline 
space &
elements &
parity &
projection &
parity subspaces\\
\hline 
$\mathcal{H}_Y$ &
$\cket{\psi_Y}$ &
&
&
$\mathcal{H}_Y^\pm$ \\
$\f{\mathcal{A}}_Y = \Lin\mathcal{H}_Y$ &
$\Idf_Y$, $\f{A}_Y$ &
$\f{T}_Y$ &
$\f{P}_Y^\pm$ &
$\f{\mathcal{A}}_Y^\pm$ \\
$\f{\mathcal{B}}_Y = \Lin\f{\mathcal{A}}_Y$ &
$\IIdf_Y$, $\f{\Omega}_Y$ &
$\f{\Theta}_Y$ &
$\f{\Pi}_Y^\pm$ &
$\f{\mathcal{B}}_Y^\pm$ \\
$\f{\mathcal{C}}_Y = \Lin\f{\mathcal{B}}_Y$ &
$\IIIdf_Y$, $\f{\mathbb{A}}_Y$ &
$\f{\mathbb{T}}_Y$ &
$\f{\mathbb{P}}_Y^\pm$ &
\\
\hline 
\end{tabular}
\end{table}

\subsection{Level I -- Hilbert space}
\label{sec:ParityMath.H}
In the Hilbert space $\mathcal{H}_Y$ of the system,
the fundamental quantity is the parity of the fermion number.
The \emph{fermion number parity} in mode subset $Y$ is given by
the phase operator $\f{T}_Y\in\f{\mathcal{A}}_Y=\Lin \mathcal{H}_Y$ as
\begin{subequations}
\begin{equation}
\label{eq:pY}
\f{T}_Y := \prodord_{i\in Y}\Gammaf_Y(p_i) 
= \bigotimesfp_{i\in Y}p_{\set{i}}
= \bigotimesf_{i\in Y}p_{\set{i}}
= \bigotimes_{i\in Y} p_i.
\end{equation}
(Recall that $p_i=\cket{\phi_i^0}\bra{\phi_i^0} - \cket{\phi_i^1}\bra{\phi_i^1} = a_ia_i^\dagger-a_i^\dagger a_i\in\mathcal{A}_i$
is the phase operator, see in Section~\ref{sec:JW.fermions}.)
Applying this twice is the identity, $\f{T}_Y^2=\Idf_Y$, so
its eigenvalues are $+1$ and $-1$,
and the projectors onto its eigensubspaces in $\mathcal{H}_Y$ are then
\begin{equation}
\label{eq:PY}
\f{P}_Y^\pm := \frac12\bigl(\Idf_Y\pm \f{T}_Y\bigr),
\end{equation}
corresponding to the subspaces of
even ($+1$) and odd ($-1$) number of fermions
\begin{equation}
\label{eq:HYpm}
\mathcal{H}_Y^\pm 
:= \Bigsset{\cket{\psi_Y}\in\mathcal{H}_Y}{\f{T}_Y\cket{\psi_Y} = \pm \cket{\psi_Y}}.
\end{equation}
(For a summary, see Table \ref{tab:spaces}.)
Since $p_i\cket{\phi_i^{\nu_i}} = (-1)^{\nu_i}\cket{\phi_i^{\nu_i}}$,
these can also be given using the standard basis \eqref{eq:phiY} as
\begin{equation}
\label{eq:HYpmSpan}
\mathcal{H}_Y^\pm 
=\Span\Bigsset{\cket{\phi_Y^{\vs{\nu}}}}{(-1)^{\esum_{i\in Y} \nu_i} = \pm1}.
\end{equation}
\end{subequations}

The subspaces $\mathcal{H}_Y^\pm$ are called the \emph{``physical subspaces''} of $\mathcal{H}_Y$, 
containing the \emph{``physically meaningful state vectors''} \cite{Wick-1952,Hegerfeldt-1968,Banuls-2007,Zimboras-2014}. 
The vectors having nonzero projections in both,
that is, superpositions of vectors of even and odd parity, are \emph{``nonphysical''}.
For more details and the relevance of physicality, see Section~\ref{sec:ParityPhys}.

The local fermion number parity subspaces (inside mode subsets) can also be formulated.
Let us have a partition $\xi = \{X_1,X_2,\dots,X_{\abs{\xi}}\}\in\Pi(Y)$ of a mode subset $Y\subseteq M$.
We give the parity of each mode subset $X\in\xi$ by the multi-index
$\vs{\epsilon}: \xi \to \{+1,-1\}$, $X\mapsto \epsilon_X$.
Then the projectors given by \eqref{eq:PY} as 
\begin{subequations}
\begin{equation}
\label{eq:Pupseps}
\f{P}_\xi^{\vs{\epsilon}} := \prod_{X\in\xi} \iotaf_{X,Y} (\f{P}_{X}^{\epsilon_X})
=\bigotimesfp_{X\in\xi}\f{P}_{X}^{\epsilon_X}
=\bigotimesf_{X\in\xi}\f{P}_{X}^{\epsilon_X},
\end{equation}
project onto the subspaces of given $\xi$-local parity $\vs{\epsilon}$ in $\mathcal{H}_Y$,
\begin{equation}
\label{eq:Hupseps}
\mathcal{H}_\xi^{\vs{\epsilon}} := 
\Bigsset{\cket{\psi_Y}\in\mathcal{H}_Y}{\forall X\in\xi: \iotaf_{X,Y}(\f{T}_X)\cket{\psi_Y} = \epsilon_X \cket{\psi_Y}}
=\bigotimes_{X\in\xi}\mathcal{H}_X^{\epsilon_X},
\end{equation}
using that $\f{T}_X\f{P}_{X}^{\epsilon_X} = \epsilon_X\f{P}_{X}^{\epsilon_X}$, see \eqref{eq:PY}.
(Note that the operators here are diagonal, therefore commuting, 
so, on the one hand, no ordering has to be referred to in \eqref{eq:Pupseps};
on the other hand, $\iotaf_{X,Y}(\f{T}_X)=\iota_{X,Y}(\f{T}_X)$,
so $\xi$-elementary tensors $\bigotimes_{X\in\xi}\cket{\psi_X}$ and linearity can be used
in the proof of the last equality in \eqref{eq:Hupseps}.)
These can also be given using the standard basis \eqref{eq:phiY} as
\begin{equation}
\label{eq:HupsepsSpan}
\mathcal{H}_\xi^{\vs{\epsilon}}
= \Span\Bigsset{\cket{\phi_Y^{\vs{\nu}}}}{\forall X\in\xi:(-1)^{\esum_{i\in X} \nu_i} = \epsilon_X}.
\end{equation}
The subspaces of well defined global parity
can be written as direct sums of the
subspaces of given local parity as
\begin{equation}
\mathcal{H}_Y^\pm = \bigoplus\Bigsset{\mathcal{H}_\xi^{\vs{\epsilon}}}{ \vs{\epsilon}: (-1)^{\abs{\sset{X\in\xi}{\epsilon_X = -1}}} = \pm1}.
\end{equation}
\end{subequations}
In particular, for a bipartition $\xi = \{X,\bar{X}\}$, we have
$\mathcal{H}_Y = 
\mathcal{H}_{X\bar{X}}^{++} \oplus 
\mathcal{H}_{X\bar{X}}^{+-} \oplus
\mathcal{H}_{X\bar{X}}^{-+} \oplus
\mathcal{H}_{X\bar{X}}^{--} =
\mathcal{H}_Y^+ \oplus \mathcal{H}_Y^-$.
The globally even and odd subspaces are
$\mathcal{H}_Y^+ = \mathcal{H}_{X\bar{X}}^{++} \oplus \mathcal{H}_{X\bar{X}}^{--}$ and
$\mathcal{H}_Y^- = \mathcal{H}_{X\bar{X}}^{+-} \oplus \mathcal{H}_{X\bar{X}}^{-+}$, respectively.
(Note that we use a simplified notation, e.g., 
$\mathcal{H}_{X\bar{X}}^{++}:=\mathcal{H}_{\set{X,\bar{X}}}^{\tuple{+,+}}$,
omitting the parentheses and colons in the writing of tuples, since this does not cause confusion.
For explicit form of these subspaces for few modes, see Appendix~\ref{appsec:Parity.ExplSpaces}.)

\subsection{Level II -- Operator algebra}
\label{sec:ParityMath.A}
In the operator algebra $\f{\mathcal{A}}_Y=\Lin\mathcal{H}_Y$ of the system,
it is fundamental to consider the parity of a fermionic operator.
The \emph{fermionic operator parity} in mode subset $Y$ is given by
the $*$-automorphism of $\f{\mathcal{A}}_Y$ as 
$\f{\Theta}_Y := \Ad_{\f{T}_Y}\in\f{\mathcal{B}}_Y:=\Lin \f{\mathcal{A}}_Y$
\cite{Alicki-2001,Araki-2003a,Araki-2003b,Moriya-2006}, that is,
\begin{subequations}
\begin{equation}
\label{eq:ThetaY}
\f{\Theta}_Y(\f{A}_Y) := \f{T}_Y \f{A}_Y \f{T}_Y^{-1}.
\end{equation}
Applying this twice is the identity, $\f{\Theta}_Y\circ\f{\Theta}_Y=\IIdf_Y$, so
its eigenvalues are $+1$ and $-1$,
and the projectors onto its eigensubspaces in $\f{\mathcal{A}}_Y$ are then
\begin{equation}
\label{eq:PiY}
\f{\Pi}_Y^\pm := \frac12\bigl(\IIdf_Y\pm \f{\Theta}_Y\bigr),
\end{equation}
corresponding to the subspaces
being the linear spans of monomials of
even ($+1$) and odd ($-1$) number of fermionic creation and annihilation operators \cite{Alicki-2001,Araki-2003a,Araki-2003b,Moriya-2006}
\begin{equation}
\label{eq:AYpm}
\f{\mathcal{A}}_Y^\pm := \Bigsset{\f{A}_Y\in\f{\mathcal{A}}_Y}{\f{\Theta}_Y(\f{A}_Y) = \pm \f{A}_Y},
\end{equation}
since $\f{\Theta}_Y$ is a $*$-automorphism,
and $\f{\Theta}_Y(\tilde{a}_{i,Y})= - \tilde{a}_{i,Y}$ for the annihilation operator.
(For a summary, see Table \ref{tab:spaces}.)
The eigensubspaces \eqref{eq:AYpm} can also be given using the basis \eqref{eq:EYf} as
\begin{equation}
\label{eq:AYpmSpan}
\f{\mathcal{A}}_Y^\pm = 
\Span\Bigsset{\f{E}\indexddu{Y}{\vs{\nu}}{\vs{\nu}'}}{(-1)^{\esum_{i\in Y} (\nu_i+\nu_i')} = \pm1 },
\end{equation}
see \eqref{eq:EYexpl} and \eqref{eq:PhiYEY}.
Note that $\f{T}_X,\f{P}_X,\Idf_X\in\f{\mathcal{A}}_X^+$.
\end{subequations}

The main point is that the definitions \eqref{eq:ThetaY} and \eqref{eq:AYpm}
lead to the commutation or anticommutation with the phase operator $\f{T}_Y$,
\begin{subequations}
\begin{equation}
\label{eq:AYpmComm}
\f{A}_Y\in \f{\mathcal{A}}_Y^\pm \quad\Longleftrightarrow\quad
\f{T}_Y \f{A}_Y = \pm \f{A}_Y \f{T}_Y.
\end{equation}
It easily follows that
even operators do not change the fermion number parity,
while odd operators do, that is,
\begin{equation}
\label{eq:AYpmComm2}
\f{A}_Y\in \f{\mathcal{A}}_Y^\pm \quad\Longleftrightarrow\quad 
\f{A}_Y: \mathcal{H}_Y^{\pm'} \longrightarrow \mathcal{H}_Y^{\pm\pm'}.
\end{equation}
\end{subequations}
(For the proof,
we have $\f{A}_Y\in \f{\mathcal{A}}_Y^\pm$ if and only if
$\f{T}_Y \f{A}_Y = \pm \f{A}_Y \f{T}_Y$ (by \eqref{eq:AYpmComm}),
which holds if and only if
$\f{T}_Y \f{A}_Y\cket{\psi_Y} = \pm \f{A}_Y \f{T}_Y\cket{\psi_Y}$ for all $\cket{\psi_Y}\in \mathcal{H}_Y^{\pm'}$
(since $\mathcal{H}_Y^+$ and $\mathcal{H}_Y^-$ span $\mathcal{H}_Y$),
which holds if and only if
$\f{T}_Y \f{A}_Y\cket{\psi_Y} = \pm \pm' \f{A}_Y\cket{\psi_Y}$ for all $\cket{\psi_Y}\in \mathcal{H}_Y^{\pm'}$ (by \eqref{eq:HYpm}),
which is equivalent to $\f{A}_Y\cket{\psi_Y}\in\mathcal{H}_Y^{\pm\pm'}$ for all $\cket{\psi_Y}\in \mathcal{H}_Y^{\pm'}$ (by \eqref{eq:HYpm}).)

The subspace $\f{\mathcal{A}}_Y^+$ is called the \emph{``even subspace''}, or \emph{``physical subspace''} of $\f{\mathcal{A}}_Y$, which is also a subalgebra,
containing the \emph{``physically meaningful operators''}  \cite{Wick-1952,Hegerfeldt-1968,Banuls-2007,Zimboras-2014,Beny-2018}, 
or \emph{``observables''}.
As we have seen in \eqref{eq:AYpmComm} and \eqref{eq:AYpmComm2},
even, or physical operators do not change the fermion number parity:
fermions can be created or annihilated only in pairs.
The subspace $\f{\mathcal{A}}_Y^-$ is called the \emph{``odd subspace''}, or \emph{``nonphysical subspace''}.
For more details and the relevance of physicality, see Section~\ref{sec:ParityPhys}.

The connection of the fermionic tensor product with the superselection (inside mode subsets) is as follows.
Let us have a partition $\xi = \{X_1,X_2,\dots,X_{\abs{\xi}}\}\in\Pi(Y)$ of a mode subset $Y\subseteq M$.
We give the parity of each mode subset $X\in\xi$ by the multi-index
$\vs{\epsilon}: \xi \to \{+1,-1\}$, $X\mapsto \epsilon_X$.
Then the projectors given by \eqref{eq:PiY} as
\begin{subequations}
\label{eq:localparity}
\begin{equation}
\label{eq:Piupseps}
\f{\Pi}_\xi^{\vs{\epsilon}} := \prod_{X\in\xi} \iotabbfp_{X,Y}(\f{\Pi}_{X}^{\epsilon_X})
\equalsref{eq:noMapLambdafp}\bigotimesfp_{X\in\xi}\f{\Pi}_{X}^{\epsilon_X}
=\prod_{X\in\xi} \iotabbf_{X,Y}(\f{\Pi}_{X}^{\epsilon_X})
\equalsref{eq:noMapLambdaf}\bigotimesf_{X\in\xi}\f{\Pi}_{X}^{\epsilon_X},
\end{equation}
project onto the subspaces of given $\xi$-local parity $\vs{\epsilon}$ in $\mathcal{A}_Y$,
\begin{equation}
\label{eq:Aupseps}
\f{\mathcal{A}}_\xi^{\vs{\epsilon}} :=  \Bigsset{\f{A}_Y\in\f{\mathcal{A}}_Y}{\forall X\in\xi: \iotabbf_{X,Y}(\f{\Theta}_X)(\f{A}_Y) = \epsilon_X \f{A}_Y}
=\bigotimesf_{X\in\xi}\f{\mathcal{A}}_X^{\epsilon_X},
\end{equation}
using that $\f{\Theta}_X\circ\f{\Pi}_{X}^{\epsilon_X} = \epsilon_X\f{\Pi}_{X}^{\epsilon_X}$, see \eqref{eq:PiY}.
(Recall that the notation on the right-hand side 
means only a linear hull of elementary fermionic tensors,
see the end of Section~\ref{sec:Tensors.TP}.
Note that there is no need for taking into account the ordering in the writing $\iotabbfp_{X,Y}(\f{\Pi}_{X}^{\epsilon_X})$,
contrary to the general case in \eqref{eq:noMapLambdafp},
since $\f{\Pi}_{X}^{\epsilon_X}$ consists of maps being conjugation by diagonal operators,
so $\f{\Pi}_\xi^{\vs{\epsilon}}$ acts in the same way on arbitrarily ordered fermionic $\ord{\xi}$-elementary products.)
These can also be given using the standard basis \eqref{eq:EYf} as
\begin{equation}
\label{eq:AupsepsSpan}
\f{\mathcal{A}}_\xi^{\vs{\epsilon}} = 
\Span\Bigsset{\f{E}\indexddu{Y}{\vs{\nu}}{\vs{\nu}'}}{\forall X\in\xi: (-1)^{\esum_{i\in X} (\nu_i+\nu_i')} = \epsilon_X },
\end{equation}
see \eqref{eq:EYexpl} and \eqref{eq:PhiYEY}.
As special case, let $\alleven: X\mapsto +1$ be the $\xi$-even parity multi-index,
with this, $\f{\mathcal{A}}_\xi^\alleven$ is called \emph{$\xi$-locally physical subspace} (or $\xi$-even subspace), 
which is also a subalgebra.
The physical and nonphysical subspaces can be written as direct sums of the 
subspaces of given local parity as
\begin{equation}
\f{\mathcal{A}}_Y^\pm = \bigoplus\Bigsset{\f{\mathcal{A}}_\xi^{\vs{\epsilon}}}{ \vs{\epsilon}: (-1)^{\abs{\sset{X\in\xi}{\epsilon_X = -1}}} = \pm1}.
\end{equation}
\end{subequations}
In particular, for a bipartition $\xi = \{X,\bar{X}\}$, we have
$\f{\mathcal{A}}_Y = 
\f{\mathcal{A}}_{X\bar{X}}^{++} \oplus 
\f{\mathcal{A}}_{X\bar{X}}^{+-} \oplus
\f{\mathcal{A}}_{X\bar{X}}^{-+} \oplus
\f{\mathcal{A}}_{X\bar{X}}^{--} =
\f{\mathcal{A}}_Y^+ \oplus \f{\mathcal{A}}_Y^-$.
The physical subspace is
$\f{\mathcal{A}}_Y^+ = \f{\mathcal{A}}_{X\bar{X}}^{++} \oplus \f{\mathcal{A}}_{X\bar{X}}^{--}$,
while the subspace
$\f{\mathcal{A}}_Y^- = \f{\mathcal{A}}_{X\bar{X}}^{+-} \oplus \f{\mathcal{A}}_{X\bar{X}}^{-+}$
is nonphysical,
and the $\xi$-locally physical subspace is $\f{\mathcal{A}}_{X\bar{X}}^\alleven\equiv\f{\mathcal{A}}_{X\bar{X}}^{++}$.
(For explicit form of these subspaces for few modes, see Appendix~\ref{appsec:Parity.ExplSpaces}.)

The one-mode anticommutation relations \eqref{eq:modeacommf},
together with the properties of $\f{\Theta}_Y$,
lead to the commutation/anticommutation of operators of disjoint mode subsets
\begin{equation}
\label{eq:ACommf}
\iotaf_{X,Y}(\f{A}_X)\iotaf_{\bar{X},Y}(\f{B}_{\bar{X}}) = \pm \iotaf_{\bar{X},Y}(\f{B}_{\bar{X}})\iotaf_{X,Y}(\f{A}_X),
\end{equation}
for all $\f{A}_X\in\f{\mathcal{A}}_X^{\epsilon_X}$ and $\f{B}_{\bar{X}}\in\f{\mathcal{A}}_{\bar{X}}^{\epsilon_{\bar{X}}}$,
where the lower sign is for the case $\epsilon_X=\epsilon_{\bar{X}}=-1$.

\begin{subequations}
Note that the trace of odd operators vanishes, 
\begin{equation}
\label{eq:Trfeven}
\Trf:\quad \f{\mathcal{A}}_Y^- \quad\longrightarrow\quad 0,
\end{equation}
see \eqref{eq:AYpmSpan} and the end of Section~\ref{sec:JW.Phi},
so the fermionic partial trace \eqref{eq:PTf} preserves the operator parity, 
\begin{equation}
\label{eq:PTrfeven}
\Trf_{Y,X} :\quad \f{\mathcal{A}}_Y^\pm \quad\longrightarrow\quad \f{\mathcal{A}}_X^\pm,
\end{equation}
see \eqref{eq:PTfEYf}. 
Also, the fermionic canonical embedding \eqref{eq:cEmbf} maps to $\f{\mathcal{A}}_X\otimesf\f{\mathcal{A}}_{\bar{X}}^+$, 
so it also preserves the parity
\begin{equation}
\label{eq:cEmbeven}
\iotaf_{X,Y} :\quad \f{\mathcal{A}}_X^\pm \quad\longrightarrow\quad \f{\mathcal{A}}_Y^\pm.
\end{equation}
\end{subequations}

\subsection{Level III -- Map algebra}
\label{sec:ParityMath.B}
In the algebra $\f{\mathcal{B}}_Y=\Lin\f{\mathcal{A}}_Y$ of the maps of the
operator algebra $\f{\mathcal{A}}_Y$ of the system,
it is fundamental to consider again a parity-like notion.
The \emph{fermionic map parity} in mode subset $Y$ is given by
the $*$-automorphism of $\f{\mathcal{B}}_Y$ as
$\f{\mathbb{T}}_Y := \Ad_{\f{\Theta}_Y}\in\f{\mathcal{C}}_Y:=\Lin \f{\mathcal{B}}_Y$, that is,
\begin{subequations}
\begin{equation}
\label{eq:TTY}
\f{\mathbb{T}}_Y(\f{\Omega}_Y) := \f{\Theta}_Y \circ \f{\Omega}_Y \circ \f{\Theta}_Y^{-1}.
\end{equation}
Applying this twice is the identity, $\f{\mathbb{T}}_Y\circ\f{\mathbb{T}}_Y=\IIIdf_Y$, so
its eigenvalues are $+1$ and $-1$,
and the projectors onto its eigensubspaces in $\f{\mathcal{B}}_Y$ are then
\begin{equation}
\label{eq:PPY}
\f{\mathbb{P}}_Y^\pm := \frac12\bigl(\IIIdf_Y\pm \f{\mathbb{T}}_Y\bigr),
\end{equation}
corresponding to the subspaces
\begin{equation}
\label{eq:BYpm}
\f{\mathcal{B}}_Y^\pm := \Bigsset{\f{\Omega}_Y\in\f{\mathcal{B}}_Y}{\f{\mathbb{T}}_Y(\f{\Omega}_Y) = \pm \f{\Omega}_Y}.
\end{equation}
(For a summary, see Table \ref{tab:spaces}.)
\end{subequations}

The main point is that the definitions \eqref{eq:TTY} and \eqref{eq:BYpm}
lead to the commutation or anticommutation with the operator parity $\f{\Theta}_Y$,
\begin{subequations}
\begin{equation}
\label{eq:BYpmComm}
\f{\Omega}_Y\in \f{\mathcal{B}}_Y^\pm \quad\Longleftrightarrow\quad
\f{\Theta}_Y\circ \f{\Omega}_Y = \pm \f{\Omega}_Y \circ \f{\Theta}_Y.
\end{equation}
It easily follows that
even maps do not change the operator parity,
while odd maps do, that is,
\begin{equation}
\label{eq:BYpmComm2}
\f{\Omega}_Y\in \f{\mathcal{B}}_Y^\pm \quad\Longleftrightarrow\quad
\f{\Omega}_Y: \f{\mathcal{A}}_Y^{\pm'} \longrightarrow \f{\mathcal{A}}_Y^{\pm\pm'}.
\end{equation}
\end{subequations}
(For the proof,
we have $\f{\Omega}_Y\in \f{\mathcal{B}}_Y^\pm$ if and only if
$\f{\Theta}_Y\circ \f{\Omega}_Y = \pm \f{\Omega}_Y \circ \f{\Theta}_Y$ (by \eqref{eq:BYpmComm}),
which holds if and only if
$(\f{\Theta}_Y\circ \f{\Omega}_Y)(\f{A}_Y) = \pm (\f{\Omega}_Y \circ \f{\Theta}_Y)(\f{A}_Y)$ for all $\f{A}_Y\in \f{\mathcal{A}}_Y^{\pm'}$
(since $\f{\mathcal{A}}_Y^+$ and $\f{\mathcal{A}}_Y^-$ span $\f{\mathcal{A}}_Y$),
which holds if and only if
$\f{\Theta}_Y(\f{\Omega}_Y(\f{A}_Y))= \pm\pm'\f{\Omega}_Y(\f{A}_Y)$ for all $\f{A}_Y\in \f{\mathcal{A}}_Y^{\pm'}$ (by \eqref{eq:AYpm}),
which is equivalent to $\f{\Omega}_Y(\f{A}_Y)\in\f{\mathcal{A}}_Y^{\pm\pm'}$ for all $\f{A}_Y\in \f{\mathcal{A}}_Y^{\pm'}$ (by \eqref{eq:AYpm}).)

The subspace $\f{\mathcal{B}}_Y^+$ is called the \emph{``even subspace''} of $\f{\mathcal{B}}_Y$, which is also a subalgebra.
As we have seen in \eqref{eq:BYpmComm} and \eqref{eq:BYpmComm2},
even maps do not change the fermionic operator parity,
that is, the parity of the number of fermionic creation and annihilation operators in an operator monomial.
In particular, it maps a physical operator to a physical one,
which is necessary for representing a physical process.
On the other hand,
physical maps can meaningfully be given only for physical operators,
so we call an even map $\f{\Omega}_X\in\f{\mathcal{B}}_X^+$ \emph{physically defined}, if it annihilates odd, or nonphysical operators,
$\f{\Omega}_X\in\f{\mathcal{B}}_X^+$ and $\f{\Omega}_X(\f{A}_X)=0$ for all $\f{A}_X\in\f{\mathcal{A}}_X^-$.
We note that this definition is an artefact, 
coming from our matrix algebraic (quantum information) point of view.
Even if the unphysical subspace is not used,
so there is no physical meaning of maps acting on that,
we would like to have the domain of a map to be the whole algebra,
and not to allow arbitrariness in the description of the map.
For more details and the relevance of physicality, see Section~\ref{sec:ParityPhys}.

The $\xi$-local properties in $\f{\mathcal{B}}_Y$ can also be given,
 using the fermionic products of maps \eqref{eq:mTPf}-\eqref{eq:mTPfp}, 
in the same way as in $\f{\mathcal{A}}_Y$, see \eqref{eq:localparity}.

\section{Physics of the parity superselection for fermions}
\label{sec:ParityPhys}

After having written out the necessary definitions in the previous section,
here we turn to the
physical reasons for the imposition of the parity superselection,
excluding the superposition of even and odd number of fermions.
With the assumption of Lorentz-invariance, 
the spin-statistics connection \cite{Fierz-1939,Pauli-1940,Schwinger-1951} revealed that
particles of fermionic statistics obey half-integer spin representation of the rotation group.
Then a superposition of even and odd number of fermions
would be changed more than an overall phase
under a $2\pi$ rotation (doing nothing) \cite{Wick-1952,Streater-1964,Weinberg-2005,Hegerfeldt-1968}.
Without the assumption of the spin-statistics connection,
only assuming that the laws of physics are the same for all different observers, together with the no-signaling principle,
the standard reasoning in the nonrelativistic context \cite{Zimboras-2014,Johansson-2016,Ding-2021} is as follows.
Consider two distant fermionic modes, $1$ and $2$, in the hands of Alice and Bob, respectively.
If Bob's mode would be in the superposition of empty and occupied states
(shortly $\frac{1}{\sqrt{2}}(\id+f_2^\dagger)\cket{00}=\cket{0}\frac{1}{\sqrt{2}}(\cket{0}+\cket{1})$, 
where $f_i^\dagger$ is the fermionic creation operator and $\id$ is the identity),
then Alice could instantaneously signal to him,
by either applying the unitary $i(f_1^\dagger-f_1)$, or doing nothing 
(for the ticks of synchronized clocks),
since the measurement of the operator $\frac12(f_2+f_2^\dagger+\id)$ by Bob
would give him $0$ or $1$, respectively, with certainty \cite{Ding-2021}.

Imposing the parity superselection rule means 
(i) the restriction to the \emph{physically meaningful state vectors} $\mathcal{H}_Y^\pm\subset\mathcal{H}_Y$,
not containing even-odd superpositions;
(ii) the restriction to the \emph{physically meaningful operators} $\f{\mathcal{A}}_Y^+\subset\f{\mathcal{A}}_Y$,
containing operators not creating even-odd superpositions \eqref{eq:AYpmComm2}; and
(iii) the restriction to the \emph{physically meaningful maps} $\f{\mathcal{B}}_Y^+\subset\f{\mathcal{B}}_Y$,
containing maps not creating odd operators out of even ones \eqref{eq:BYpmComm2}.
In this section, we write out the tensor product structure arising in the $\xi$-locally physical subalgebra,
making possible to elaborate the notions of locality of operations, as well as independence, correlation and entanglement of mode subsets.

\subsection{Tensor product structure in the \texorpdfstring{$\xi$}{xi}-locally physical subalgebra}
\label{sec:ParityPhys.TPSA}

The most important consequence of the local parity superselection is that,
similarly to the qubit case,
the physical operators of disjoint mode subsets commute,
\begin{subequations}
\begin{align}
\label{eq:Comm}
\forall A_X\in\mathcal{A}_X, B_{\bar{X}}\in\mathcal{A}_{\bar{X}}: \quad 
&\bigl[\iota_{X,Y}(A_X), \iota_{\bar{X},Y}(B_{\bar{X}}) \bigr] = 0,\\
\label{eq:Commf}
\forall \f{A}_X\in\f{\mathcal{A}}_X^+, \f{B}_{\bar{X}}\in\f{\mathcal{A}}_{\bar{X}}^+: \quad 
&\bigl[\iotaf_{X,Y}(\f{A}_X), \iotaf_{\bar{X},Y}(\f{B}_{\bar{X}}) \bigr] = 0,
\end{align}
\end{subequations}
recalling \eqref{eq:ACommf}.
This is because the physical subspace of each mode subset consists of
operators of linear combinations of monomials of \emph{even} number of fermionic operators, see \eqref{eq:EYfexpl} and \eqref{eq:AYpmSpan}.
Here we write out the tensor product structure
in the $\xi$-locally physical subalgebra $\f{\mathcal{A}}_\xi^\alleven\subseteq\f{\mathcal{A}}_Y^+$ explicitly,
leading to this commutation.
(For the treatment of the $\xi$-locally physical subalgebra in our formalism, see Appendix~\ref{appsec:Parity.LocPhys}.)

For the partition $\xi= \{X_1,X_2,\dots,X_{\abs{\xi}}\}\in\Pi(Y)$
of the mode subset $Y\subseteq M$,
the tensor product structure is given 
by the unitary $\f{U}_{\ord{\xi}}\in\LieGrp{U}(\mathcal{H}_Y)$,
by which
for $\f{A}_X\in\f{\mathcal{A}}_X^+$ for all $X\in\xi$,
we can write
\begin{equation}
\label{eq:TPSA}
\bigotimesfp_{X\in\xi} \f{A}_X 
= \f{U}_{\ord{\xi}} \Bigl( \bigotimes_{X\in\xi}\f{A}_X \Bigr) \f{U}_{\ord{\xi}}^\dagger.
\end{equation}
(For the proof, by the construction of $\f{U}_{\ord{\xi}}$, see Appendix~\ref{appsec:Parity.LocPhysTPS}.
For the explicit form for few modes, see Appendix~\ref{appsec:Parity.Explu}.)
Note that on the left-hand side we have $\bigotimesfp_{X\in\xi} \f{A}_X\equiv \prod_{X\in\xi} \iotaf_{X,Y}(\f{A}_X)$ 
in the $\xi$-locally physical subalgebra,
by applying the commutativity to \eqref{eq:TPfp}.
It turns out that $\f{U}_{X\bar{X}}$, appearing in \eqref{eq:iotaU}
is just a special case of $\f{U}_{\ord{\xi}}$ for $\ord{\xi}=\tuple{X,\bar{X}}$ see in \eqref{eq:uXXb} and \eqref{eq:uupsilon},
however, the definition \eqref{eq:iotaU}, leading to $\f{U}_{X\bar{X}}$ is meaningful for the whole algebra $\f{\mathcal{A}}_Y$,
but the definition \eqref{eq:TPSA} leading to $\f{U}_{\ord{\xi}}$ is meaningful only in the $\xi$-locally physical subalgebra $\f{\mathcal{A}}_{\xi}^\alleven$.
Note also that although $\f{U}_{\ord{\xi}}$, acting on $\mathcal{H}_Y$, depends on the ordered partition ${\ord{\xi}}$,
this is just an artifact of the representation:
the map $\f{A}_Y\mapsto \f{U}_{\ord{\xi}} \f{A}_Y \f{U}_{\ord{\xi}}^\dagger$, acting on $\f{\mathcal{A}}_Y$,
is the same for all orderings of the parts of $\ord{\xi}$ in the $\xi$-locally physical subalgebra $\f{A}_Y\in\f{\mathcal{A}}_{\xi}^\alleven$.
(See at the end of Appendix~\ref{appsec:Parity.LocPhysTPS}.)
This also justifies the use of the unordered product (arbitrarily ordered product of commuting elements) on the left-hand side of \eqref{eq:TPSA}.
In particular, for the map $\f{\Lambda}_{\ord{\xi}}$, given in \eqref{eq:Lambda}, we have
\begin{equation}
\f{\Lambda}_{\ord{\xi}}\bigrestrict{\f{\mathcal{A}}_{\xi}^\alleven} 
= \f{\Lambda}_{\ord{\xi}'}\bigrestrict{\f{\mathcal{A}}_{\xi}^\alleven},
\end{equation}
where $\ord{\xi}$ and $\ord{\xi}'$ are any two differently ordered partitions
containing the same parts.

The immediate consequence of \eqref{eq:TPSA} is that,
for $\f{A}_X,\f{B}_X\in\f{\mathcal{A}}_X^+$ for all $X\in\xi$,
we have for the product and $*$-operation that
\begin{equation}
\label{eq:allevenPHom}
\bigotimesfp_{X\in\xi}\f{A}_X \bigotimesfp_{X'\in\xi}\f{B}_{X'} = \bigotimesfp_{X\in\xi}\f{A}_X\f{B}_X,\qquad
\bigotimesfp_{X\in\xi}\f{A}_X^\dagger = \Bigl(\bigotimesfp_{X\in\xi}\f{A}_X\Bigr)^\dagger.
\end{equation}
In particular, from the first equation in \eqref{eq:allevenPHom}, the commutativity \eqref{eq:Commf} follows directly.
From \eqref{eq:allevenPHom}, it directly follows that 
\emph{products of positive physical operators are positive,}
\begin{equation}
\label{eq:allevenPPos}
0\leq\f{A}_X\in \f{\mathcal{A}}_X^+
\arrthen
\bigotimesfp_{X\in\xi}\f{A}_X\geq0,
\end{equation}
by using \eqref{eq:posabssquare}.
The important point is that 
this is not necessarily true
without the parity superselection, 
the products of canonical embeddings of positive (so self-adjoint) operators can be not even self-adjoint (so not positive).
(For explicit examples, see Appendix~\ref{appsec:Parity.examples}.)
We will come back to this issue in Section~\ref{sec:ParityPhys.Indep}, and show some detailed results.

With the \eqref{eq:TPSA} tensor product structure in the $\xi$-locally physical subalgebra,
we can also write out explicitly the fermionic embedding of maps \eqref{eq:cmEmbfp} as,
for $\f{A}_X\in\f{\mathcal{A}}_X^+$, $\f{B}_{\bar{X}}\in\f{\mathcal{A}}_{\bar{X}}^+$,
$\f{\Omega}_X\in\f{\mathcal{B}}_X^+$
and $\f{\Xi}_X\in\f{\mathcal{B}}_{\bar{X}}^+$,
\begin{subequations}
\label{eq:mIdExtfpAe}
\begin{align}
\label{eq:mIdExtfpAe1}
\iotabbfp_{X,Y}(\f{\Omega}_X)\bigl(\f{A}_X\otimesfp\f{B}_{\bar{X}}\bigr) &= \f{\Omega}_X(\f{A}_X)\otimesfp\f{B}_{\bar{X}},\\ 
\label{eq:mIdExtfpAe2}
\iotabbfp_{X,Y}(\f{\Omega}_X)\bigl(\f{B}_{\bar{X}}\otimesfp\f{A}_X\bigr) &= \f{B}_{\bar{X}}\otimesfp\f{\Omega}_X(\f{A}_X),\\
\label{eq:mIdExtfpAe3}
\iotabbfp_{\bar{X},Y}(\f{\Xi}_{\bar{X}})\bigl(\f{A}_X\otimesfp\f{B}_{\bar{X}}\bigr) &= \f{A}_X\otimesfp\f{\Xi}_{\bar{X}}(\f{B}_{\bar{X}}).
\end{align}
(This follows simply by the commutativity $\f{A}_X\otimesfp\f{B}_{\bar{X}}=\f{B}_{\bar{X}}\otimesfp\f{A}_X$,
which holds for the physical operators, applying to the definition \eqref{eq:cmEmbfp}.)
So $\iotabbfp_{X,Y}(\f{\Omega}_X)$ is a strong extension in this restricted manner,
both equalities hold in \eqref{eq:mIdExtfp} for the $\set{X,\bar{X}}$-locally physical case.
\end{subequations}

Note also that in the case when
$\ord{\xi} = \tuple{X_1,X_2,\dots,X_{\abs{\ord{\xi}}}}$ is such that
$X_s<X_r$ (elementwisely) for all $s<r$,
that is, the mode subsets $X\in\ord{\xi}$ contain modes neighboring with respect to $Y$,
and are ordered in $\ord{\xi}$ accordingly to the Jordan--Wigner ordering of the modes,
then the effect of $\f{U}_{\ord{\xi}} $ is trivial.
(This can directly be read off from the phase factors \eqref{eq:uupsilon} in Appendix~\ref{appsec:Parity.LocPhysTPS}.)
So, for the $\xi$-locally physical subalgebra,
for these very special partitions,
one can use the usual Kronecker product of matrices in explicit numerical or symbolical calculations;
while for more general partitions, which contain mode subsets
of non-neighboring modes,
the phase factors cannot be neglected,
that is, Kronecker product can be used only after mode reordering.
Neither the latter hold without parity superselection in general. 
(For explicit example, see Appendix~\ref{appsec:Parity.examples}.)

\subsection{Tensor product structure for vectors for the \texorpdfstring{$\xi$}{xi}-locally physical operators}
\label{sec:ParityPhys.TPSH}

If the $\xi$-locally physical subalgebra is considered only, 
then we can have a tensor product structure also in the Hilbert space, 
written as,
for $\f{A}_X\in\f{\mathcal{A}}_X^+$, 
\begin{subequations}
\begin{equation}
\label{eq:TPSHA}
\Bigl( \bigotimesfp_{X\in\xi}\f{A}_X \Bigr)
\Bigl( \bigotimesfp_{X'\in\xi}\cket{\psi_{X'}} \Bigr) 
:=: \bigotimesfp_{X\in\xi} \bigl( \f{A}_X\cket{\psi_X} \bigr),
\end{equation}
leading to
\begin{equation}
\label{eq:TPSH}
\bigotimesfp_{X\in\xi}\cket{\psi_X} = \f{U}_{\ord{\xi}} \Bigl( \bigotimes_{X\in\xi}\cket{\psi_X} \Bigr)
\end{equation}
\end{subequations}
by \eqref{eq:TPSA}.
Note that, contrary to the algebra,
the ordering in $\ord{\xi}$ is a parameter of the tensor product structure in the Hilbert space:
the transformation $\cket{\psi_Y}\mapsto\f{U}_{\ord{\xi}}\cket{\psi_Y}$, acting on $\mathcal{H}_Y$,
depends on the ordering of the parts of $\ord{\xi}$.
However, such ordering $\ord{\xi}$ can be fixed for every partition $\xi$,
and then can be used for the tensor product structure \eqref{eq:TPSA}
of the $\xi$-locally physical subalgebra $\f{\mathcal{A}}_{\xi}^\alleven$.
Without the parity superselection, this could not be done in the whole algebra $\f{\mathcal{A}}_Y$ for all orderings.
Another, related point is
that the fermionic canonical embeddings \eqref{eq:cEmbf}
of $\f{\mathcal{A}}_X^+$ and $\f{\mathcal{A}}_{\bar{X}}^+$
can be realized by the same unitary,
e.g., 
fixing $\ord{\xi}=\tuple{X,\bar{X}}$, we have
\begin{subequations}
\begin{align}
\iotaf_{X,Y}\bigl(\f{A}_X\bigr)
&=\f{U}_{X\bar{X}}\bigl(\f{A}_X\otimes\Idf_{\bar{X}}\bigr)\f{U}_{X\bar{X}}^\dagger,\\
\iotaf_{\bar{X},Y}\bigl(\f{B}_{\bar{X}}\bigr)
&=\f{U}_{\bar{X}X}\bigl(\Idf_X\otimes\f{B}_{\bar{X}}\bigr)\f{U}_{\bar{X}X}^\dagger
=\f{U}_{X\bar{X}}\bigl(\Idf_X\otimes\f{B}_{\bar{X}}\bigr)\f{U}_{X\bar{X}}^\dagger, 
\end{align}
\end{subequations}
for $\f{A}_X\in\f{\mathcal{A}}_X^+$ and $\f{B}_{\bar{X}}\in\f{\mathcal{A}}_{\bar{X}}^+$,
see $\eqref{eq:iotaU}$,
so these act identically 
on $\mathcal{H}_{\bar{X}}$, respectively $\mathcal{H}_X$ 
of the $\f{U}_{X\bar{X}}$-transformed $\mathcal{H}_Y$,
\begin{subequations}
\begin{align}
\iotaf_{X,Y}\bigl(\f{A}_X\bigr)\f{U}_{X\bar{X}}\bigl(\cket{\psi_X}\otimes\cket{\psi_{\bar{X}}}\bigr)
 &= \f{U}_{X\bar{X}} \bigl(\f{A}_X\cket{\psi_X}\otimes\cket{\psi_{\bar{X}}}\bigr),\\
\iotaf_{\bar{X},Y}\bigl(\f{B}_{\bar{X}}\bigr)\f{U}_{X\bar{X}}\bigl(\cket{\psi_X}\otimes\cket{\psi_{\bar{X}}}\bigr)
 &= \f{U}_{X\bar{X}} \bigl(\cket{\psi_X}\otimes\f{B}_{\bar{X}}\cket{\psi_{\bar{X}}}\bigr).
\end{align}
\end{subequations}

Considering pure states, that is, rank-$1$ projectors,
we have that the parity superselection rule for \emph{pure states} is
\begin{equation}
\label{eq:pureSSR}
\cket{\psi_X}\bra{\psi_X}\in\f{\mathcal{A}}_X^+ 
\arriff
\cket{\psi_X}\in\mathcal{H}_X^\pm, 
\end{equation}
(this follows simply by \eqref{eq:HYpmSpan} and \eqref{eq:AYpmSpan}),
that is, a physical pure state is given by a state vector with well-defined fermion number parity \eqref{eq:HYpm}. 
It is also easy to see by \eqref{eq:pureSSR} that 
\emph{the product of canonical embeddings of pure physical states is pure},
\begin{subequations}
\begin{equation}
\label{eq:allevenPpures}
\forall X\in\xi, \cket{\psi_X}\in\mathcal{H}_X^{\epsilon_X}
\arrthen
\bigotimesfp_{X\in\xi} \cket{\psi_X}\bra{\psi_X} = \cket{\psi_Y}\bra{\psi_Y} \in \f{\mathcal{A}}_{\xi}^\alleven \subseteq \f{\mathcal{A}}_Y^+,
\end{equation}
where the joint vector $\cket{\psi_Y}$ can be given by \eqref{eq:TPSA} and \eqref{eq:TPSH} as
\begin{equation}
\label{eq:allevenPvecs}
\cket{\psi_Y} 
= \f{U}_{\ord{\xi}} \Bigl(\bigotimes_{X\in\xi}\cket{\psi_X}\Bigr)
= \bigotimesfp_{X\in\xi}\cket{\psi_X}. 
\end{equation}
\end{subequations}
The important point is that 
such property would not hold
without the parity superselection rule, 
the products of canonical embeddings of pure states can be non-pure \cite{Araki-2003b}.
(It can be nonpositive, so not even a state. 
For explicit example, see Appendix~\ref{appsec:Parity.examples}.)

\subsection{Independence of mode subsets}
\label{sec:ParityPhys.Indep}

Here we present some detailed results on the properties of
fermionic $\ord{\xi}$-elementary product operators in the general case,
that is, without superselection, which illustrate how superselection fits in the general picture.
Let us consider a set of operators $\sset{\f{A}_X\neq0}{\f{A}_X\in\f{\mathcal{A}}_X,X\in\xi}$,
and let us denote with 
$m_+:=\bigabs{\sset{X\in\xi}{\f{A}_X\in\f{\mathcal{A}}_X^+}}$, 
$m_-:=\bigabs{\sset{X\in\xi}{\f{A}_X\in\f{\mathcal{A}}_X^-}}$ and 
$m_0:=\bigabs{\sset{X\in\xi}{\f{A}_X\in\f{\mathcal{A}}_X\setminus(\f{\mathcal{A}}_X^+\cup\f{\mathcal{A}}_X^-)}}$
the number of even, odd and neither-even-nor-odd ones
(the latter ones have nonvanishing projection in both subspaces).

First, for self-adjoint operators of well-defined parity,
 $\f{A}_X^\dagger=\f{A}_X\in \f{\mathcal{A}}_X^{\epsilon_X}$, $m_0=0$,
we have
\begin{subequations}
\begin{equation}
\label{eq:prod.sapm}
\Bigl(\bigotimesfp_{X\in\ord{\xi}}\f{A}_X\Bigr)^\dagger = \bigotimesfp_{X\in\ord{\xi}}\f{A}_X
\arriff
m_- = (\text{$0$ or $1$}) \pmod{4}.
\end{equation}
Second, for self-adjoint operators of non-defined parity, 
$\f{A}_X^\dagger=\f{A}_X\in \f{\mathcal{A}}_X$,
we have
\begin{equation}
\label{eq:prod.sa}
\Bigl(\bigotimesfp_{X\in\ord{\xi}}\f{A}_X\Bigr)^\dagger = \bigotimesfp_{X\in\ord{\xi}}\f{A}_X
\arriff
\begin{aligned}
&\bigl(m_0=0 \;\text{ and }\; m_- = (0 \text{ or } 1) \pmod{4} \bigr)\\
\text{or }
&\bigl( m_0=1 \;\text{ and }\; m_- = 0 \pmod{4} \bigr)
\end{aligned}
\end{equation}
and, if additionally $\f{A}_X\geq0$, then $m_-=0$, and
\begin{equation}
\label{eq:prod.pos}
\bigotimesfp_{X\in\ord{\xi}}\f{A}_X \geq 0
\arriff
m_0 = \text{ $0$ or $1$}.
\end{equation}
\end{subequations}
From the latter we can recover the known result that
product state extension of local states 
exists if and only if all of the local states are physical with at most one exception,
proven originally in \cite{Araki-2003b,Moriya-2006}, from which the main idea of the proof of \eqref{eq:prod.pos} is taken.

For the proof of \eqref{eq:prod.sapm},
$\bigl(\bigotimesfp_{X\in\ord{\xi}}\f{A}_X\bigr)^\dagger = \bigotimesfp_{X\in\ordback{\xi}}\f{A}_X^\dagger$
(where $\ordback{\xi}$ is the reverse ordered $\ord{\xi}$)
which equals to $\bigotimesfp_{X\in\ord{\xi}}\f{A}_X^\dagger$
if and only if no overall $-1$ sign appears when the ordering of operators is reversed, see \eqref{eq:ACommf}.
Even operators do not count when swapping neighboring terms \eqref{eq:ACommf},
 the overall sign comes from swapping odd operators.
Having $m_-$ odd operators, the number of those swappings is
$(m_--1)+(m_--2)+\dots+2+1 = \sum_{k=1}^{m_--1}k = \frac12m_-(m_--1)$,
which is even, if and only if 
$m_-(m_--1) = 0 \pmod{4}$, if and only if
$m_- = (\text{$0$ or $1$}) \pmod{4}$. 

For the proof of \eqref{eq:prod.sa},
having $m_+$, $m_-$ and $m_0$ even, odd and neither-even-nor-odd operators, 
the product of them $\bigotimesfp_{X\in\ord{\xi}}\f{A}_X^\dagger$
has components in subspaces of local parity $m_+'$ times even and $m_-'$ times odd,
such that $(m_+',m_-') = (m_++m_0-k,m_-+k)$ for $k=0,1,\dots,m_0$.
Since these are orthogonal linear subspaces in $\f{\mathcal{A}}_Y$,
the condition \eqref{eq:prod.sapm} has to hold for all such local parities:
$m_-'=m_-+k := (\text{$0$ or $1$}) \pmod{4}$ for all $k=0,1,\dots,m_0$,
that is, we have the condition
\begin{equation*}
\sset{m_-+k}{k=0,1,\dots,m_0}\subseteq\sset{0+4l,1+4l}{l=0,1,\dots}.
\end{equation*}
If $m_0=0$, every operator has well-defined parity,
then $k=0$, and we get back the previous case, $m_- = (\text{$0$ or $1$}) \pmod{4}$.
If $m_0=1$,
then $k=0,1$, and the condition is
$\set{m_-,m_-+1}\subseteq\sset{0+4l,1+4l}{l=0,1,\dots}$
which holds if and only if $m_- = 0 \pmod{4}$.
If $m_0\geq2$,
then $k=0,1,2,\dots,m_0$, and
$\set{m_-,m_-+1,m_-+2,\dots ,m_-+m_0}$ contains at least three consecutive numbers,
so the condition does not hold.

For the proof of \eqref{eq:prod.pos},
first note that if $\f{A}_X \geq 0$, then $\Trf(\f{A}_X) > 0$,
so positive semidefinite operators always have even component, see \eqref{eq:Trfeven} (that is, $m_-=0$);
and the question is, how many of them can have odd component too (that is, what are the possible values of $m_0$).
Without loss of generality, assume that the first $m_0$ operators have odd component,
$\f{A}_{X_k}\notin\f{\mathcal{A}}_{X_k}^+$ for $k=1,2,\dots,m_0$.
Take the odd parts of the first two,
$\f{B}_{X_1}:=\f{A}_{X_1}^-:=\f{\Pi}_{X_1}^-(\f{A}_{X_1})$ and 
$\f{B}_{X_2}:=\f{A}_{X_2}^-:=\f{\Pi}_{X_2}^-(\f{A}_{X_2})$,
and $\f{B}_{X_k}:=\Idf_{X_k}$ for $k = 3,4,\dots,\abs{\ord{\xi}}$,
by which form $\f{B}_Y:=\bigotimesfp_{X\in\ord{\xi}}\f{B}_X=
\f{A}_{X_1}^-\otimesfp\f{A}_{X_2}^-\otimesfp\Idf_{Y\setminus(X_1\cup X_2)}$.
In general, $\f{A}_X^\dagger=\f{A}_X$ if and only if 
$(\f{A}_X^+)^\dagger = \f{A}_X^+$ and
$(\f{A}_X^-)^\dagger = \f{A}_X^-$ 
(because $\f{\Pi}_X^\pm(\f{A}_X)^\dagger = \f{\Pi}_X^\pm(\f{A}_X^\dagger)$, see \eqref{eq:PiY}).
So 
\begin{align*}
\f{B}_X^\dagger &= \f{B}_X,\quad \forall X\in\ord{\xi},\\
\f{B}_Y^\dagger &= \Idf_{Y\setminus(X_1\cup X_2)}\otimesfp\f{A}_{X_2}^-\otimesfp\f{A}_{X_1}^-=-\f{B}_Y,
\end{align*}
see \eqref{eq:ACommf}.
In general, $C^\dagger=\pm C$ if and only if
$\skalpHS{C}{D}\in\field{R}$ for all $D^\dagger=\pm D$ for the same sign.
($C^\dagger=\pm C$ if and only if $\skalpHS{C}{D}=\pm \skalpHS{C^\dagger}{D}$ for all $D^\dagger=D$  or $D^\dagger=-D$,
since both self-adjoint and skew-self-adjoint operators span the finite dimensional operator algebra;
then
$\skalpHS{C}{D}=\pm \skalpHS{C^\dagger}{D}=\skalpHS{C^\dagger}{\pm D}=\skalpHS{C^\dagger}{D^\dagger}=\skalpHS{C}{D}^*$ for all $D^\dagger=\pm D$ for the same sign.)
So, forming the inner product 
\begin{equation*}
\BigskalpHS{\bigotimesfp_{X\in\ord{\xi}}\f{A}_X}{\f{B}_Y} = 
\bigskalpHS{\f{A}_{X_1}}{\f{B}_{X_1}}\bigskalpHS{\f{A}_{X_2}}{\f{B}_{X_2}} \prod_{k=3}^{\abs{\ord{\xi}}} \Trf(\f{A}_{X_k})
\end{equation*}
(see \eqref{eq:TPfpHS}, and $\bigskalpHS{\f{A}_{X_k}}{\f{B}_{X_k}}=\Trf(\f{A}_{X_k})>0$ for $k = 3,4,\dots,\abs{\ord{\xi}}$),
we have that the right-hand side is real,
so $\bigl(\bigotimesfp_{X\in\ord{\xi}}\f{A}_X\bigr)^\dagger = -\bigotimesfp_{X\in\ord{\xi}}\f{A}_X$ on the left-hand side,
leading to that,
to have self-adjoint product $\bigotimesfp_{X\in\ord{\xi}}\f{A}_X$,
there cannot be two operators $\f{A}_X$ having odd component (so $m_0\leq1$).
On the other hand, if none or one $\f{A}_X$ has odd component ($m_0\leq1$),
then the operators $\iotaf_{X,Y}(\f{A}_X)$ for all $X\in\xi$ are commuting pairwise, see \eqref{eq:PiY},
and, due to the existence of a common set of spectral projectors,
their product $\prodord_{X\in\ord{\xi}}\iotaf_{X,Y}(\f{A}_X)\equiv\bigotimesfp_{X\in\ord{\xi}}\f{A}_X$ is also positive.

The mode subsets $X\in\xi$ are called \emph{algebraically independent},
if the subalgebras describing them commute with each other.
As we have seen, CAR subalgebras 
$\iotaf_{X,Y}(\f{\mathcal{A}}_X)$
of disjoint mode subsets $X\in\xi$
do not commute with each other \eqref{eq:ACommf},
leading to that the mode subsets $X\in\xi$ are not algebraically independent.
There is also a weaker concept.
The mode subsets $X\in\xi$ are called \emph{statistically independent}
(also called \emph{$\mathrm{C}^*$-independent} \cite{Florig-1997,Summers-1990,Moriya-2002}),
if, for all states of the mode subsets, there exist joint state extensions;
that is, for all $X\in\xi$ for all $\f{\rho}_X\in\f{\mathcal{D}}_X$
there exists $\f{\omega}_Y\in\f{\mathcal{D}}_Y$, 
such that $\Trf_{Y,X}(\f{\omega}_Y)=\f{\rho}_X$ for all $X\in\xi$.
As we have seen, the product extension
$\f{\omega}_Y=\bigotimesfp_{X\in\xi}\f{\rho}_X$ would be a joint extension,
however, not a joint \emph{state} extension, 
since it is not a state in general,
it is not positive semidefinite if $\f{\rho}_X\in\f{\mathcal{D}}_X$ is not physical for at least two $X\in\xi$, see \eqref{eq:prod.pos}.
(For explicit example, see Appendix~\ref{appsec:Parity.examples}.)
Moreover, it can be shown \cite{Araki-2003b} that
if $\f{\rho}_X\in\f{\mathcal{D}}_X$ are pure for all $X\in\xi$,
then there exists no joint state extension
if (and only if) at least two $\f{\rho}_X\in\f{\mathcal{D}}_X$ are not physical.

\subsection{Locality of maps}
\label{sec:ParityPhys.Loc}

After constructing maps on mode subset $Y$ from maps on mode subsets $X\in\xi$,
acting properly on fermionic $\ord{\vs{\xi}}$-elementary products \eqref{eq:mTPfpTPfp} 
at the end of Section~\ref{sec:Tensors.mTP}, we mentioned that the same construction would not act properly on 
$\ord{\xi}'$-elementary products, ordered differently than $\ord{\xi}$.
Having the anticommutation \eqref{eq:ACommf}, we can now illustrate this,
by taking two anticommuting elements, 
$\f{A}_{X_1}\in \f{\mathcal{A}}_{X_1}^-$,
$\f{A}_{X_2}\in \f{\mathcal{A}}_{X_2}^-$,
a parity changing map $\f{\Omega}_{X_1}\in\f{\mathcal{B}}_{X_1}^-$,
a parity preserving map $\f{\Omega}_{X_2}\in\f{\mathcal{B}}_{X_2}^+$,
and $\ord{\xi}'=\tuple{X_2,X_1,X_3,\dots,X_{\abs{\xi}}}$,
interchanging the first two mode subsets only,
leading to
\begin{equation*}
\begin{split}
\Bigl(\bigotimesfp_{X\in\ord{\xi}} \f{\Omega}_X\Bigr)
\Bigl( \bigotimesfp_{X\in\ord{\xi}'}\f{A}_X \Bigr)
&=\Bigl(\bigotimesfp_{X\in\ord{\xi}} \f{\Omega}_X\Bigr)
\bigl(\f{A}_{X_2}\otimesfp\f{A}_{X_1}\otimesfp\f{A}_{X_3}\otimesfp\dots\otimesfp\f{A}_{X_{\abs{\xi}}}\bigr)\\
&=\Bigl(\bigotimesfp_{X\in\ord{\xi}} \f{\Omega}_X\Bigr)
\bigl(-\f{A}_{X_1}\otimesfp\f{A}_{X_2}\otimesfp\f{A}_{X_3}\otimesfp\dots\otimesfp\f{A}_{X_{\abs{\xi}}}\bigr)\\
&=-\f{\Omega}_{X_1}(\f{A}_{X_1})\otimesfp\f{\Omega}_{X_2}(\f{A}_{X_2})
\otimesfp\f{\Omega}_{X_3}(\f{A}_{X_3})\otimesfp\dots\otimesfp\f{\Omega}_{\abs{\xi}}(\f{A}_{\abs{\xi}})\\
&=-\f{\Omega}_{X_2}(\f{A}_{X_2})\otimesfp\f{\Omega}_{X_1}(\f{A}_{X_1})
\otimesfp\f{\Omega}_{X_3}(\f{A}_{X_3})\otimesfp\dots\otimesfp\f{\Omega}_{\abs{\xi}}(\f{A}_{\abs{\xi}})\\
&=-\bigotimesfp_{X\in\ord{\xi}'}\f{\Omega}_X(\f{A}_X).
\end{split}
\end{equation*}
From this, it is also clear that for general maps $\f{\Omega}_X$ for all $X\in\xi$,
one cannot form $\f{\Omega}_Y$, which would act as
$\f{\Omega}_Y\Bigl( \bigotimesfp_{X\in\ord{\xi}'}\f{A}_X \Bigr)
=\bigotimesfp_{X\in\ord{\xi}}\f{\Omega}_X(\f{A}_X)$ for all $\ord{\xi}$ ordering of $\xi$.
However, it can be done for even maps.

A special case of the above is that of the fermionic map embedding \eqref{eq:cmEmbfp}.
At the end of Section~\ref{sec:Tensors.mcEmb}, we mentioned that it is not a strong extension,
the second equality does not hold in \eqref{eq:mIdExtfp}.
Having the anticommutation \eqref{eq:ACommf}, we can now illustrate this,
by taking two anticommuting elements,
$\f{A}_{X}\in \f{\mathcal{A}}_{X}^-$,
$\f{B}_{\bar{X}}\in \f{\mathcal{A}}_{\bar{X}}^-$,
and a parity changing map $\f{\Omega}_{X}\in\f{\mathcal{B}}_X^-$,
leading to
\begin{equation*}
\begin{split}
\iotabbfp_{X,Y}(\f{\Omega}_X)(\f{B}_{\bar{X}}\otimesfp\f{A}_X)
=(\f{\Omega}_X\otimesfp\IIdf_{\bar{X}})(-\f{A}_X\otimesfp\f{B}_{\bar{X}})
=-\f{\Omega}_X(\f{A}_X)\otimesfp\f{B}_{\bar{X}}
=-\f{B}_{\bar{X}}\otimesfp\f{\Omega}_X(\f{A}_X).
\end{split}
\end{equation*}
From this, it is also clear that for general maps $\f{\Omega}_X$,
one cannot form strong extension over the whole algebra.

Turning to even maps, first,
the product $\bigotimesfp_{X\in\ord{\xi}} \f{\Omega}_X$
of the even maps $\f{\Omega}_X\in\f{\mathcal{B}}_X^+$
acts properly on $\ord{\xi}'$-elementary products
of different ordering from $\ord{\xi}$,
\begin{equation}
\label{eq:mTPfpBe}
\Bigl(\bigotimesfp_{X\in\ord{\xi}} \f{\Omega}_X\Bigr)
\Bigl( \bigotimesfp_{X\in\ord{\xi}'}\f{A}_X \Bigr)
=\bigotimesfp_{X\in\ord{\xi}'}\f{\Omega}_X(\f{A}_X).
\end{equation}
Indeed, for operators of well-defined parity $\f{A}_X\in\f{\mathcal{A}}_X^{\epsilon_X}$,
the same overall sign appears in the two steps in the calculation
\begin{equation*}
\begin{split}
\Bigl(\bigotimesfp_{X\in\ord{\xi}} \f{\Omega}_X\Bigr)
\Bigl( \bigotimesfp_{X\in\ord{\xi}'}\f{A}_X \Bigr)
=\Bigl(\bigotimesfp_{X\in\ord{\xi}} \f{\Omega}_X\Bigr)
\Bigl(\pm \bigotimesfp_{X\in\ord{\xi}}\f{A}_X \Bigr)
=\pm\bigotimesfp_{X\in\ord{\xi}} \f{\Omega}_X(\f{A}_X)
=(\pm1)^2\bigotimesfp_{X\in\ord{\xi}'}\f{\Omega}_X(\f{A}_X),
\end{split}
\end{equation*}
since $\f{\Omega}_X$ does not change the parity.
Then \eqref{eq:mTPfpBe} follows by linearity.

As a special case of the above, 
the fermionic embedding $\iotabbfp_{X,Y}(\f{\Omega}_X)$ 
of the \emph{even} map $\f{\Omega}_X\in\f{\mathcal{B}}_X^+$ 
is a \emph{strong extension} for the whole algebra $\f{A}_Y$ 
(not only on the $\set{X,\bar{X}}$-physical subalgebra $\f{\mathcal{A}}_{X\bar{X}}^\alleven$, 
as we have seen \eqref{eq:mIdExtfpAe1}-\eqref{eq:mIdExtfpAe2} in Section~\ref{sec:ParityPhys.TPSA}),
since in this case
not only \eqref{eq:mIdExtfp1} holds, but there is equality also in \eqref{eq:mIdExtfp2},
that is,
for $\f{A}_X\in\f{\mathcal{A}}_X$, $\f{B}_{\bar{X}}\in\f{\mathcal{A}}_{\bar{X}}$,
and for $\f{\Omega}_X\in\f{\mathcal{B}}_X^+$,
\begin{subequations}
\label{eq:mIdExtfpBe}
\begin{align}
\label{eq:mIdExtfpBe1}
\iotabbfp_{X,Y}(\f{\Omega}_X)\bigl(\f{A}_X\otimesfp\f{B}_{\bar{X}}\bigr) &= \f{\Omega}_X(\f{A}_X)\otimesfp\f{B}_{\bar{X}},\\ 
\label{eq:mIdExtfpBe2}
\iotabbfp_{X,Y}(\f{\Omega}_X)\bigl(\f{B}_{\bar{X}}\otimesfp\f{A}_X\bigr) &= \f{B}_{\bar{X}}\otimesfp\f{\Omega}_X(\f{A}_X).
\end{align}
\end{subequations}
Indeed, for $\f{A}_X\in\f{\mathcal{A}}_X^{\epsilon_X}$, $\f{B}_{\bar{X}}\in\f{\mathcal{A}}_{\bar{X}}^{\epsilon_{\bar{X}}}$, 
we have
\begin{equation*}
\iotabbfp_{X,Y}(\f{\Omega}_X)\bigl(\f{B}_{\bar{X}}\otimesfp\f{A}_X\bigr)
=\pm\iotabbfp_{X,Y}(\f{\Omega}_X)\bigl(\f{A}_X\otimesfp\f{B}_{\bar{X}}\bigr)
=\pm\f{\Omega}_X(\f{A}_X)\otimesfp\f{B}_{\bar{X}}
=(\pm1)^2\f{B}_{\bar{X}}\otimesfp\f{\Omega}_X(\f{A}_X),
\end{equation*}
where the lower sign is for the case $\epsilon_X=\epsilon_{\bar{X}}=-1$
(two nonphysical operators anticommute),
and $\f{\Omega}_X\in\f{\mathcal{B}}_X^+$ does not change the parity \eqref{eq:BYpmComm2},
$\f{\Omega}_X(\f{A}_X)\in\f{\mathcal{A}}_X^{\epsilon_X}$.

As we have mentioned in Section~\ref{sec:ParityMath.B},
in a strict sense,
a map can be defined physically meaningfully only for physical operators $\f{\mathcal{A}}_X^+$.
On the other hand, by the embedding $\iotabbfp_{X,Y}$, due to \eqref{eq:mIdExtfpBe}, 
an even map $\f{\Omega}_X\in\f{\mathcal{B}}_X^+$ 
can act on odd-odd parts of a physical operator of a larger mode subset,
as $\f{\mathcal{A}}_Y^+ = \f{\mathcal{A}}_{X\bar{X}}^{++} \oplus \f{\mathcal{A}}_{X\bar{X}}^{--}$,
and, in principle, its effect on odd-odd operators can be completely independent of 
its effect on even-even operators, which has consequences for the notion of local maps.

In the theory of correlation and entanglement,
the characterization with respect to local maps is essential.
In the laboratory, \emph{only physically defined maps can be performed},
for which we have the embedding above, working as a strong extension \eqref{eq:mIdExtfpBe}.
However, using a single physically defined $\f{\Omega}_X$ can be too restrictive for the definition of locality,
so we call a map \emph{$X$-local} for a subset of modes $X\subseteq Y$,
if it is of the form 
\begin{subequations}
\begin{equation}
\label{eq:Xlocal}
\f{\Omega}_Y= \iotabbfp_{X,Y}(\f{\Omega}_X)\circ \f{\Pi}_Y^+ + \f{\Xi}_Y,
\end{equation}
with the physically defined map $\f{\Omega}_X$
and the physically defined map $\f{\Xi}_Y$,
such that $\f{\Xi}_Y(\f{A}_Y)=0$ for all $\f{A}_Y\in\f{\mathcal{A}}_{X\bar{X}}^{++}$;
and
we call a map \emph{$\xi$-local} for a partition $\xi$ of $Y$,
if it is of the form 
\begin{equation}
\label{eq:xilocal}
\f{\Omega}_Y=\bigotimesfp_{X\in\xi}\f{\Omega}_X + \f{\Xi}_Y,
\end{equation}
\end{subequations}
with physically defined maps $\f{\Omega}_X\in\f{\mathcal{B}}_X^+$,
and the physically defined map $\f{\Xi}_Y$, 
such that $\f{\Xi}_Y(\f{A}_Y)=0$ for all $\f{A}_Y\in\f{\mathcal{A}}_\xi^\alleven$.
With these we have that $\xi$-local maps are generated by $X$-local maps (where $X\in\xi$),
and $X$-local maps are the special cases of $\xi$-local maps (where $X\in\xi$).
The maps $\f{\Xi}_Y$ are somewhat arbitrary, containing the operations which we cannot handle locally,
since these act on operators having no physical parts locally.
The definition of $X$-local maps leads to that these are also strong extensions if this is understood only for locally physical operators.
Special cases are when $\f{\Xi}_Y=0$ (such $X$-local maps annihilate $\f{\mathcal{A}}_{X\bar{X}}^{--}$),
or when $\f{\Xi}_Y=\IIdf_Y\restrict{\f{\mathcal{A}}_{X\bar{X}}^{--}}$ (such $X$-local maps do not change $\f{\mathcal{A}}_{X\bar{X}}^{--}$).
An example for nontrivial action on $\f{\mathcal{A}}_{X\bar{X}}^{--}$ is
an important special case, 
when $\f{U}_Y=\iotaf_{X,Y}(\f{U}_X)=\f{U}_X\otimesfp\Idf_{\bar{X}}$ 
(with the physical unitary $\f{U}_X\in \f{\mathcal{A}}_X^+$) acts as
$\f{\Omega}_Y(\f{A}_Y)=\f{U}_Y \f{A}_Y \f{U}_Y^\dagger$, that is, $\f{\Omega}_Y=\Ad_{\f{U}_Y}$;
then $\f{\Omega}_X=\f{\Pi}_X^+\circ\Ad_{\f{U}_X}\circ\f{\Pi}_X^+=\Ad_{\f{U}_X}\circ\f{\Pi}_X^+$,
and $\f{\Xi}_Y = \f{\Pi}_{X \bar{X}}^{--} \circ \Ad_{\f{U}_Y} \circ \f{\Pi}_{X \bar{X}}^{--}=\Ad_{\f{U}_Y} \circ \f{\Pi}_{X \bar{X}}^{--}$.
Similarly, when $\f{U}_Y=\bigotimesfp_{X\in\xi}\f{U}_X$ 
(product of physical unitaries $\f{U}_X\in \f{\mathcal{A}}_X^+$) acts as
$\f{\Omega}_Y=\Ad_{\f{U}_Y}$;
then $\f{\Omega}_X=\Ad_{\f{U}_X}\circ\f{\Pi}_X^+$,
and $\f{\Xi}_Y = \Ad_{\f{U}_Y} \circ (\f{\Pi}_Y^+-\f{\Pi}_{\xi}^\alleven)$.

Trace preserving completely positive (TPCP) $\xi$-local maps are 
called \emph{$\xi$-local operations} ($\xi$-LO).
\emph{$\xi$-local operations and classical communications} ($\xi$-LOCC)
can also be formulated in the usual way.
That is,
when the application of $\xi$-LOs 
may depend on outcomes of measurements performed locally,
that is, described by $X$-local TPCP maps $\f{\Omega}_Y=\sum_m\f{\Omega}_{Y,m}$,
where the measurement outcomes
are given by $X$-local trace nonincreasing CP maps $\f{\Omega}_{Y,m}$.

\subsection{Fermionic correlation and entanglement under parity superselection}
\label{sec:ParityPhys.Corrent}

Turning to quantum states, let us use the notations
$\f{\mathcal{D}}_Y^+ := \f{\mathcal{D}}_Y\cap\f{\mathcal{A}}_Y^+$
for the \emph{physical states}, and
$\f{\mathcal{D}}_\xi^\alleven := \f{\mathcal{D}}_Y\cap\f{\mathcal{A}}_\xi^\alleven$
for the \emph{$\xi$-locally physical states}
for the partition $\xi\in\Pi(Y)$ of the mode subset $Y\subseteq M$.

Making use of the parity superselection,
for the definition of $\xi$-uncorrelated mode subsets,
we expect factorizing expectation values again, however,
this time not for all $\ord{\xi}$-elementary product operators,
but only for $\xi$-locally physical ones.
Expressing these with the Hilbert--Schmidt inner product \eqref{eq:HS}
and the state $\f{\rho}_Y\in\f{\mathcal{D}}_Y^+$,
we have the following definition for the \emph{$\xi$-uncorrelated physical states of mode subset $Y$} 
(states of the $\xi$-uncorrelated mode subset $Y$)
\begin{equation}
\label{eq:defuncssfp}
\forall X\in\xi,
\forall \f{A}_X\in\f{\mathcal{A}}_X^+: \quad
\BigskalpHS{\f{\rho}_Y}{\bigotimesfp_{X\in\xi} \f{A}_X}
:= \prod_{X\in\xi}\bigskalpHS{\f{\rho}_X}{\f{A}_X},
\end{equation}
where $\f{\rho}_X=\f{\Tr}_{Y,X}(\f{\rho}_Y)$ is the fermionic reduced state \eqref{eq:defreducedf}, \eqref{eq:reducedfPTf}.
The difference between this and \eqref{eq:defuncfp} is in that
this condition is given only for physical operators,
so only for the $\xi$-even subspace $\f{\mathcal{A}}_\xi^\alleven$.
Because of this, 
the step from \eqref{eq:derivuncfp} to \eqref{eq:uncprodfp} 
in the derivation in Section~\ref{sec:CorrEnt.fermion}
cannot be done, we have the result on the $\xi$-even subspace only.
This leads to
the set of \emph{$\xi$-uncorrelated physical states of mode subset $Y$} as
\begin{subequations}
\begin{equation}
\label{eq:Dfpupsunc}
\f{\mathcal{D}}_{\text{$\xi$-unc}}^+
:= \Bigsset{\f{\rho}_Y\in\f{\mathcal{D}}_Y^+}{\f{\Pi}_\xi^\alleven(\f{\rho}_Y) = \bigotimesfp_{X\in\xi} \Trf_{Y,X}(\f{\rho}_Y) },
\end{equation}
using the projector \eqref{eq:Piupseps}.
(This is the multipartite generalization of the state set $\mathcal{P}1_\pi$ in \cite{Banuls-2007}.)
Note that if $\f{\rho}_Y\in\f{\mathcal{D}}_Y^+$,
then $\Trf_{Y,X}(\f{\rho}_Y)\in\f{\mathcal{D}}_X^+$ for all $X\in\xi$ because of \eqref{eq:PTrfeven},
then $\bigotimesfp_{X\in\xi}\Trf_{Y,X}(\f{\rho}_Y)\in\f{\mathcal{D}}_\xi^\alleven$, 
because of \eqref{eq:cEmbeven} and \eqref{eq:TPfp},
this is why $\f{\Pi}_\xi^\alleven$ could be dropped on the right-hand side of the condition.
On the other hand, it is also important to note that 
$\bigotimesfp_{X\in\xi}\Trf_{Y,X}\bigl(\f{\Pi}_\xi^\alleven(\f{\rho}_Y)\bigr)
= \bigotimesfp_{X\in\xi}\f{\Pi}_X^+\bigl(\Trf_{Y,X}(\f{\rho}_Y)\bigr)
= \f{\Pi}_\xi^\alleven\bigl(\bigotimesfp_{X\in\xi}\Trf_{Y,X}(\f{\rho}_Y)\bigr)
= \bigotimesfp_{X\in\xi} \Trf_{Y,X}(\f{\rho}_Y)$, by \eqref{eq:Piupseps},
so these are the physical states which are product in the $\xi$-locally physical subspace.
There is no restriction on the other $\xi$-local parity subspaces of $\f{\mathcal{D}}_Y^+$.
Although these states cannot be prepared from pure product physical states (product physical states are $\xi$-locally physical) 
by $\xi$-LO in general,
the correlations in these states cannot be accessed by $\xi$-LO.
The other states are \emph{$\xi$-correlated physical} ones, contained in $\f{\mathcal{D}}_Y^+\setminus\f{\mathcal{D}}^+_{\text{$\xi$-unc}}$.

We can write also
the states which can be prepared from pure product physical states by $\xi$-LO,
leading to the set of \emph{$\xi$-product physical states of mode subset $Y$},
which can also be called \emph{strongly $\xi$-uncorrelated physical states of mode subset $Y$} 
(states of the strongly $\xi$-uncorrelated mode subset $Y$) as
\begin{align}
\label{eq:Dupsprodfp}
\f{\mathcal{D}}_{\text{$\xi$-unc}}^\alleven
:= \Bigsset{\f{\rho}_Y\in\f{\mathcal{D}}_Y^+}{\f{\rho}_Y = \bigotimesfp_{X\in\xi} \Trf_{Y,X}(\f{\rho}_Y) }
\subset \f{\mathcal{D}}_\xi^\alleven,
\end{align}
\end{subequations}
thanks to the tensor product structure in $\f{\mathcal{A}}_\xi^\alleven$, see Section~\ref{sec:ParityPhys.TPSA}.
(This is the multipartite generalization of the state set $\mathcal{P}3_\pi$ in \cite{Banuls-2007}.)
The inclusion follows from the same reasoning as above.
It can also be seen that $\xi$-LO-preparable ($\xi$-product) states are $\xi$-uncorrelated,
\begin{equation}
\label{eq:Dfpupsuncinc}
\f{\mathcal{D}}_{\text{$\xi$-unc}}^\alleven \subseteq
\f{\mathcal{D}}_{\text{$\xi$-unc}}^+.
\end{equation}
Note that if $\f{\rho}_Y\in\f{\mathcal{D}}_Y^+$,
and $\f{\rho}_Y\in\f{\mathcal{D}}_{\text{$\xi$-unc}}$, see \eqref{eq:DfpupsuncwoSSR},
then $\f{\rho}_Y\in\f{\mathcal{D}}_{\text{$\xi$-unc}}^\alleven$
because of \eqref{eq:prod.pos},
so we have
$\f{\mathcal{D}}_{\text{$\xi$-unc}}^\alleven=\f{\mathcal{D}}_{\text{$\xi$-unc}}\cap\f{\mathcal{D}}_Y^+$.
The difference between $\f{\mathcal{D}}_{\text{$\xi$-unc}}^\alleven$ and $\f{\mathcal{D}}_{\text{$\xi$-unc}}$
is restricted to nonphysical subspaces $\f{\mathcal{A}}_\xi^{\vs{\epsilon}}$ which have exactly one nonphysical part,
see \eqref{eq:prod.pos},
that is, when $\vs{\epsilon}$ is such that $\abs{\sset{X\in\xi}{\epsilon_X=-1}}=1$.

We can write the set of statistical mixtures of $\xi$-uncorrelated physical states,
which can also be called \emph{weakly $\xi$-separable physical states of mode subset $Y$} 
(states of the weakly $\xi$-separable mode subset $Y$) as
\begin{subequations}
\begin{equation}
\label{eq:Dfpupsconv}
\f{\mathcal{D}}_{\text{$\xi$-sep}}^+ := \Conv \f{\mathcal{D}}_{\text{$\xi$-unc}}^+.
\end{equation}
(This is the multipartite generalization of the state set $\mathcal{S}1_\pi$ in \cite{Banuls-2007}.)

A mode subset $Y\subseteq M$ is \emph{$\xi$-separable}
that is, separable with respect to the partition $\xi\in\Pi(Y)$,
if it can be prepared by the use of $\xi$-LOCC only
from uncorrelated $\xi$-local sources.
So the set of \emph{$\xi$-separable physical states of mode subset $Y$} 
(states of the $\xi$-separable mode subset $Y$) is
\begin{equation}
\label{eq:Dfpupssep}
\f{\mathcal{D}}_{\text{$\xi$-sep}}^\alleven := \Conv \f{\mathcal{D}}_{\text{$\xi$-unc}}^\alleven.
\end{equation}
\end{subequations}
(This is the multipartite generalization of the state set $\mathcal{S}3_\pi$ in \cite{Banuls-2007}.)
The other states are \emph{$\xi$-entangled physical} ones, contained in $\f{\mathcal{D}}^+_Y\setminus\f{\mathcal{D}}_{\text{$\xi$-sep}}^\alleven$;
to prepare them, some quantum communication (quantum interaction) is needed among the mode subsets.
It can also be seen that $\xi$-LOCC-preparable states are mixtures of $\xi$-uncorrelated states,
\begin{equation}
\label{eq:Dfpupssepinc}
\f{\mathcal{D}}_{\text{$\xi$-sep}}^\alleven \subseteq
\f{\mathcal{D}}_{\text{$\xi$-sep}}^+,
\end{equation}
following from \eqref{eq:Dfpupsuncinc}.

Note that for the so called Level~I notions of partial correlation or entanglement
\cite{Szalay-2015b,Szalay-2017,Szalay-2018,Szalay-2019},
it is enough to consider one partition at a time,
which would allow the use of
standard partial trace and Kronecker product \emph{after particular mode reordering}
to write conditions like \eqref{eq:Dfpupsunc}, \eqref{eq:Dupsprodfp}, \eqref{eq:Dfpupsconv} or \eqref{eq:Dfpupssep}
\emph{in terms of matrix elements}, as we have seen.
For Level~II notions, 
such as $k$-partitionability, $k$-producibility or $k$-stretchability of correlation or entanglement
\cite{Szalay-2015b,Szalay-2017,Szalay-2018,Szalay-2019},
one has to consider many incompatible partitions simultaneously,
and the use of many mode reorderings in parallel would be highly inconvenient.
By contrast, 
applying phase factors in the use of $\Trf_{Y,X}$ and $\otimesfp$
is straightforward, and much simpler than mode reordering. 
The same holds when functions of reduced states are evaluated,
which is typically the case of correlation and entanglement \emph{measures}.

Summing up, we have seen that
the two characteristic properties of correlation and of entanglement are split up \cite{Banuls-2007}.
For correlation, factorizing expectation values of $\xi$-locally physical quantities (no correlation) \eqref{eq:Dfpupsunc}
leads to weaker condition than $\xi$-LO-preparability (productness) \eqref{eq:Dupsprodfp}.
Similarly for entanglement,
statistical mixtures of uncorrelated modes \eqref{eq:Dfpupsconv}
leads to a weaker condition than $\xi$-LOCC-preparability (separability) \eqref{eq:Dfpupssep}.

\section{Summary and remarks}
\label{sec:Summ}

The operator algebra of $n$ fermionic modes is isomorphic to that of $n$-qubit systems,
the difference between them is twofold:
the \emph{embedding of subalgebras} corresponding to mode subsets and multiqubit subsystems
on the one hand,
and the \emph{parity superselection} for fermions
on the other.
In this work, these two points were discussed extensively,
and illustrated in the Jordan--Wigner representation.
Summing up, we emphasize the main points here.

In Section~\ref{sec:JW},
we recalled the Jordan--Wigner representation of fermionic modes. 
Because of the anticommutation of the fermionic operators,
the fermionic occupation cannot be encoded \emph{locally} into qubits,
as can be seen also in the form $\Gammaf_Y$ of Jordan--Wigner representation \eqref{eq:EYf}.
Two bases \eqref{eq:bases}
are defined in the operator algebra,
which are natural in the qubit and fermionic cases \eqref{eq:basesexpl},
leading to two isomorphic algebras $\mathcal{A}_Y$ and $\f{\mathcal{A}}_Y$.
The map $\Phi_Y$ connects the two bases \eqref{eq:PhiY},
it is unitary \eqref{eq:PhiU}, and although it is not a $*$-algebra isomorphism,
it connects the products of qubit and (ordered) fermionic operators \eqref{eq:EYXEYXftraf},
so it respects the subsystem/mode-subset structure.
One can think of the effect of $\Phi_Y^{-1}$ as transforming out the fermionic nature of the operators,
leaving only the quantum information about the occupation structure.
However, one should be careful with this interpretation, since $\Phi_Y$ is not positive, 
and not even self-adjointness preserving.

In Section~\ref{sec:Tensors},
we have constructed some advanced tools for the Jordan--Wigner representation.
The way of this turned out to be mainly
(i)~transforming out the fermionic nature of the operators by $\Phi_Y^{-1}$,
so that only the occupation properties of the modes remain,
(ii)~then applying the standard tool for the occupation properties,
using the natural tensor product structure for qubits,
(iii)~then transforming back by $\Phi_Y$.
Based on analogy with the standard basis,
by the fermionic basis we have constructed 
the \emph{fermionic tensor product} \eqref{eq:EYfXs}
through the unitary map $\f{\Psi}_\xi$ \eqref{eq:Psiupsilon};
the \emph{fermionic canonical embedding} \eqref{eq:cEmbf},
being the generalization of $\Gammaf_Y$ \eqref{eq:GammafTPf};
the \emph{fermionic partial trace} \eqref{eq:TPfPTf}, \eqref{eq:PTffp},
being the $\Phi_Y$-adjoined version of the qubit partial trace \eqref{eq:PTf};
the \emph{fermionic product of maps} \eqref{eq:mTPfTPf}-\eqref{eq:mTPfpTPfp};
and the \emph{fermionic embedding of maps} \eqref{eq:mapInjf}-\eqref{eq:mapInjfp}.

Fermionic systems are usually treated in the algebraic way,
when the algebra of the whole system is considered, then subalgebras describing the subsystems are characterized.
Here we have followed the opposite way,
when the algebra of the system is built up from the algebras of the subsystems,
analogously to the qubit case,
where this is made possible by the tensor product structure, being present naturally in qubit systems.
So it is important to emphasize again that
the operation which we call ``fermionic tensor product'' for convenience 
is not a proper tensor product,
it obeys only the linear but not the algebraic properties.
Also, the fermionic tensor product $\otimesf$ \eqref{eq:TPf} and the usual product of canonical embeddings $\otimesfp$ \eqref{eq:TPfp}
are different in general, connected by the unitary map $\f{\Lambda}_{\ord{\xi}}$ \eqref{eq:Lambda}.
The exception is the case of 
partitions containing subsets of neighboring modes, ordered accordingly to the Jordan--Wigner ordering,
considered usually in the literature,
when $\f{\Lambda}_{\ord{\xi}}$ is the identity map  \eqref{eq:specialmulti}.
Another important point is that,
since there is no proper tensor product structure compatible with the mode subalgebra structure,
the fermionic product and embedding of maps constructed are not canonical. 
Nevertheless, the practical importance of these constructions is that 
they provide explicit ways of writing maps acting on mode subsets \eqref{eq:mTPf}-\eqref{eq:mTPfp}, \eqref{eq:mapInjf}-\eqref{eq:mapInjfp},
by $\f{\Psi}_\xi$ and $\f{\Lambda}_{\ord{\xi}}$.

From a technical point of view, 
this formalism, by $\Phi_Y$, $\f{\Psi}_\xi$ and $\f{\Lambda}_{\ord{\xi}}$,
provides a very convenient way of book-keeping fermionic phase factors,
coming from the anticommutation of the fermionic creation and annihilation operators.
It allows to handle arbitrary mode subsets and arbitrary partitionings in a uniform way,
without the need for ad hoc reordering of the modes.
The resulting formulas \eqref{eq:phasefY}, \eqref{eq:hups} and \eqref{eq:lups} are easy to implement also in numerical program packages,
where concrete matrices and Kronecker products are employed instead of abstract generated algebras.
Note that 
standard partial trace and Kronecker product of matrices, i.e., without phase factors, can be used
for special mode subsets or partitions only,
which can be reached by applying mode reordering in general
(Sections~\ref{sec:Tensors.Lambda}, \ref{sec:States.redfmx}, \ref{sec:ParityPhys.TPSA} and \ref{sec:ParityPhys.Corrent}).
In many cases, mode reordering is inconvenient,
while the formalism presented is straightforward.

In Section~\ref{sec:States},
we have considered states and reduced states in the qubit and fermionic cases.
We have shown that,
as the \emph{qubit reduced states} \eqref{eq:defreduced} are given by the usual \emph{partial trace} \eqref{eq:TPPT},
the \emph{fermionic reduced states} \eqref{eq:defreducedf} are given by the \emph{fermionic partial trace} \eqref{eq:TPfPTf}, \eqref{eq:reducedfPTf}.
This is because the state reduction, given by the partial traces, 
is the adjoint map of the state extension, given by the canonical embeddings in both cases \eqref{eq:cembPTadjgen}.
An important, out-of-the-box result 
is the calculation of the \emph{fermionic reduced density matrices} \eqref{eq:rhoXstd} 
(Appendix \ref{appsec:States.ExplR}),
using the phase factors \eqref{eq:phasefY}.
The fermionic partial trace turns out then to be a trace preserving completely positive map,
since the state reduction \eqref{eq:defreducedf} itself has this property.
Note that states and reduced states, as positive normalized linear functionals,
could be written without the parity superselection.
Superselection is needed for joint state extensions, independence of mode subsets and locality of maps.

In Section~\ref{sec:CorrEnt},
we have considered correlation and entanglement 
of qubits and of fermionic modes
without the parity superselection. 
For qubits, we recalled that
factorizing expectation values of local operators (no correlation), productness and LO-preparability coincide \eqref{eq:Dunc},
as well as
mixability from uncorrelated systems and LOCC-preparability (separability) coincide \eqref{eq:Dsep}.
For fermionic systems,
factorizing expectation values of operators of disjoint mode subsets (no correlation) and productness \eqref{eq:DfpupsuncwoSSR},
as well as
mixability from uncorrelated systems \eqref{eq:DfpupssepwoSSR} could be defined;
however, these could not be interpreted as LO- or LOCC-preparability,
since locality could not be defined at that point, without parity superselection.

In Section~\ref{sec:ParityMath},
we have written out some tools for the parity superselection.
We have done this on the level of Hilbert spaces, operator algebras and map algebras.
The superselection leads to
subspaces $\mathcal{H}_Y^\pm$ of vectors of well-defined \emph{fermion-number parity} in the Hilbert space \eqref{eq:HYpm};
subspaces $\f{\mathcal{A}}_Y^\pm$ of operators of well-defined \emph{fermion-operator parity} in the operator algebra \eqref{eq:AYpm},
containing operators which preserve (even) or alter (odd) the fermion-number parity \eqref{eq:AYpmComm2};
subspaces $\f{\mathcal{B}}_Y^\pm$ of maps of well-defined \emph{fermion-map parity} in the map algebra \eqref{eq:BYpm},
containing maps which preserve (even) or alter (odd) the fermion-operator parity \eqref{eq:BYpmComm2}.
On the level of maps, we have made another distinction, being important conceptionally:
the even maps annihilating odd operators are the \emph{physically defined} maps.
This restriction is given because a map can be given physically meaningfully only for even operators.
Note that different embeddings into physically defined maps of larger mode subsets 
make possible the action on globally even, locally odd operators of $\f{\mathcal{A}}_{X\bar{X}}^{--}$.

In Section~\ref{sec:ParityPhys},
we have considered the consequences of the parity superselection.
We have written out the tensor product structure in the $\xi$-locally physical subalgebra explicitly \eqref{eq:TPSA}, \eqref{eq:TPSH},
the independence of mode subsets, the locality of maps and the notions of correlation and entanglement.

Imposing the parity superselection,
which establishes the commutation of subalgebras describing disjoint mode subsets,
leads to the tensor product structure \eqref{eq:TPSA}
in the $\xi$-locally physical subalgebra $\f{\mathcal{A}}_\xi^\alleven$.
This also restores the $*$-algebraic properties of $\otimesfp$ \eqref{eq:allevenPHom}, \eqref{eq:allevenPPos},
and makes the fermionic map embedding \eqref{eq:mapInjfp} a strong extension \eqref{eq:mIdExtfpBe} for even maps for the whole algebra.
The tensor product structure in the algebra $\f{\mathcal{A}}_\xi^\alleven$
gives rise to a tensor product structure in the Hilbert space $\mathcal{H}_Y$ \eqref{eq:TPSH},
which works only if the action of the $\xi$-locally physical subalgebra $\f{\mathcal{A}}_\xi^\alleven$ is considered.
In this case, one can also express the joint state vector \eqref{eq:allevenPvecs} of pure states given by local state vectors \eqref{eq:allevenPpures}.

Without the parity superselection, the \emph{algebraic independence} (commutativity)
of CAR subalgebras of disjoint mode subsets does not hold.
Neither does the weaker \emph{statistical independence} hold,
which means that joint state extensions do not exist,
since
there exist states of disjoint mode subsets,
for which there exists no product state extension \cite{Moriya-2002}.
We have given a slightly modified proof of the result \eqref{eq:prod.pos} that 
the existence of joint extension is guaranteed, if all the states are physical, with at most one exception \cite{Araki-2003b}.
It is also known that
two nonphysical states may not have symmetric purification (pure state extension of the same non-zero part of the spectrum) 
\cite{Moriya-2005}.
We note that these nonindependence results recalled here
have far-reaching consequences for entropic quantities, which are often used
for the quantification of correlation and entanglement.
For example,
the triangle inequality for the von Neumann entropy does not hold in general \cite{Moriya-2002},
and, although the strong subadditivity (SSA) of the von Neumann entropy still holds for the general case \cite{Araki-2003a},
the so called MONO-SSA, which is equivalent by purification to SSA for the qubit case, does not hold \cite{Moriya-2005}.
Such phenomena were also noticed later in the literature \cite{Montero-2011,Bradler-2012,Montero-2012,Bradler-2011,Amosov-2016},
and can be avoided by imposing the parity superselection \cite{Bradler-2011,Amosov-2016}, establishing not only statistical but also algebraic independence.

In the theory of correlation and entanglement, 
the characterization of these notions with respect to local maps is essential. 
The tensor product structure \eqref{eq:TPSA} is given 
only in the $\xi$-locally physical subalgebra
$\f{\mathcal{A}}_\xi^\alleven\subseteq\f{\mathcal{A}}_Y^+$ of the whole physical subalgebra.
The appearance of subspaces like 
$\f{\mathcal{A}}_{X_1X_2X_3\dots X_{\abs{\xi}}}^{--+\dots+}\subseteq\f{\mathcal{A}}_Y^+$
leads to conceptional novelties of the notions of fermionic correlation and entanglement,
even for globally physical states.
By the superselection, 
$X$-local \eqref{eq:Xlocal} and $\xi$-local \eqref{eq:xilocal} maps can be defined 
as the embeddings \eqref{eq:mapInjfp}, \eqref{eq:mIdExtfpBe} and products \eqref{eq:mTPfpTPfp} \eqref{eq:mTPfpBe} of physically defined maps,
supplemented by physically defined maps acting on globally even operators having locally odd components.
In our point of view the maps are given on the whole algebra,
so sticking to physically defined maps (even maps annihilating odd operators) is important,
since maps acting on odd operators is not meaningful physically.

We have also considered correlation and entanglement of fermionic modes
with the parity superselection.
We recovered that the two characteristic properties of correlation and of entanglement are split up \cite{Banuls-2007}.
For correlation, factorizing expectation values of $\xi$-locally physical quantities (no correlation) \eqref{eq:Dfpupsunc}
leads to a weaker condition than $\xi$-LO-preparability (productness) \eqref{eq:Dupsprodfp}.
Similarly for entanglement,
statistical mixtures of uncorrelated modes \eqref{eq:Dfpupsconv}
lead to a weaker condition than $\xi$-LOCC-preparability (separability) \eqref{eq:Dfpupssep}.

Regarding correlation and entanglement \cite{Horodecki-2009,Guhne-2009},
the material covered here is just the beginning of the story for fermions,
the mere definition of correlation and entanglement (uncorrelated and separable states) \cite{Banuls-2007}.
We mention that, for fermionic modes,
entropic entanglement measures \cite{Zanardi-2002,Gittings-2002,Larsson-2006}
and negativity \cite{Shapourian-2019} were applied;
SLOCC invariant polynomials and entanglement measures were constructed for few mode systems \cite{Johansson-2016b};
and separability, locality, canonical forms, and maximally entangled state sets were given for fermionic \emph{Gaussian states} \cite{Spee-2018}.
We note in a nutshell that 
in a scenario when the distant laboratories of Alice and Bob 
can share multiple copies of a state (an even state, having locally even-even and odd-odd parts), 
the locally odd parts, 
which could not be reached by local physical operations when only a single copy is present,
become accessible,
since the locally odd parts of an even number of copies will be even in each laboratory,
affecting the definition of entanglement.
Such situation was already considered for the case 
when superselection rules are imposed on qubit systems 
\cite{Schuch-2004a,Schuch-2004b}.

Finally, it is natural to ask,
at least for typographic reasons,
what if $\otimesf$ would be used instead of $\otimesfp$ in the definitions of correlation and entanglement?
That is, can these be formulated in terms of fermionic tensors, instead of the usual algebraic point of view, used above?
The short answer is no.
Although the definitions in Section~\ref{sec:ParityPhys.Corrent} 
and in Section~\ref{sec:CorrEnt.fermion} 
could be repeated with $\otimesf$ instead of $\otimesfp$,
the problem is that
for $\f{A}_X\geq0$,
the fermionic tensor product $\bigotimesf_{X\in\xi}(\f{A}_X)$ is not necessarily positive, 
not even for $\f{A}_X\in\f{\mathcal{A}}_X^+$.
Although this could be handled by redefining positivity with $\f{\Lambda}_{\ord{\xi}}$,
but this would be $\ord{\xi}$-dependent in general, and $\xi$-dependent in $\f{\mathcal{A}}_\xi^\alleven$,
so it could hardly carry any meaning more than a transformation of the original approach.

In conclusion, we would like to emphasize again
the nonlocal properties of CAR algebraic description of fermionic systems and
the conceptual relevance of the parity superselection,
which could be out of scope of some parts of the quantum information community.
These fundamental concepts are inevitable for the physically correct investigation of fermionic systems,
and future research in this topic should necessarily rely on these.

\section*{Acknowledgment}
This research project was supported
by the {National Research, Development and Innovation Fund of Hungary}
 within the \textit{Researcher-initiated Research Program} (Sz.Sz., {\"O}.L~and G.B., project Nr:~NKFIH-K120569, NKFIH-K134983, Z.Z., project Nr:~NKFIH-K124152, NKFIH-K124176, NKFIH-KH129601)
 within the \textit{Quantum Technology National Excellence Program} (Sz.Sz., Z.Z., {\"O}.L.~and G.B., project Nr:~2017-1.2.1-NKP-2017-00001)
 and within the \textit{Quantum Information National Laboratory of Hungary} ({\"O}.L.);
by the {Hungarian Academy of Sciences}
 within the \textit{``Lend{\"u}let'' Program} (Sz.Sz., {\"O}.L.~and G.B.),
 and within the \textit{J{\'a}nos Bolyai Research Scholarship} (Sz.Sz., Z.Z.~and G.B.);
by the {Ministry for Innovation and Technology}
 within the \textit{{\'U}NKP-19-3, {\'U}NKP-19-4 and {\'U}NKP-20-5 New National Excellence Program} (M.M., Sz.Sz.~and Z.Z.);
by the {Center for Scalable and Predictive Methods for Excitation and Correlated Phenomena (SPEC)},
 which is funded from the Computational Chemical Sciences Program by the U.S. Department of Energy (DOE), at Pacific Northwest National Laboratory ({\"O}.L.);
by the {Deutsche Forschungsgemeinschaft (DFG, German Research Foundation)}
 within Grant SCHI 1476/1-1 (C.S.);
by the {Munich Center for Quantum Science and Technology} (C.S.);
and by the Wolfson College Oxford (C.S.).
The research project was carried out partially during a stay in the
inspiring working environment of the
Erwin Schr{\"o}dinger International Institute for Mathematics and Physics (ESI) of the University of Vienna (Sz.Sz.~and Z.Z.).
The support of various coffee machines around the world is gratefully acknowledged (Sz.Sz.~and Z.Z.).

\newpage
\appendix
\section{On the Jordan--Wigner representation}
\label{appsec:JW}

\subsection{Identities for the Jordan--Wigner representation}
\label{appsec:JW.identities}
First, for mode $i\in M$, we have for the phase operator that
\begin{equation}
\label{eq:Ep}
\begin{split}
E\indexddu{i}{\nu}{\nu'} p_i^\mu
&\equalsref{eq:Ei}
  \cket{\phi_i^{\nu}}\bra{\phi_i^{\nu'}}
\bigl( \cket{\phi_i^0}\bra{\phi_i^0} + (-1)^\mu \cket{\phi_i^1}\bra{\phi_i^1} \bigr)\\
&\equals (-1)^{\nu'\mu} \cket{\phi_i^{\nu}}\bra{\phi_i^{\nu'}} 
 = (-1)^{\nu'\mu} E\indexddu{i}{\nu}{\nu'}.
\end{split}
\end{equation}
Using this, we begin with deriving some basic identities as finger-exercises,
gradually from simpler to involved.
We will use the multi-index notation, introduced in the main text.

For the whole subsystem/mode subset $Y\subseteq M$, we have
\begin{subequations}
\begin{equation}
\label{eq:usefulEY}
E\indexddu{Y}{\vs{\mu}}{\vs{\mu}'} 
\equalsref{eq:EY}
   \prod_{i\in Y} \Gamma_Y(E\indexddu{i}{\mu_i}{\mu_i'})
\equalsref{eq:Gamma}
   \prod_{i\in Y} \biggl( \bigotimes_{\substack{k\in Y\\ k<i}}\Id_k
   \otimes E\indexddu{i}{\mu_i}{\mu_i'} 
   \otimes \bigotimes_{\substack{k\in Y\\ i<k}}\Id_k \biggr)
\equals
   \bigotimes_{i\in Y} E\indexddu{i}{\mu_i}{\mu_i'},
\end{equation}
being just the standard multipartite basis, and, for the fermionic case,
\begin{equation}
\label{eq:usefulEYf}
\begin{split}
&\f{E}\indexddu{Y}{\vs{\mu}}{\vs{\mu}'} 
\equalsref{eq:EYf}
   \prodord_{i\in Y} \Gammaf_Y(E\indexddu{i}{\mu_i}{\mu_i'})
\equalsref{eq:Gammaf}
   \prodord_{i\in Y} \biggl( \bigotimes_{\substack{k\in Y\\ k<i}} p_k^{\mu_i+\mu_i'}
   \otimes E\indexddu{i}{\mu_i}{\mu_i'} 
   \otimes \bigotimes_{\substack{k\in Y\\ i<k}}\Id_k \biggr)\\
&\equals
   \bigotimes_{i\in Y} \biggl(E\indexddu{i}{\mu_i}{\mu_i'} \prod_{\substack{k\in Y\\i<k}} p_i^{\mu_k+\mu_k'}\biggr) 
\equals
   \bigotimes_{i\in Y} \biggl(E\indexddu{i}{\mu_i}{\mu_i'}  p_i^{\esum_{k\in Y,i<k}(\mu_k+\mu_k')}\biggr) \\
&\equalsref{eq:Ep}
   \bigotimes_{i\in Y} \biggl(E\indexddu{i}{\mu_i}{\mu_i'}  (-1)^{\mu_i'\esum_{k\in Y,i<k}(\mu_k+\mu_k')}\biggr)  
\equals
   (-1)^{\esum_{i\in Y}\mu_i'\esum_{k\in Y,i<k}(\mu_k+\mu_k')} \bigotimes_{i\in Y} E\indexddu{i}{\mu_i}{\mu_i'}.
\end{split}
\end{equation}
On the other hand, for a subsystems/mode subsets $X\subseteq Y\subseteq M$, we have
\begin{equation}
\label{eq:usefulEYX}
\prod_{j\in X} \Gamma_Y(E\indexddu{j}{\mu_j}{\mu_j'})
\equalsref{eq:Gamma}
   \prod_{j\in X} \biggl( \bigotimes_{\substack{k\in Y\\k<j}}\Id_k
   \otimes E\indexddu{j}{\mu_j}{\mu_j'}
   \otimes \bigotimes_{\substack{k\in Y\\j<k}}\Id_k  \biggr)
\equals
   \bigotimes_{j\in X} E\indexddu{j}{\mu_j}{\mu_j'} \otimes \bigotimes_{k\in \bar{X}} \Id_k,
\end{equation}
and, for the fermionic case,
\begin{equation}
\label{eq:usefulEYXf}
\begin{split}
&\prodord_{j\in X} \Gammaf_Y(E\indexddu{j}{\mu_j}{\mu_j'})
\equalsref{eq:Gammaf}
   \prodord_{j\in X} \biggl( \bigotimes_{\substack{k\in Y\\k<j}} p_k^{\mu_j+\mu_j'}
   \otimes E\indexddu{j}{\mu_j}{\mu_j'}
   \otimes \bigotimes_{\substack{k\in Y\\j<k}}\Id_k \biggr)\\
&\equals
   \bigotimes_{j\in X} \biggl(E\indexddu{j}{\mu_j}{\mu_j'} \prod_{\substack{l\in X\\j<l}} p_j^{\mu_l+\mu_l'}\biggr) 
   \otimes \bigotimes_{k\in \bar{X}} \biggl(\Id_k \prod_{\substack{l\in X\\k<l}} p_k^{\mu_l+\mu_l'}\biggr)\\
&\equals
   \bigotimes_{j\in X} \biggl(E\indexddu{j}{\mu_j}{\mu_j'} p_j^{\esum_{l\in X,j<l} (\mu_l+\mu_l')}\biggr) 
   \otimes \bigotimes_{k\in \bar{X}}  p_k^{\esum_{l\in X,k<l} (\mu_l+\mu_l')}\\
&\equalsref{eq:Ep}
   \bigotimes_{j\in X} \biggl(E\indexddu{j}{\mu_j}{\mu_j'} (-1)^{\mu_j'\esum_{l\in X,j<l} (\mu_l+\mu_l')}\biggr) 
   \otimes \bigotimes_{k\in \bar{X}}  p_k^{\esum_{l\in X,k<l} (\mu_l+\mu_l')}\\
&\equals
   (-1)^{\esum_{j\in X}\mu_j'\esum_{l\in X,j<l} (\mu_l+\mu_l')}
   \bigotimes_{j\in X} E\indexddu{j}{\mu_j}{\mu_j'}
   \otimes \bigotimes_{k\in \bar{X}}  p_k^{\esum_{l\in X,k<l} (\mu_l+\mu_l')}.
\end{split}
\end{equation}
\end{subequations}
We use the convention that 
the empty sum is zero (the additive identity),
the empty product is the (multiplicative) identity,
and the empty tensor product is the identity over the one-dimensional Hilbert space (the tensor product identity).

\newpage
Now we obtain the expansion coefficients in the standard basis \eqref{eq:EY}.
For the whole subsystem/mode subset $Y\subseteq M$, we have
\begin{subequations}
\begin{equation}
\label{eq:skalpEYEY}
\begin{split}
\BigskalpHS{E\indexddu{Y}{\vs{\nu}}{\vs{\nu}'}}{E\indexddu{Y}{\vs{\mu}}{\vs{\mu}'}}
&\equalsref{eq:usefulEY}
\BigskalpHS{\bigotimes_{i\in Y} E\indexddu{i}{\nu_i}{\nu_i'}}{\bigotimes_{j\in Y} E\indexddu{j}{\mu_j}{\mu_j'}}\\
&\equalsref{eq:TPHS}
\prod_{i\in Y}\BigskalpHS{E\indexddu{i}{\nu_i}{\nu_i'}}{E\indexddu{j}{\mu_j}{\mu_j'}}
\equals
\delta^{\vs{\nu},\vs{\mu}}\delta^{\vs{\nu}',\vs{\mu}'},
\end{split}
\end{equation}
which is just the orthonormality of the standard multipartite basis,
and, for the fermionic case,
\begin{equation}
\label{eq:skalpEYEYf}
\begin{split}
\BigskalpHS{E\indexddu{Y}{\vs{\nu}}{\vs{\nu}'}}{\f{E}\indexddu{Y}{\vs{\mu}}{\vs{\mu}'}}
&\equalsreff{eq:usefulEY}{eq:usefulEYf}
   (-1)^{\esum_{j\in Y}\mu_j'\esum_{k\in Y,j<k}(\mu_k+\mu_k')}
   \BigskalpHS{\bigotimes_{i\in Y} E\indexddu{i}{\nu_i}{\nu_i'}}{\bigotimes_{j\in Y} E\indexddu{j}{\mu_j}{\mu_j'}}\\
&\equalsref{eq:skalpEYEY}
   (-1)^{\esum_{i\in Y} \mu_i'\esum_{k\in Y,i<k}(\mu_k+\mu_k')} \delta^{\vs{\nu},\vs{\mu}}\delta^{\vs{\nu}',\vs{\mu}'}.
\end{split}
\end{equation}
On the other hand, for a subsystems/mode subsets $X\subseteq Y\subseteq M$, we have
\begin{equation}
\label{eq:skalpEYEYX}
\begin{split}
&\BigskalpHS{E\indexddu{Y}{\vs{\nu}}{\vs{\nu}'}}{\prod_{j\in X} \Gamma_Y(E\indexddu{j}{\mu_j}{\mu_j'})}
\equalsreff{eq:usefulEY}{eq:usefulEYX}
   \BigskalpHS{\bigotimes_{i\in Y} E\indexddu{i}{\nu_i}{\nu_i'}}
     {\bigotimes_{j\in X} E\indexddu{j}{\mu_j}{\mu_j'} \otimes \bigotimes_{k\in \bar{X}} \Id_k}\\
&\equalsref{eq:TPHS}
   \prod_{j\in X}\BigskalpHS{E\indexddu{j}{\nu_j}{\nu_j'}}{E\indexddu{j}{\mu_j}{\mu_j'}}
   \prod_{k\in \bar{X}}\BigskalpHS{E\indexddu{k}{\nu_k}{\nu_k'}}{\Id_k}
\equals
   \delta^{\vs{\nu}_X,\vs{\mu}_X}\delta^{\vs{\nu}_X',\vs{\mu}_X'}
   \delta\indexdu{\vs{\nu}_{\bar{X}}}{\vs{\nu}_{\bar{X}}'},
\end{split}
\end{equation}
and, for the fermionic case,
\begin{equation}
\label{eq:skalpEYEYXf}
\begin{split}
&\BigskalpHS{E\indexddu{Y}{\vs{\nu}}{\vs{\nu}'}}{\prodord_{j\in X} \Gammaf_Y(E\indexddu{j}{\mu_j}{\mu_j'})}\\
&\equalsreff{eq:usefulEY}{eq:usefulEYXf}
   (-1)^{\esum_{j\in X}\mu_j'\esum_{l\in X,j<l} (\mu_l+\mu_l')}
   \BigskalpHS{\bigotimes_{i\in Y} E\indexddu{i}{\nu_i}{\nu_i'}}
     {\bigotimes_{j\in X} E\indexddu{j}{\mu_j}{\mu_j'} 
    \otimes \bigotimes_{k\in \bar{X}}  p_k^{\esum_{l\in X,k<l} (\mu_l+\mu_l')}}\\
&\equalsref{eq:TPHS}
   (-1)^{\esum_{j\in X}\mu_j'\esum_{l\in X,j<l} (\mu_l+\mu_l')}
   \prod_{j\in X} \BigskalpHS{E\indexddu{j}{\nu_j}{\nu_j'}}{E\indexddu{j}{\mu_j}{\mu_j'}}
   \prod_{k\in \bar{X}}\BigskalpHS{E\indexddu{k}{\nu_k}{\nu_k'}}{p_k^{\esum_{l\in X,k<l} (\mu_l+\mu_l')}}\\
&\equalsref{eq:Ep}
   (-1)^{\esum_{j\in X}\mu_j'\esum_{l\in X,j<l} (\mu_l+\mu_l')}   \delta^{\vs{\nu}_X,\vs{\mu}_X}\delta^{\vs{\nu}_X',\vs{\mu}_X'}
   (-1)^{\esum_{k\in \bar{X}}\nu_k'\esum_{l\in X,k<l} (\mu_l+\mu_l')} \delta\indexdu{\vs{\nu}_{\bar{X}}}{\vs{\nu}_{\bar{X}}'}\\
&\equals
   (-1)^{\esum_{i\in Y}\nu_i'\esum_{l\in X,i<l}(\nu_l+\nu_l')} 
   \delta^{\vs{\nu}_X,\vs{\mu}_X}\delta^{\vs{\nu}_X',\vs{\mu}_X'}
   \delta\indexdu{\vs{\nu}_{\bar{X}}}{\vs{\nu}_{\bar{X}}'}\\
&\equals
   (-1)^{\esum_{i\in Y}\nu_i'\esum_{k\in Y,i<k}(\nu_k+\nu_k')} 
   \delta^{\vs{\nu}_X,\vs{\mu}_X}\delta^{\vs{\nu}_X',\vs{\mu}_X'}
   \delta\indexdu{\vs{\nu}_{\bar{X}}}{\vs{\nu}_{\bar{X}}'}.
\end{split}
\end{equation}
\end{subequations}

Summarizing the findings, we have
\begin{subequations}
\begin{align}
\label{eq:skalpEYEYp}
\BigskalpHS{E\indexddu{Y}{\vs{\nu}}{\vs{\nu}'}}{E\indexddu{Y}{\vs{\mu}}{\vs{\mu}'}}
&\equalsref{eq:skalpEYEY}
   \delta^{\vs{\nu},\vs{\mu}}\delta^{\vs{\nu}',\vs{\mu}'},\\
\label{eq:skalpEYEYfp}
\BigskalpHS{E\indexddu{Y}{\vs{\nu}}{\vs{\nu}'}}{\f{E}\indexddu{Y}{\vs{\mu}}{\vs{\mu}'}}
&\equalsref{eq:skalpEYEYf}
   f\indexddu{Y}{\vs{\nu}}{\vs{\nu}'}
   \delta^{\vs{\nu},\vs{\mu}}\delta^{\vs{\nu}',\vs{\mu}'},\\
\label{eq:skalpEYEYXp}
\BigskalpHS{E\indexddu{Y}{\vs{\nu}}{\vs{\nu}'}}{\prod_{j\in X} \Gamma_Y(E\indexddu{j}{\mu_j}{\mu_j'})}
&\equalsref{eq:skalpEYEYX}
   \delta^{\vs{\nu}_X,\vs{\mu}_X}\delta^{\vs{\nu}_X',\vs{\mu}_X'}
   \delta\indexdu{\vs{\nu}_{\bar{X}}}{\vs{\nu}_{\bar{X}}'},\\
\label{eq:skalpEYEYXfp}
\BigskalpHS{E\indexddu{Y}{\vs{\nu}}{\vs{\nu}'}}{\prodord_{j\in X} \Gammaf_Y(E\indexddu{j}{\mu_j}{\mu_j'})}
&\equalsref{eq:skalpEYEYXf}
   f\indexddu{Y}{\vs{\nu}}{\vs{\nu}'}
   \delta^{\vs{\nu}_X,\vs{\mu}_X}\delta^{\vs{\nu}_X',\vs{\mu}_X'}
   \delta\indexdu{\vs{\nu}_{\bar{X}}}{\vs{\nu}_{\bar{X}}'},
\end{align}
\end{subequations}
with the phase factor
\begin{subequations}
\begin{equation}
\label{eq:fY}
f\indexddu{Y}{\vs{\nu}}{\vs{\nu}'} 
= (-1)^{\esum_{i\in Y}\nu_i'\esum_{k\in Y,i<k}(\nu_k+\nu_k')},
\end{equation}
which can also be written for all partitions $\xi\in\Pi(Y)$ as
\begin{equation}
\label{eq:fYalt}
f\indexddu{Y}{\vs{\nu}}{\vs{\nu}'} 
= (-1)^{\esum_{X,X'\in\xi}  \esum_{i\in X}\nu_i'\esum_{k\in X',i<k}(\nu_k+\nu_k')}.
\end{equation}
\end{subequations}

\subsection{Phase factors for few modes}
\label{appsec:JW.Explf}
Here we show the phase factors $f\indexddu{Y}{\vs{\nu}}{\vs{\nu}'}$, given in \eqref{eq:fY}, for small mode subsets $Y\subseteq M$.
These are written in matrices indexed with multi-indices
$\vs{\nu}_Y=\tuple{\nu_1,\nu_2,\dots}$
ordered lexicographically with respect to the Jordan--Wigner ordering of the modes.
For $\abs{Y}=2$, we have
\begin{equation*}
f_{\set{1,2}} =
\left[\begin{smallmatrix}
 + & + & + & - \\
 + & + & - & + \\
 + & + & + & - \\
 + & + & - & +
\end{smallmatrix}\right].
\end{equation*}
For $\abs{Y}=3$, we have
\begin{equation*}
f_{\set{1,2,3}} = 
\left[\begin{smallmatrix}
 + & + & + & - & + & - & - & - \\
 + & + & - & + & - & + & - & - \\
 + & + & + & - & - & + & + & + \\
 + & + & - & + & + & - & + & + \\
 + & + & + & - & + & - & - & - \\
 + & + & - & + & - & + & - & - \\
 + & + & + & - & - & + & + & + \\
 + & + & - & + & + & - & + & +
\end{smallmatrix}\right].
\end{equation*}
For $\abs{Y}=4$, we have
\begin{equation*}
f_{\set{1,2,3,4}} = 
\left[\begin{smallmatrix}
 + & + & + & - & + & - & - & - & + & - & - & - & - & - & - & + \\
 + & + & - & + & - & + & - & - & - & + & - & - & - & - & + & - \\
 + & + & + & - & - & + & + & + & - & + & + & + & - & - & - & + \\
 + & + & - & + & + & - & + & + & + & - & + & + & - & - & + & - \\
 + & + & + & - & + & - & - & - & - & + & + & + & + & + & + & - \\
 + & + & - & + & - & + & - & - & + & - & + & + & + & + & - & + \\
 + & + & + & - & - & + & + & + & + & - & - & - & + & + & + & - \\
 + & + & - & + & + & - & + & + & - & + & - & - & + & + & - & + \\
 + & + & + & - & + & - & - & - & + & - & - & - & - & - & - & + \\
 + & + & - & + & - & + & - & - & - & + & - & - & - & - & + & - \\
 + & + & + & - & - & + & + & + & - & + & + & + & - & - & - & + \\
 + & + & - & + & + & - & + & + & + & - & + & + & - & - & + & - \\
 + & + & + & - & + & - & - & - & - & + & + & + & + & + & + & - \\
 + & + & - & + & - & + & - & - & + & - & + & + & + & + & - & + \\
 + & + & + & - & - & + & + & + & + & - & - & - & + & + & + & - \\
 + & + & - & + & + & - & + & + & - & + & - & - & + & + & - & +
\end{smallmatrix}\right].
\end{equation*}

\subsection{On the explicit formula of the basis transformation}
\label{appsec:JW.Phi}

Here we show the derivation of \eqref{eq:EYEYftraf}-\eqref{eq:phasefY}, which is simply
\begin{equation}
\f{E}\indexddu{Y}{\vs{\nu}}{\vs{\nu}'}
\equals
   \sum_{\vs{\mu},\vs{\mu}'}
   \BigskalpHS{E\indexddu{Y}{\vs{\mu}}{\vs{\mu}'}}{\f{E}\indexddu{Y}{\vs{\nu}}{\vs{\nu}'}}
   E\indexddu{Y}{\vs{\mu}}{\vs{\mu}'} 
\equalsref{eq:skalpEYEYfp}
  f\indexddu{Y}{\vs{\nu}}{\vs{\nu}'}E\indexddu{Y}{\vs{\nu}}{\vs{\nu}'}.
\end{equation}

\subsection{Product of fermionic basis elements}
\label{appsec:JW.prod}

Here we show the derivation of the product of two fermionic basis elements \eqref{eq:EYf}, as
\begin{equation*}
\begin{split}
\f{E}\indexddu{Y}{\vs{\nu}}{\vs{\nu}'}\f{E}\indexddu{Y}{\vs{\mu}}{\vs{\mu}'}
&\equalsref{eq:PhiYEY}
   f\indexddu{Y}{\vs{\nu}}{\vs{\nu}'} f\indexddu{Y}{\vs{\mu}}{\vs{\mu}'}
   E\indexddu{Y}{\vs{\nu}}{\vs{\nu}'} E\indexddu{Y}{\vs{\mu}}{\vs{\mu}'}\\
&\equals
   f\indexddu{Y}{\vs{\nu}}{\vs{\nu}'} f\indexddu{Y}{\vs{\mu}}{\vs{\mu}'}
   \delta^{\vs{\nu}',\vs{\mu}} E\indexddu{Y}{\vs{\nu}}{\vs{\mu}'}\\
&\equalsref{eq:PhiYEY}
   \delta^{\vs{\nu}',\vs{\mu}} 
   f\indexddu{Y}{\vs{\nu}}{\vs{\nu}'} f\indexddu{Y}{\vs{\mu}}{\vs{\mu}'}
   f\indexddu{Y}{\vs{\nu}}{\vs{\mu}'} \f{E}\indexddu{Y}{\vs{\nu}}{\vs{\mu}'}  \\
&\equalsref{eq:phasefY}
   \delta^{\vs{\nu}',\vs{\mu}} (-1)^{
    \esum_{i\in Y}\nu_i'\esum_{k\in Y,i<k}(\nu_k+\nu_k')
   +\esum_{i\in Y}\mu_i'\esum_{k\in Y,i<k}(\mu_k+\mu_k')
   +\esum_{i\in Y}\mu_i'\esum_{k\in Y,i<k}(\nu_k+\mu_k')  }
   \f{E}\indexddu{Y}{\vs{\nu}}{\vs{\mu}'}\\
&\equals
   \delta^{\vs{\nu}',\vs{\mu}} (-1)^{
    \esum_{i\in Y}\nu_i'\esum_{k\in Y,i<k}(\nu_k+\nu_k')
   +\esum_{i\in Y}\mu_i'\esum_{k\in Y,i<k}(\mu_k+\nu_k)   }
   \f{E}\indexddu{Y}{\vs{\nu}}{\vs{\mu}'}.
\end{split}
\end{equation*}
Exploiting the Kronecker-delta, we have
\begin{equation}
\label{eq:Efprod}
\begin{split}
\f{E}\indexddu{Y}{\vs{\nu}}{\vs{\nu}'}\f{E}\indexddu{Y}{\vs{\mu}}{\vs{\mu}'}
&= \delta^{\vs{\nu}',\vs{\mu}}
  (-1)^{ \esum_{i\in Y}(\mu_i+\mu_i')\esum_{k\in Y,i<k}(\nu_k+\nu_k') }
  \f{E}\indexddu{Y}{\vs{\nu}}{\vs{\mu}'}\\
&= \delta^{\vs{\nu}',\vs{\mu}}
  (-1)^{ \esum_{i\in Y}(\nu_i'+\mu_i')\esum_{k\in Y,i<k}(\nu_k+\mu_k) }
  \f{E}\indexddu{Y}{\vs{\nu}}{\vs{\mu}'}.
\end{split}
\end{equation}
Note that the first form here tells us that if at least one of the operators is diagonal, 
the arising phase factor is the trivial $+1$.

\subsection{On the product property of the basis transformation}
\label{appsec:JW.morphism}

Here we show the derivation of \eqref{eq:EYXEYXftraf},
that is, for the mode subsets $X\subseteq Y\subseteq M$,
\begin{align*}
\forall\bigsset{A_j}{j\in X}:\qquad
\Phi_Y \Bigl( \prod_{j\in X} \Gamma_Y (A_j) \Bigr) &=
\prodord_{j\in X} \Phi_Y \bigl(\Gamma_Y (A_j) \bigr),\\
\intertext{which holds if and only if
it holds for the basis elements \eqref{eq:Ei},}
\forall\vs{\mu}_X,\vs{\mu}_X':\qquad
\Phi_Y \Bigl( \prod_{j\in X} \Gamma_Y (E\indexddu{j}{\mu_j}{\mu_j'}) \Bigr) &=
\prodord_{j\in X} \Phi_Y \bigl(\Gamma_Y (E\indexddu{j}{\mu_j}{\mu_j'}) \bigr),\\
\end{align*}
because of linearity.
Let us start with the left-hand side
\begin{equation*}
\begin{split}
\Phi_Y \Bigl( \prod_{j\in X} \Gamma_Y(E\indexddu{j}{\mu_j}{\mu_j'}) \Bigr)
&\equalsref{eq:skalpEYEYXp}
   \Phi_Y \Bigl( \sum_{\vs{\nu},\vs{\nu}'} \delta^{\vs{\nu}_X,\vs{\mu}_X}\delta^{\vs{\nu}_X',\vs{\mu}_X'}
   \delta\indexdu{\vs{\nu}_{\bar{X}}}{\vs{\nu}_{\bar{X}}'} 
   E\indexddu{Y}{\vs{\nu}}{\vs{\nu}'} \Bigr)\\
&\equals
   \sum_{\vs{\nu},\vs{\nu}'} \delta^{\vs{\nu}_X,\vs{\mu}_X}\delta^{\vs{\nu}_X',\vs{\mu}_X'}
   \delta\indexdu{\vs{\nu}_{\bar{X}}}{\vs{\nu}_{\bar{X}}'} 
   \Phi_Y ( E\indexddu{Y}{\vs{\nu}}{\vs{\nu}'} )\\
&\equalsref{eq:EYEYftraf}
   \sum_{\vs{\nu},\vs{\nu}'} \delta^{\vs{\nu}_X,\vs{\mu}_X}\delta^{\vs{\nu}_X',\vs{\mu}_X'}
   \delta\indexdu{\vs{\nu}_{\bar{X}}}{\vs{\nu}_{\bar{X}}'} 
   \f{E}\indexddu{Y}{\vs{\nu}}{\vs{\nu}'}\\
&\equalsref{eq:EYf}
   \sum_{\vs{\nu},\vs{\nu}'} \delta^{\vs{\nu}_X,\vs{\mu}_X}\delta^{\vs{\nu}_X',\vs{\mu}_X'}
   \delta\indexdu{\vs{\nu}_{\bar{X}}}{\vs{\nu}_{\bar{X}}'} 
   \prodord_{i\in Y} \Gammaf_Y(E\indexddu{i}{\nu_i}{\nu_i'}) \\
&\equals
   \sum_{\vs{\nu},\vs{\nu}'}
   \prodord_{i\in Y}
   \left\{\begin{aligned}
    \delta^{\nu_i,\mu_i}\delta^{\nu_i',\mu_i'} \Gammaf_Y(E\indexddu{i}{\nu_i}{\nu_i'}) \quad&\text{if $i\in X$}\\
    \delta\indexdu{\nu_i}{\nu_i'} \Gammaf_Y(E\indexddu{i}{\nu_i}{\nu_i'}) \quad&\text{if $i\in \bar{X}$}
   \end{aligned}\right\}\\
&\equals
   \prodord_{i\in Y}
   \left\{\begin{aligned}
    \sum_{\nu_i,\nu_i'}
    \delta^{\nu_i,\mu_i}\delta^{\nu_i',\mu_i'} \Gammaf_Y(E\indexddu{i}{\nu_i}{\nu_i'}) \quad&\text{if $i\in X$}\\
    \sum_{\nu_i,\nu_i'}
    \delta\indexdu{\nu_i}{\nu_i'} \Gammaf_Y(E\indexddu{i}{\nu_i}{\nu_i'})  \quad&\text{if $i\in \bar{X}$}
   \end{aligned}\right\}\\
&\equals
   \prodord_{i\in Y}
   \left\{\begin{aligned}
    \Gammaf_Y(E\indexddu{i}{\mu_i}{\mu_i'}) \quad&\text{if $i\in X$}\\
    \sum_{\nu_i}\Gammaf_Y(E\indexddu{i}{\nu_i}{\nu_i})  \quad&\text{if $i\in \bar{X}$}
   \end{aligned}\right\}\\
&\equals
   \prodord_{j\in X}
   \Gammaf_Y(E\indexddu{j}{\mu_j}{\mu_j'}),
\end{split}
\end{equation*}
where we have used
$\sum_{\nu_i}\Gammaf_Y(E\indexddu{i}{\nu_i}{\nu_i}) = \Gammaf_Y(\Id_i) = \Idf_Y$
in the last step.
So we have
\begin{subequations}
\begin{equation}
\Phi_Y \Bigl( \prod_{j\in X} \Gamma_Y(E\indexddu{j}{\mu_j}{\mu_j'}) \Bigr)
\equals
   \prodord_{j\in X}
   \Gammaf_Y(E\indexddu{j}{\mu_j}{\mu_j'}),
\end{equation}
which holds also for $X=\set{j}$, 
\begin{equation}
\Phi_Y \Bigl( \Gamma_Y(E\indexddu{j}{\mu_j}{\mu_j'}) \Bigr)
\equals
   \Gammaf_Y(E\indexddu{j}{\mu_j}{\mu_j'}),
\end{equation}
\end{subequations}
together leading to our claim.

\section{On fermionic tensors}
\label{appsec:Tensors}

\subsection{Fermionic tensor product}
\label{appsec:Tensors.TP}

By the definition \eqref{eq:TPf} and \eqref{eq:EYEYftraf},
the fermionic tensor product of the basis elements is given as
\begin{subequations}
\begin{equation}
\label{eq:hupsdef}
\bigotimesf_{X\in\xi}\f{E}\indexddu{X}{\vs{\nu}_X}{\vs{\nu}_X'}
 = f\indexddu{Y}{\vs{\nu}}{\vs{\nu}'}
\bigotimes_{X\in\xi}f\indexddu{X}{\vs{\nu}_X}{\vs{\nu}_X'} 
\f{E}\indexddu{X}{\vs{\nu}_X}{\vs{\nu}_X'}
 = h\indexddu{\xi}{\vs{\nu}}{\vs{\nu}'} \bigotimes_{X\in\xi}\f{E}\indexddu{X}{\vs{\nu}_X}{\vs{\nu}_X'},
\end{equation}
with the phase factor
\begin{equation}
\label{eq:hups}
h\indexddu{\xi}{\vs{\nu}}{\vs{\nu}'} = 
f\indexddu{Y}{\vs{\nu}}{\vs{\nu}'}
\prod_{X\in\xi} f\indexddu{X}{\vs{\nu}_X}{\vs{\nu}_X'}
=(-1)^{\esum_{X,X'\in\xi, X\neq X'} \esum_{i\in X}\nu_i'\esum_{k\in X',i<k}(\nu_k+\nu_k') },
\end{equation}
using the form \eqref{eq:fYalt} of the phase factors.
\end{subequations}

\subsection{Phase factors for few modes}
\label{appsec:Tensors.Explh}
Here we show the phase factors $h\indexddu{\xi}{\vs{\nu}}{\vs{\nu}'}$,
given in \eqref{eq:hups}, for small mode subsets $Y\subseteq M$.
These are written in matrices indexed with multi-indices
$\vs{\nu}_Y=\tuple{\nu_1,\nu_2,\dots}$
ordered lexicographically with respect to the Jordan--Wigner ordering of the modes.
(Note that we use a simplified notation,
$\set{1}\set{2}\set{3}:=\set{\set{1},\set{2},\set{3}}$,
omitting the curly brackets and colons in the writing of partitions, since this does not cause confusion.)
For $\abs{Y}=2$, we have
\begin{equation*}
h_{\set{1}\set{2}} =
\left[\begin{smallmatrix}
 + & + & + & - \\
 + & + & - & + \\
 + & + & + & - \\
 + & + & - & +
\end{smallmatrix}\right].
\end{equation*}
For $\abs{Y}=3$, we have
\begin{align*}
h_{\set{1,2}\set{3}} &= 
\left[\begin{smallmatrix}
 + & + & + & - & + & - & + & + \\
 + & + & - & + & - & + & + & + \\
 + & + & + & - & + & - & + & + \\
 + & + & - & + & - & + & + & + \\
 + & + & + & - & + & - & + & + \\
 + & + & - & + & - & + & + & + \\
 + & + & + & - & + & - & + & + \\
 + & + & - & + & - & + & + & +
\end{smallmatrix}\right],\qquad &
h_{\set{1,3}\set{2}} &= 
\left[\begin{smallmatrix}
 + & + & + & - & + & + & - & + \\
 + & + & - & + & + & + & + & - \\
 + & + & + & - & - & - & + & - \\
 + & + & - & + & - & - & - & + \\
 + & + & + & - & + & + & - & + \\
 + & + & - & + & + & + & + & - \\
 + & + & + & - & - & - & + & - \\
 + & + & - & + & - & - & - & +
\end{smallmatrix}\right],\\
h_{\set{1}\set{2,3}} &= 
\left[\begin{smallmatrix}
 + & + & + & + & + & - & - & + \\
 + & + & + & + & - & + & + & - \\
 + & + & + & + & - & + & + & - \\
 + & + & + & + & + & - & - & + \\
 + & + & + & + & + & - & - & + \\
 + & + & + & + & - & + & + & - \\
 + & + & + & + & - & + & + & - \\
 + & + & + & + & + & - & - & +
\end{smallmatrix}\right],\qquad &
h_{\set{1}\set{2}\set{3}} &= 
\left[\begin{smallmatrix}
 + & + & + & - & + & - & - & - \\
 + & + & - & + & - & + & - & - \\
 + & + & + & - & - & + & + & + \\
 + & + & - & + & + & - & + & + \\
 + & + & + & - & + & - & - & - \\
 + & + & - & + & - & + & - & - \\
 + & + & + & - & - & + & + & + \\
 + & + & - & + & + & - & + & +
\end{smallmatrix}\right].
\end{align*}

\subsection{Fermionic canonical embedding}
\label{appsec:Tensors.cEmb}

Here we show the derivation of that 
the standard \eqref{eq:cEmb} and fermionic \eqref{eq:cEmbf} canonical embeddings
are unitarily equivalent \eqref{eq:iotaU}.
\begin{equation*}
\begin{split}
\iotaf_{X,Y}(\f{E}\indexddu{X}{\vs{\nu}_X}{\vs{\nu}_X'})
&\equalsref{eq:cEmbf}
 \f{E}\indexddu{X}{\vs{\nu}_X}{\vs{\nu}_X'} \otimesf \Idf_{\bar{X}}\\
&\equals
 \sum_{\vs{\nu}_{\bar{X}}, \vs{\nu}_{\bar{X}}'} \delta\indexud{\vs{\nu}_{\bar{X}}}{\vs{\nu}_{\bar{X}}'} 
\f{E}\indexddu{X}{\vs{\nu}_X}{\vs{\nu}_X'} \otimesf \f{E}\indexddu{\bar{X}}{\vs{\nu}_{\bar{X}}}{\vs{\nu}_{\bar{X}}'}\\
&\equalsref{eq:hupsdef}
 \sum_{\vs{\nu}_{\bar{X}}, \vs{\nu}_{\bar{X}}'} \delta\indexud{\vs{\nu}_{\bar{X}}}{\vs{\nu}_{\bar{X}}'}
h\indexddu{X\bar{X}}{\vs{\nu}_X\vs{\nu}_{\bar{X}}}{\vs{\nu}_X'\vs{\nu}_{\bar{X}}'}
\f{E}\indexddu{X}{\vs{\nu}_X}{\vs{\nu}_X'} \otimes \f{E}\indexddu{\bar{X}}{\vs{\nu}_{\bar{X}}}{\vs{\nu}_{\bar{X}}'}\\
&\equalsref{eq:hdelta}
 \sum_{\vs{\nu}_{\bar{X}}, \vs{\nu}_{\bar{X}}'} 
 u_{X\bar{X}}^{\vs{\nu}} u_{X\bar{X}}^{\vs{\nu}'}
 \delta\indexud{\vs{\nu}_{\bar{X}}}{\vs{\nu}_{\bar{X}}'}
 \f{E}\indexddu{X}{\vs{\nu}_X}{\vs{\nu}_X'} \otimes \f{E}\indexddu{\bar{X}}{\vs{\nu}_{\bar{X}}}{\vs{\nu}_{\bar{X}}'}\\
&\equalsref{eq:PhiYEY}
 \sum_{\vs{\nu}_{\bar{X}}, \vs{\nu}_{\bar{X}}'} 
u_{X\bar{X}}^{\vs{\nu}} u_{X\bar{X}}^{\vs{\nu}'}
 \delta\indexud{\vs{\nu}_{\bar{X}}}{\vs{\nu}_{\bar{X}}'}
 f\indexddu{X}{\vs{\nu}_X}{\vs{\nu}_X'} f\indexddu{\bar{X}}{\vs{\nu}_{\bar{X}}}{\vs{\nu}_{\bar{X}}'}
 E\indexddu{Y}{\vs{\nu}}{\vs{\nu}'}\\
&\equalsref{eq:EYexpl}
 \sum_{\vs{\mu}} u_{X\bar{X}}^{\vs{\mu}} E\indexddu{Y}{\vs{\mu}}{\vs{\mu}}
 \sum_{\vs{\nu}_{\bar{X}}, \vs{\nu}_{\bar{X}}'}\delta\indexud{\vs{\nu}_{\bar{X}}}{\vs{\nu}_{\bar{X}}'}
 f\indexddu{X}{\vs{\nu}_X}{\vs{\nu}_X'} f\indexddu{\bar{X}}{\vs{\nu}_{\bar{X}}}{\vs{\nu}_{\bar{X}}'}
 E\indexddu{Y}{\vs{\nu}}{\vs{\nu}'}
 \sum_{\vs{\mu}'} u_{X\bar{X}}^{\vs{\mu}'} E\indexddu{Y}{\vs{\mu}'}{\vs{\mu}'}\\
&\equalsref{eq:UYXXb}
 \f{U}_{X\bar{X}} \bigl( \f{E}\indexddu{X}{\vs{\nu}_X}{\vs{\nu}_X'} \otimes \Idf_{\bar{X}} \bigr)  \f{U}_{X\bar{X}}^\dagger\\
&\equalsref{eq:cEmb}
 \f{U}_{X\bar{X}}\iota_{X,Y}(\f{E}\indexddu{X}{\vs{\nu}_X}{\vs{\nu}_X'}) \f{U}_{X\bar{X}}^\dagger.
\end{split}
\end{equation*}
Here the fourth equality is
\begin{equation}
\label{eq:hdelta}
\begin{split}
 h\indexddu{X\bar{X}}{\vs{\nu}}{\vs{\nu}'}
 \delta\indexud{\vs{\nu}_{\bar{X}}}{\vs{\nu}_{\bar{X}}'}
&\equalsref{eq:hups}
 f\indexddu{Y}{\vs{\nu}}{\vs{\nu}'}
 f\indexddu{X}{\vs{\nu}_X}{\vs{\nu}_X'}
 f\indexddu{\bar{X}}{\vs{\nu}_{\bar{X}}}{\vs{\nu}_{\bar{X}}'}
 \delta\indexud{\vs{\nu}_{\bar{X}}}{\vs{\nu}_{\bar{X}}'}\\
&\equalsref{eq:fY}
 f\indexddu{Y}{\vs{\nu}}{\vs{\nu}'}
 f\indexddu{X}{\vs{\nu}_X}{\vs{\nu}_X'}
 \delta\indexud{\vs{\nu}_{\bar{X}}}{\vs{\nu}_{\bar{X}}'}\\
&\equalsref{eq:fY}
 (-1)^{\esum_{i\in Y}\nu_i'\esum_{k\in Y,i<k}(\nu_k+\nu_k')}
 (-1)^{\esum_{j\in X}\nu_j'\esum_{l\in X,j<l}(\nu_l+\nu_l')}
\delta\indexud{\vs{\nu}_{\bar{X}}}{\vs{\nu}_{\bar{X}}'}\\
&\equals
 (-1)^{\esum_{i\in \bar{X}}\nu_i'\esum_{l\in X,i<l}(\nu_l+\nu_l')}
 \delta\indexud{\vs{\nu}_{\bar{X}}}{\vs{\nu}_{\bar{X}}'}\\
&\equals
 u_{X\bar{X}}^{\vs{\nu}} u_{X\bar{X}}^{\vs{\nu}'}
\delta\indexud{\vs{\nu}_{\bar{X}}}{\vs{\nu}_{\bar{X}}'},
\end{split}
\end{equation}
with 
\begin{subequations}
\begin{equation}
\label{eq:uXXb}
u_{X\bar{X}}^{\vs{\nu}} = (-1)^{\esum_{i\in \bar{X}}\nu_i\esum_{l\in X,i<l}\nu_l}.
\end{equation}
That is, we managed to write $h\indexddu{X\bar{X}}{\vs{\nu}}{\vs{\nu}'}\delta\indexud{\vs{\nu}_{\bar{X}}}{\vs{\nu}_{\bar{X}}'}$
as a product of two factors, depending on the unprimed or primed indices only.
Now let us have the operator
\begin{equation}
\label{eq:UYXXb}
\f{U}_{X\bar{X}} = \sum_{\vs{\nu}} u_{X\bar{X}}^{\vs{\nu}} E\indexddu{Y}{\vs{\nu}}{\vs{\nu}}
= \sum_{\vs{\nu}} u_{X\bar{X}}^{\vs{\nu}} \f{E}\indexddu{Y}{\vs{\nu}}{\vs{\nu}},
\end{equation}
\end{subequations}
which is diagonal in the standard basis \eqref{eq:phiY}, see \eqref{eq:EYexpl}, with entries $\pm1$, so it is unitary.

\subsection{Fermionic product operators for two mode subsets}
\label{appsec:Tensors.Lambda2}

Here we show the derivation of the effect of the map $\f{\Lambda}_{X\bar{X}}$ given in \eqref{eq:Lambda2}.
\begin{equation*}
\begin{split}
\f{\Lambda}_{X\bar{X}} \bigl( \f{E}\indexddu{Y}{\vs{\nu}}{\vs{\nu}'} \bigr)
&\equalsref{eq:EYfXs} 
   \f{\Lambda}_{X\bar{X}} \bigl( \f{E}\indexddu{X}{\vs{\nu}_X}{\vs{\nu}_X'} \otimesf \f{E}\indexddu{\bar{X}}{\vs{\nu}_{\bar{X}}}{\vs{\nu}_{\bar{X}}'}\bigr)\\
&\equalsref{eq:Lambda2}
   \bigl( \f{E}\indexddu{X}{\vs{\nu}_X}{\vs{\nu}_X'} \otimesf \Idf_{\bar{X}} \bigr)
   \bigl( \Idf_X \otimesf \f{E}\indexddu{\bar{X}}{\vs{\nu}_{\bar{X}}}{\vs{\nu}_{\bar{X}}'} \bigr)\\
&\equals
   \sum_{\vs{\mu},\vs{\mu}'} \delta^{\vs{\mu}_X,\vs{\mu}_X'}\delta^{\vs{\mu}_{\bar{X}},\vs{\mu}_{\bar{X}}'}
   \bigl( \f{E}\indexddu{X}{\vs{\nu}_X}{\vs{\nu}_X'} \otimesf \f{E}\indexddu{\bar{X}}{\vs{\mu}_{\bar{X}}}{\vs{\mu}_{\bar{X}}'} \bigr)
   \bigl( \f{E}\indexddu{X}{\vs{\mu}_X}{\vs{\mu}_X'} \otimesf \f{E}\indexddu{\bar{X}}{\vs{\nu}_{\bar{X}}}{\vs{\nu}_{\bar{X}}'} \bigr)\\
&\equalsref{eq:EYfXs}
   \sum_{\vs{\mu},\vs{\mu}'} \delta^{\vs{\mu},\vs{\mu}'}
   \f{E}\indexddu{Y}{\vs{\nu}_X\vs{\mu}_{\bar{X}}}{\vs{\nu}_X'\vs{\mu}_{\bar{X}}'}
   \f{E}\indexddu{Y}{\vs{\mu}_X\vs{\nu}_{\bar{X}}}{\vs{\mu}_X'\vs{\nu}_{\bar{X}}'} \\
&\equalsref{eq:PhiYEY}
   \sum_{\vs{\mu},\vs{\mu}'} \delta^{\vs{\mu},\vs{\mu}'}
   f\indexddu{Y}{\vs{\nu}_X\vs{\mu}_{\bar{X}}}{\vs{\nu}_X'\vs{\mu}_{\bar{X}}'}
   f\indexddu{Y}{\vs{\mu}_X\vs{\nu}_{\bar{X}}}{\vs{\mu}_X'\vs{\nu}_{\bar{X}}'}  
   E\indexddu{Y}{\vs{\nu}_X\vs{\mu}_{\bar{X}}}{\vs{\nu}_X'\vs{\mu}_{\bar{X}}'}
   E\indexddu{Y}{\vs{\mu}_X\vs{\nu}_{\bar{X}}}{\vs{\mu}_X'\vs{\nu}_{\bar{X}}'} \\
&\equalsref{eq:EYexpl}
   \sum_{\vs{\mu},\vs{\mu}'} \delta^{\vs{\mu},\vs{\mu}'}
   f\indexddu{Y}{\vs{\nu}_X\vs{\mu}_{\bar{X}}}{\vs{\nu}_X'\vs{\mu}_{\bar{X}}'}
   f\indexddu{Y}{\vs{\mu}_X\vs{\nu}_{\bar{X}}}{\vs{\mu}_X'\vs{\nu}_{\bar{X}}'}  
   \delta^{\vs{\nu}_X',\vs{\mu}_X}\delta^{\vs{\mu}_{\bar{X}}',\vs{\nu}_{\bar{X}}}
   E\indexddu{Y}{\vs{\nu}_X\vs{\mu}_{\bar{X}}}{\vs{\mu}_X'\vs{\nu}_{\bar{X}}'} \\
&\equals \sum_{\vs{\mu}} 
   f\indexddu{Y}{\vs{\nu}_X\vs{\mu}_{\bar{X}}}{\vs{\nu}_X'\vs{\mu}_{\bar{X}}}
   f\indexddu{Y}{\vs{\mu}_X\vs{\nu}_{\bar{X}}}{\vs{\mu}_X\vs{\nu}_{\bar{X}}'} 
   \delta^{\vs{\nu}_X',\vs{\mu}_X}\delta^{\vs{\mu}_{\bar{X}},\vs{\nu}_{\bar{X}}}
   E\indexddu{Y}{\vs{\nu}_X\vs{\mu}_{\bar{X}}}{\vs{\mu}_X\vs{\nu}_{\bar{X}}'} \\
&\equals
   f\indexddu{Y}{\vs{\nu}_X\vs{\nu}_{\bar{X}}}{\vs{\nu}_X'\vs{\nu}_{\bar{X}}}
   f\indexddu{Y}{\vs{\nu}_X'\vs{\nu}_{\bar{X}}}{\vs{\nu}_X'\vs{\nu}_{\bar{X}}'} 
   E\indexddu{Y}{\vs{\nu}_X\vs{\nu}_{\bar{X}}}{\vs{\nu}_X'\vs{\nu}_{\bar{X}}'} \\
&\equalsref{eq:PhiYEY}
   f\indexddu{Y}{\vs{\nu}_X\vs{\nu}_{\bar{X}}}{\vs{\nu}_X'\vs{\nu}_{\bar{X}}}
   f\indexddu{Y}{\vs{\nu}_X'\vs{\nu}_{\bar{X}}}{\vs{\nu}_X'\vs{\nu}_{\bar{X}}'} 
   f\indexddu{Y}{\vs{\nu}}{\vs{\nu}'} \f{E}\indexddu{Y}{\vs{\nu}}{\vs{\nu}'}.
\end{split}
\end{equation*}
(Here we use the notation that
$\vs{\nu}_X\vs{\mu}_{\bar{X}}:Y\to\set{0,1}$ is the multi-index
taking values $\vs{\nu}_X:X\to\set{0,1}$ on $X$
and $\vs{\mu}_{\bar{X}}:\bar{X}\to\set{0,1}$ on $\bar{X}$.)
From this, we can read off the form
\begin{subequations}
\label{eq:Lambda2exp}
\begin{equation}
\label{eq:Lambda2def}
\f{\Lambda}_{X\bar{X}}(\f{E}\indexddu{Y}{\vs{\nu}}{\vs{\nu}'} )
= l\indexddu{X\bar{X}}{\vs{\nu}}{\vs{\nu}'}\f{E}\indexddu{Y}{\vs{\nu}}{\vs{\nu}'}
\end{equation}
with the phase factors
\begin{equation}
\label{eq:lXXb}
l\indexddu{X\bar{X}}{\vs{\nu}}{\vs{\nu}'}
\equals f\indexddu{Y}{\vs{\nu}_X\vs{\nu}_{\bar{X}}}{\vs{\nu}_X'\vs{\nu}_{\bar{X}}}
   f\indexddu{Y}{\vs{\nu}_X'\vs{\nu}_{\bar{X}}}{\vs{\nu}_X'\vs{\nu}_{\bar{X}}'} 
   f\indexddu{Y}{\vs{\nu}}{\vs{\nu}'} 
\equalsref{eq:stf} (-1)^{\esum_{i\in \bar{X}}(\nu_i+\nu_i')\esum_{k\in X,i<k}(\nu_k+\nu_k')}.
\end{equation}
\end{subequations}
Here the last equality is 
\begin{equation}
\label{eq:stf}
\begin{split}
&f\indexddu{Y}{\vs{\nu}_X\vs{\nu}_{\bar{X}}}{\vs{\nu}_X'\vs{\nu}_{\bar{X}}}
f\indexddu{Y}{\vs{\nu}_X'\vs{\nu}_{\bar{X}}}{\vs{\nu}_X'\vs{\nu}_{\bar{X}}'} 
f\indexddu{Y}{\vs{\nu}}{\vs{\nu}'}\\
&= (-1)^{
 \esum_{i\in     X }\nu_i'\esum_{k\in     X ,i<k}(\nu_k +\nu_k')
+\esum_{i\in     X }\nu_i'\esum_{l\in\bar{X},i<l}(\nu_l +\nu_l )
+\esum_{j\in\bar{X}}\nu_j'\esum_{k\in     X ,j<k}(\nu_k +\nu_k')
+\esum_{j\in\bar{X}}\nu_j'\esum_{l\in\bar{X},j<l}(\nu_l +\nu_l )}\\
&\times(-1)^{
 \esum_{i\in     X }\nu_i'\esum_{k\in     X ,i<k}(\nu_k'+\nu_k')
+\esum_{i\in     X }\nu_i'\esum_{l\in\bar{X},i<l}(\nu_l +\nu_l')
+\esum_{j\in\bar{X}}\nu_j'\esum_{k\in     X ,j<k}(\nu_k'+\nu_k')
+\esum_{j\in\bar{X}}\nu_j'\esum_{l\in\bar{X},j<l}(\nu_l +\nu_l')}\\
&\times(-1)^{
 \esum_{i\in     X }\nu_i'\esum_{k\in     X ,i<k}(\nu_k +\nu_k')
+\esum_{i\in     X }\nu_i'\esum_{l\in\bar{X},i<l}(\nu_l +\nu_l')
+\esum_{j\in\bar{X}}\nu_j'\esum_{k\in     X ,j<k}(\nu_k +\nu_k')
+\esum_{j\in\bar{X}}\nu_j'\esum_{l\in\bar{X},j<l}(\nu_l +\nu_l')},
\end{split}
\end{equation}
where all but two terms in the exponents are 
either even 
(second and fourth terms of the first phase factor,
and first and third terms of the second phase factor)
or cancelling each other 
(first terms of the first and the third phase factors,
second terms of the second and the third phase factors,
and fourth terms of the second and the third phase factors).

\subsection{Fermionic product operators for any number of mode subsets}
\label{appsec:Tensors.Lambda}

After having the explicit form \eqref{eq:Lambda2exp} 
of the bipartite $\f{\Lambda}_{X\bar{X}}$ map \eqref{eq:Lambda2} in hand,
we are able to derive the explicit form 
of the multipartite $\f{\Lambda}_{\ord{\xi}}$ map \eqref{eq:Lambda}.

First, note that because of \eqref{eq:Lambda2}, \eqref{eq:TPfAssoc} and \eqref{eq:mTPfTPf},
we can write the multipartite map as compositions of extensions of the bipartite map as
\begin{align*}
\f{\Lambda}_{(X_1,X_2,X_3)} &= \f{\Lambda}_{(X_1\cup X_2,X_3)} \circ \bigl(\f{\Lambda}_{(X_1,X_2)}\otimesf \IIdf_{X_3} \bigr),\\
\f{\Lambda}_{(X_1,X_2,X_3,X_4)} &= \f{\Lambda}_{(X_1\cup X_2\cup X_3,X_4)} 
\circ \bigl( \f{\Lambda}_{(X_1\cup X_2,X_3)}\otimesf \IIdf_{X_4} \bigr) \circ \bigl(\f{\Lambda}_{(X_1,X_2)}\otimesf \IIdf_{X_3\cup X_4} \bigr).\\
\vdots
\end{align*}
(This can be checked by applying it to elementary fermionic tensors $\bigotimesf_{X\in\xi}\f{A}_X$.)
For an ordered partition $\ord{\xi}=\tuple{X_1,X_2,\dots,X_{\abs{\ord{\xi}}}}$, 
this can be written as
\begin{equation}
\f{\Lambda}_{\ord{\xi}} = \prodordback_{r=2,\dots,\abs{\ord{\xi}}}
 \Bigl( \f{\Lambda}_{(W_{r-1}, X_r)} \otimesf \IIdf_{\bar{W}_r}  \Bigr),
\end{equation}
where $\prodordback$ stands for the reverse ordered composition, and
we use the notations $W_r:=\bigcup_{s=1}^{r} X_s$, being nested subsets, 
and $\bar{W}_r=Y\setminus W_r = \bigcup_{s=r+1}^{\abs{\ord{\xi}}} X_s$ for the complement with respect to $Y$, as usual.
From this, we can read off the form
\begin{subequations}
\label{eq:Lambdaexp}
\begin{equation}
\f{\Lambda}_{\ord{\xi}}(\f{E}\indexddu{Y}{\vs{\nu}}{\vs{\nu}'})
= l\indexddu{\ord{\xi}}{\vs{\nu}}{\vs{\nu}'}\f{E}\indexddu{Y}{\vs{\nu}}{\vs{\nu}'},
\end{equation}
with the phase factors
\begin{equation}
\label{eq:lups}
l\indexddu{\ord{\xi}}{\vs{\nu}}{\vs{\nu}'}
 = \prod_{r=2}^{\abs{\ord{\xi}}} l\indexddu{W_{r-1}, X_r}{\vs{\nu}_{W_r}}{\vs{\nu}'_{W_r}}
 = (-1)^{\esum_{r=2}^{\abs{\ord{\xi}}} \esum_{s=1}^{r-1} \esum_{i\in X_r}(\nu_i+\nu_i')\esum_{k\in X_s,i<k}(\nu_k+\nu_k')}
 = \prod_{1\leq s<r\leq\abs{\ord{\xi}}}l\indexddu{X_s, X_r}{\vs{\nu}_{X_{s,r}}}{\vs{\nu}'_{X_{s,r}}},
\end{equation}
with the notation $X_{s,r}:= X_s\cup X_r$.
\end{subequations}

\subsection{Phase factors for few modes}
\label{appsec:Tensors.Expll}
Here we show the phase factors $l\indexddu{\ord{\xi}}{\vs{\nu}}{\vs{\nu}'}$, given in \eqref{eq:lups}, for small mode subsets $Y$.
These are written in matrices indexed with multi-indices
$\vs{\nu}_Y=\tuple{\nu_1,\nu_2,\dots}$
ordered lexicographically with respect to the Jordan--Wigner ordering of the modes.
(Note that we use a simplified notation,
$\set{1}\set{2}\set{3}:=\tuple{\set{1},\set{2},\set{3}}$,
omitting the round brackets and colons in the writing of ordered partitions, since this does not cause confusion.)
For $\abs{Y}=2$, we have
\begin{equation*} 
l_{\set{1}\set{2}} = 
\left[\begin{smallmatrix}
 + & + & + & + \\
 + & + & + & + \\
 + & + & + & + \\
 + & + & + & +
\end{smallmatrix}\right], \qquad
l_{\set{2}\set{1}} = 
\left[\begin{smallmatrix}
 + & + & + & - \\
 + & + & - & + \\
 + & - & + & + \\
 - & + & + & +
\end{smallmatrix}\right].
\end{equation*}
For $\abs{Y}=3$, we have
\begin{align*} 
l_{\set{1}\set{2,3}} &= 
\left[\begin{smallmatrix}
 + & + & + & + & + & + & + & + \\
 + & + & + & + & + & + & + & + \\
 + & + & + & + & + & + & + & + \\
 + & + & + & + & + & + & + & + \\
 + & + & + & + & + & + & + & + \\
 + & + & + & + & + & + & + & + \\
 + & + & + & + & + & + & + & + \\
 + & + & + & + & + & + & + & +
\end{smallmatrix}\right],\qquad &
l_{\set{1,2}\set{3}} &= 
\left[\begin{smallmatrix}
 + & + & + & + & + & + & + & + \\
 + & + & + & + & + & + & + & + \\
 + & + & + & + & + & + & + & + \\
 + & + & + & + & + & + & + & + \\
 + & + & + & + & + & + & + & + \\
 + & + & + & + & + & + & + & + \\
 + & + & + & + & + & + & + & + \\
 + & + & + & + & + & + & + & +
\end{smallmatrix}\right],\\
l_{\set{2}\set{1,3}} &= 
\left[\begin{smallmatrix}
 + & + & + & + & + & + & - & - \\
 + & + & + & + & + & + & - & - \\
 + & + & + & + & - & - & + & + \\
 + & + & + & + & - & - & + & + \\
 + & + & - & - & + & + & + & + \\
 + & + & - & - & + & + & + & + \\
 - & - & + & + & + & + & + & + \\
 - & - & + & + & + & + & + & +
\end{smallmatrix}\right],\qquad &
l_{\set{1,3}\set{2}} &= 
\left[\begin{smallmatrix}
 + & + & + & - & + & + & + & - \\
 + & + & - & + & + & + & - & + \\
 + & - & + & + & + & - & + & + \\
 - & + & + & + & - & + & + & + \\
 + & + & + & - & + & + & + & - \\
 + & + & - & + & + & + & - & + \\
 + & - & + & + & + & - & + & + \\
 - & + & + & + & - & + & + & +
\end{smallmatrix}\right],\\
l_{\set{3}\set{1,2}} &= 
\left[\begin{smallmatrix}
 + & + & + & - & + & - & + & + \\
 + & + & - & + & - & + & + & + \\
 + & - & + & + & + & + & + & - \\
 - & + & + & + & + & + & - & + \\
 + & - & + & + & + & + & + & - \\
 - & + & + & + & + & + & - & + \\
 + & + & + & - & + & - & + & + \\
 + & + & - & + & - & + & + & +
\end{smallmatrix}\right],\qquad &
l_{\set{2,3}\set{1}} &= 
\left[\begin{smallmatrix}
 + & + & + & + & + & - & - & + \\
 + & + & + & + & - & + & + & - \\
 + & + & + & + & - & + & + & - \\
 + & + & + & + & + & - & - & + \\
 + & - & - & + & + & + & + & + \\
 - & + & + & - & + & + & + & + \\
 - & + & + & - & + & + & + & + \\
 + & - & - & + & + & + & + & +
\end{smallmatrix}\right],\\
l_{\set{1}\set{2}\set{3}} &= 
\left[\begin{smallmatrix}
 + & + & + & + & + & + & + & + \\
 + & + & + & + & + & + & + & + \\
 + & + & + & + & + & + & + & + \\
 + & + & + & + & + & + & + & + \\
 + & + & + & + & + & + & + & + \\
 + & + & + & + & + & + & + & + \\
 + & + & + & + & + & + & + & + \\
 + & + & + & + & + & + & + & +
\end{smallmatrix}\right],\qquad &
l_{\set{1}\set{3}\set{2}} &= 
\left[\begin{smallmatrix}
 + & + & + & - & + & + & + & - \\
 + & + & - & + & + & + & - & + \\
 + & - & + & + & + & - & + & + \\
 - & + & + & + & - & + & + & + \\
 + & + & + & - & + & + & + & - \\
 + & + & - & + & + & + & - & + \\
 + & - & + & + & + & - & + & + \\
 - & + & + & + & - & + & + & +
\end{smallmatrix}\right],\\
l_{\set{2}\set{1}\set{3}} &= 
\left[\begin{smallmatrix}
 + & + & + & + & + & + & - & - \\
 + & + & + & + & + & + & - & - \\
 + & + & + & + & - & - & + & + \\
 + & + & + & + & - & - & + & + \\
 + & + & - & - & + & + & + & + \\
 + & + & - & - & + & + & + & + \\
 - & - & + & + & + & + & + & + \\
 - & - & + & + & + & + & + & +
\end{smallmatrix}\right],\qquad &
l_{\set{3}\set{1}\set{2}} &= 
\left[\begin{smallmatrix}
 + & + & + & - & + & - & + & + \\
 + & + & - & + & - & + & + & + \\
 + & - & + & + & + & + & + & - \\
 - & + & + & + & + & + & - & + \\
 + & - & + & + & + & + & + & - \\
 - & + & + & + & + & + & - & + \\
 + & + & + & - & + & - & + & + \\
 + & + & - & + & - & + & + & +
\end{smallmatrix}\right],\\
l_{\set{2}\set{3}\set{1}} &= 
\left[\begin{smallmatrix}
 + & + & + & + & + & - & - & + \\
 + & + & + & + & - & + & + & - \\
 + & + & + & + & - & + & + & - \\
 + & + & + & + & + & - & - & + \\
 + & - & - & + & + & + & + & + \\
 - & + & + & - & + & + & + & + \\
 - & + & + & - & + & + & + & + \\
 + & - & - & + & + & + & + & +
\end{smallmatrix}\right],\qquad &
l_{\set{3}\set{2}\set{1}} &= 
\left[\begin{smallmatrix}
 + & + & + & - & + & - & - & - \\
 + & + & - & + & - & + & - & - \\
 + & - & + & + & - & - & + & - \\
 - & + & + & + & - & - & - & + \\
 + & - & - & - & + & + & + & - \\
 - & + & - & - & + & + & - & + \\
 - & - & + & - & + & - & + & + \\
 - & - & - & + & - & + & + & +
\end{smallmatrix}\right].
\end{align*}

\subsection{Qubit and fermionic partial trace}
\label{appsec:Tensors.ptr}

Here we show that 
the effect \eqref{eq:rhoXstd} of the fermionic partial trace \eqref{eq:PTf}
to the particular mode subset $X=\set{1,2,\dots,m}$
coincides with that of the standard partial trace \eqref{eq:TPPT}.
Using the phase factors \eqref{eq:fY}, we have in general that
\begin{equation*}
\begin{split}
f\indexddu{Y}{\vs{\nu}}{\vs{\nu}'} \delta^{\vs{\nu}_{\bar{X}},\vs{\nu}_{\bar{X}}'} 
&= (-1)^{\esum_{i\in Y}\nu_i'\esum_{k\in Y,i<k}(\nu_k+\nu_k')}
   \prod_{l\in\bar{X}} \delta^{\nu_l,\nu_l'}\\
&= (-1)^{\esum_{i\in Y}\nu_i'\esum_{k\in X,i<k}(\nu_k+\nu_k')}
   \prod_{l\in\bar{X}} \delta^{\nu_l,\nu_l'}\\
&= (-1)^{\esum_{i\in \bar{X}}\nu_i'\esum_{k\in X,i<k}(\nu_k+\nu_k')}
   (-1)^{\esum_{i\in X}\nu_i'\esum_{k\in X,i<k}(\nu_k+\nu_k')}
   \prod_{l\in\bar{X}} \delta^{\nu_l,\nu_l'}\\
&= (-1)^{\esum_{i\in \bar{X}}\nu_i'\esum_{k\in X,i<k}(\nu_k+\nu_k')}
f\indexddu{X}{\vs{\nu}_X}{\vs{\nu}_X'}
   \delta^{\vs{\nu}_{\bar{X}},\vs{\nu}_{\bar{X}}'}.
\end{split}
\end{equation*}
Now, if $X=\set{1,2,\dots,m}$ then there are no $i\in \bar{X}$ and $k\in X$ such that $i<k$,
so the first phase factor in the last line is $+1$,
and we have $f\indexddu{Y}{\vs{\nu}}{\vs{\nu}'} \delta^{\vs{\nu}_{\bar{X}},\vs{\nu}_{\bar{X}}'}
=f\indexddu{X}{\vs{\nu}_X}{\vs{\nu}_X'} \delta^{\vs{\nu}_{\bar{X}},\vs{\nu}_{\bar{X}}'}$,
leading to that the two phase factors in \eqref{eq:rhoXstd} cancel each other.

\subsection{Reduced density matrices for few modes}
\label{appsec:States.ExplR}
The density matrices $R\indexdud{Y}{\vs{\nu}}{\vs{\nu}'} = \bigskalpHS{E\indexddu{Y}{\vs{\nu}}{\vs{\nu}'}}{\f{\rho}_Y}$,
being the expansion coefficients \eqref{eq:rhoYstd}
of the density operators $\f{\rho}_Y$ in the standard basis \eqref{eq:EY},
and their reduced density matrices $R\indexdud{X}{\vs{\nu}_X}{\vs{\nu}_X'}$,
obtained by \eqref{eq:rhoXstd}, 
are shown for small mode subsets $X\subseteq Y \subseteq M$.
Note that all the minus signs in the matrices below
follow from the fermionic nature of the reduction \eqref{eq:rhoXstd},
containing the phase factors \eqref{eq:fY}, which are shown explicitly in Section~\ref{appsec:JW.Explf}.

For $Y=\{1,2\}$, we have
\begin{equation*}
R_{\{1,2\}} =
\left[\begin{smallmatrix}
 R\indexud{00}{00} & R\indexud{00}{01} & R\indexud{00}{10} &R\indexud{00}{11} \\
 R\indexud{01}{00} & R\indexud{01}{01} & R\indexud{01}{10} &R\indexud{01}{11} \\
 R\indexud{10}{00} & R\indexud{10}{01} & R\indexud{10}{10} &R\indexud{10}{11} \\
 R\indexud{11}{00} & R\indexud{11}{01} & R\indexud{11}{10} &R\indexud{11}{11}
\end{smallmatrix}\right],
\end{equation*}
\begin{equation*}
R_{\{1\}} =
\left[\begin{smallmatrix}
 R\indexud{00}{00} + R\indexud{01}{01} &R\indexud{00}{10} + R\indexud{01}{11} \\
 R\indexud{10}{00} + R\indexud{11}{01} &R\indexud{10}{10} + R\indexud{11}{11}
\end{smallmatrix}\right],
\end{equation*}
\begin{equation*}
R_{\{2\}} =
\left[\begin{smallmatrix}
 R\indexud{00}{00} + R\indexud{10}{10} &R\indexud{00}{01} - R\indexud{10}{11} \\
 R\indexud{01}{00} - R\indexud{11}{10} &R\indexud{01}{01} + R\indexud{11}{11}
\end{smallmatrix}\right].
\end{equation*}

For $Y=\{1,2,3\}$, we have
\begin{equation*}
R_{\{1,2,3\}} =
\left[\begin{smallmatrix}
 R\indexud{000}{000} & R\indexud{000}{001} & R\indexud{000}{010} & R\indexud{000}{011} & R\indexud{000}{100} & R\indexud{000}{101} & R\indexud{000}{110} &R\indexud{000}{111} \\
 R\indexud{001}{000} & R\indexud{001}{001} & R\indexud{001}{010} & R\indexud{001}{011} & R\indexud{001}{100} & R\indexud{001}{101} & R\indexud{001}{110} &R\indexud{001}{111} \\
 R\indexud{010}{000} & R\indexud{010}{001} & R\indexud{010}{010} & R\indexud{010}{011} & R\indexud{010}{100} & R\indexud{010}{101} & R\indexud{010}{110} &R\indexud{010}{111} \\
 R\indexud{011}{000} & R\indexud{011}{001} & R\indexud{011}{010} & R\indexud{011}{011} & R\indexud{011}{100} & R\indexud{011}{101} & R\indexud{011}{110} &R\indexud{011}{111} \\
 R\indexud{100}{000} & R\indexud{100}{001} & R\indexud{100}{010} & R\indexud{100}{011} & R\indexud{100}{100} & R\indexud{100}{101} & R\indexud{100}{110} &R\indexud{100}{111} \\
 R\indexud{101}{000} & R\indexud{101}{001} & R\indexud{101}{010} & R\indexud{101}{011} & R\indexud{101}{100} & R\indexud{101}{101} & R\indexud{101}{110} &R\indexud{101}{111} \\
 R\indexud{110}{000} & R\indexud{110}{001} & R\indexud{110}{010} & R\indexud{110}{011} & R\indexud{110}{100} & R\indexud{110}{101} & R\indexud{110}{110} &R\indexud{110}{111} \\
 R\indexud{111}{000} & R\indexud{111}{001} & R\indexud{111}{010} & R\indexud{111}{011} & R\indexud{111}{100} & R\indexud{111}{101} & R\indexud{111}{110} &R\indexud{111}{111}
\end{smallmatrix}\right],
\end{equation*}
\begin{equation*}
R_{\{2,3\}} =
\left[\begin{smallmatrix}
 R\indexud{000}{000} + R\indexud{100}{100} & R\indexud{000}{001} - R\indexud{100}{101} & R\indexud{000}{010} - R\indexud{100}{110} &R\indexud{000}{011} + R\indexud{100}{111} \\
 R\indexud{001}{000} - R\indexud{101}{100} & R\indexud{001}{001} + R\indexud{101}{101} & R\indexud{001}{010} + R\indexud{101}{110} &R\indexud{001}{011} - R\indexud{101}{111} \\
 R\indexud{010}{000} - R\indexud{110}{100} & R\indexud{010}{001} + R\indexud{110}{101} & R\indexud{010}{010} + R\indexud{110}{110} &R\indexud{010}{011} - R\indexud{110}{111} \\
 R\indexud{011}{000} + R\indexud{111}{100} & R\indexud{011}{001} - R\indexud{111}{101} & R\indexud{011}{010} - R\indexud{111}{110} &R\indexud{011}{011} + R\indexud{111}{111}
\end{smallmatrix}\right],
\end{equation*}
\begin{equation*}
R_{\{1,3\}} =
\left[\begin{smallmatrix}
 R\indexud{000}{000} + R\indexud{010}{010} & R\indexud{000}{001} - R\indexud{010}{011} & R\indexud{000}{100} + R\indexud{010}{110} &R\indexud{000}{101} - R\indexud{010}{111} \\
 R\indexud{001}{000} - R\indexud{011}{010} & R\indexud{001}{001} + R\indexud{011}{011} & R\indexud{001}{100} - R\indexud{011}{110} &R\indexud{001}{101} + R\indexud{011}{111} \\
 R\indexud{100}{000} + R\indexud{110}{010} & R\indexud{100}{001} - R\indexud{110}{011} & R\indexud{100}{100} + R\indexud{110}{110} &R\indexud{100}{101} - R\indexud{110}{111} \\
 R\indexud{101}{000} - R\indexud{111}{010} & R\indexud{101}{001} + R\indexud{111}{011} & R\indexud{101}{100} - R\indexud{111}{110} &R\indexud{101}{101} + R\indexud{111}{111}
\end{smallmatrix}\right],
\end{equation*}
\begin{equation*}
R_{\{1,2\}} =
\left[\begin{smallmatrix}
 R\indexud{000}{000} + R\indexud{001}{001} & R\indexud{000}{010} + R\indexud{001}{011} & R\indexud{000}{100} + R\indexud{001}{101} &R\indexud{000}{110} + R\indexud{001}{111} \\
 R\indexud{010}{000} + R\indexud{011}{001} & R\indexud{010}{010} + R\indexud{011}{011} & R\indexud{010}{100} + R\indexud{011}{101} &R\indexud{010}{110} + R\indexud{011}{111} \\
 R\indexud{100}{000} + R\indexud{101}{001} & R\indexud{100}{010} + R\indexud{101}{011} & R\indexud{100}{100} + R\indexud{101}{101} &R\indexud{100}{110} + R\indexud{101}{111} \\
 R\indexud{110}{000} + R\indexud{111}{001} & R\indexud{110}{010} + R\indexud{111}{011} & R\indexud{110}{100} + R\indexud{111}{101} &R\indexud{110}{110} + R\indexud{111}{111}
\end{smallmatrix}\right],
\end{equation*}
\begin{equation*}
R_{\{1\}} =
\left[\begin{smallmatrix}
 R\indexud{000}{000} + R\indexud{001}{001} + R\indexud{010}{010} + R\indexud{011}{011} &R\indexud{000}{100} + R\indexud{001}{101} + R\indexud{010}{110} + R\indexud{011}{111} \\
 R\indexud{100}{000} + R\indexud{101}{001} + R\indexud{110}{010} + R\indexud{111}{011} &R\indexud{100}{100} + R\indexud{101}{101} + R\indexud{110}{110} + R\indexud{111}{111}
\end{smallmatrix}\right],
\end{equation*}
\begin{equation*}
R_{\{2\}} =
\left[\begin{smallmatrix}
 R\indexud{000}{000} + R\indexud{001}{001} + R\indexud{100}{100} + R\indexud{101}{101} &R\indexud{000}{010} + R\indexud{001}{011} - R\indexud{100}{110} - R\indexud{101}{111} \\
 R\indexud{010}{000} + R\indexud{011}{001} - R\indexud{110}{100} - R\indexud{111}{101} &R\indexud{010}{010} + R\indexud{011}{011} + R\indexud{110}{110} + R\indexud{111}{111}
\end{smallmatrix}\right],
\end{equation*}
\begin{equation*}
R_{\{3\}} =
\left[\begin{smallmatrix}
 R\indexud{000}{000} + R\indexud{010}{010} + R\indexud{100}{100} + R\indexud{110}{110} &R\indexud{000}{001} - R\indexud{010}{011} - R\indexud{100}{101} + R\indexud{110}{111} \\
 R\indexud{001}{000} - R\indexud{011}{010} - R\indexud{101}{100} + R\indexud{111}{110} &R\indexud{001}{001} + R\indexud{011}{011} + R\indexud{101}{101} + R\indexud{111}{111}
\end{smallmatrix}\right].
\end{equation*}

It can also be seen that there is no $-1$ phase factor
when the reduction is taken to the first consecutive modes,
$R_{\{1\}}$, $R_{\{1,2\}}$, see Appendix~\ref{appsec:Tensors.ptr}.

\section{On the parity superselection}
\label{appsec:Parity}

\subsection{\texorpdfstring{$\xi$}{xi}-local parity eigenspaces for few modes}
\label{appsec:Parity.ExplSpaces} 

Because the phase operator \eqref{eq:pY},
which defines the parity superselection rule,
is diagonal in the natural occupation number basis \eqref{eq:phiY} we use,
the linear constraints determining the parity subspaces 
simplify to the vanishing or non-vanishing of vector or matrix elements
in the occupation number basis \eqref{eq:phiY}, \eqref{eq:EY} and \eqref{eq:EYf}.
(Note that we use a simplified notation,
$\set{1}\set{2}\set{3}:=\set{\set{1},\set{2},\set{3}}$,
omitting the curly brackets and colons in the writing of partitions, since this does not cause confusion.)

First, consider the Hilbert space $\mathcal{H}_Y$ and \eqref{eq:HYpmSpan}, \eqref{eq:HupsepsSpan}.
For $Y=\set{1}$,
the subspaces of vectors of even, respectively odd number of fermions are
\begin{equation*}
\mathcal{H}_{\set{1}}^+ :
\left[\begin{smallmatrix}
v^0\\
0
\end{smallmatrix}\right], \qquad 
\mathcal{H}_{\set{1}}^- :
\left[\begin{smallmatrix}
0\\
v^1
\end{smallmatrix}\right], 
\end{equation*}
For $Y=\set{1,2}$,
the subspaces of vectors of well-defined local fermion number parities are
\begin{equation*}
\mathcal{H}_{\set{1}\set{2}}^{++} :
\left[\begin{smallmatrix}
v^{00}\\
0\\
0\\
0
\end{smallmatrix}\right], \qquad 
\mathcal{H}_{\set{1}\set{2}}^{+-} :
\left[\begin{smallmatrix}
0\\
v^{01}\\
0\\
0
\end{smallmatrix}\right], \qquad 
\mathcal{H}_{\set{1}\set{2}}^{-+} :
\left[\begin{smallmatrix}
0\\
0\\
v^{10}\\
0
\end{smallmatrix}\right], \qquad 
\mathcal{H}_{\set{1}\set{2}}^{--} :
\left[\begin{smallmatrix}
0\\
0\\
0\\
v^{11}
\end{smallmatrix}\right], 
\end{equation*}
and the subspaces of vectors of even, respectively odd number of fermions are
\begin{equation*}
\mathcal{H}_{\set{1,2}}^{+}=\mathcal{H}_{\set{1}\set{2}}^{++}\oplus\mathcal{H}_{\set{1}\set{2}}^{--} :
\left[\begin{smallmatrix}
v^{00}\\
0\\
0\\
v^{11}
\end{smallmatrix}\right], \qquad 
\mathcal{H}_{\set{1,2}}^{-}=\mathcal{H}_{\set{1}\set{2}}^{+-}\oplus\mathcal{H}_{\set{1}\set{2}}^{-+} :
\left[\begin{smallmatrix}
0\\
v^{01}\\
v^{10}\\
0
\end{smallmatrix}\right].
\end{equation*}
Using a shorthand notation for the parity subspaces, we have the following pattern
for $Y=\set{1}$, $Y=\set{1,2}$ and $Y=\set{1,2,3}$,
\begin{equation*}
\left[\begin{smallmatrix}
+\\
-
\end{smallmatrix}\right], \qquad
\left[\begin{smallmatrix}
++\\
+-\\
-+\\
--
\end{smallmatrix}\right], \qquad
\left[\begin{smallmatrix}
+++\\
++-\\
+-+\\
+--\\
-++\\
-+-\\
--+\\
---
\end{smallmatrix}\right].
\end{equation*}

Second, consider the operator algebra $\mathcal{A}_Y$ and \eqref{eq:AYpmSpan}, \eqref{eq:AupsepsSpan}.
For $Y=\set{1}$, 
the subspaces of even, respectively odd operators are
\begin{equation*}
\f{\mathcal{A}}_{\{1\}}^{+}:
\left[\begin{smallmatrix}
 A\indexud{0}{0} & 0                \\
 0               & A\indexud{1}{1}  
\end{smallmatrix}\right], \qquad
\f{\mathcal{A}}_{\{1\}}^{-}:
\left[\begin{smallmatrix}
 0               & A\indexud{0}{1}  \\
 A\indexud{1}{0} & 0                
\end{smallmatrix}\right].
\end{equation*}
For $Y=\set{1,2}$,
the subspaces of operators of well-defined local parities are
\begin{align*}
\f{\mathcal{A}}_{\set{1}\set{2}}^{++}:
&\left[\begin{smallmatrix}
 A\indexud{00}{00} & 0                 & 0                 & 0                 \\
 0                 & A\indexud{01}{01} & 0                 & 0                 \\
 0                 & 0                 & A\indexud{10}{10} & 0                 \\
 0                 & 0                 & 0                 & A\indexud{11}{11}
\end{smallmatrix}\right], \qquad
\f{\mathcal{A}}_{\set{1}\set{2}}^{+-}:
\left[\begin{smallmatrix}
 0                 & A\indexud{00}{01} & 0                 & 0                 \\
 A\indexud{01}{00} & 0                 & 0                 & 0                 \\
 0                 & 0                 & 0                 & A\indexud{10}{11} \\
 0                 & 0                 & A\indexud{11}{10} & 0                
\end{smallmatrix}\right],\\
\f{\mathcal{A}}_{\set{1}\set{2}}^{-+}:
&\left[\begin{smallmatrix}
 0                 & 0                 & A\indexud{00}{10} & 0                 \\
 0                 & 0                 & 0                 & A\indexud{01}{11} \\
 A\indexud{10}{00} & 0                 & 0                 & 0                 \\
 0                 & A\indexud{11}{01} & 0                 & 0                
\end{smallmatrix}\right], \qquad
\f{\mathcal{A}}_{\set{1}\set{2}}^{--}:
\left[\begin{smallmatrix}
 0                 & 0                 & 0                 & A\indexud{00}{11} \\
 0                 & 0                 & A\indexud{01}{10} & 0                 \\
 0                 & A\indexud{10}{01} & 0                 & 0                 \\
 A\indexud{11}{00} & 0                 & 0                 & 0                
\end{smallmatrix}\right],
\end{align*}
and the subspaces of even, respectively odd operators are
\begin{equation*}
\f{\mathcal{A}}_{\set{1,2}}^{+}:
\left[\begin{smallmatrix}
 A\indexud{00}{00} & 0                 & 0                 & A\indexud{00}{11} \\
 0                 & A\indexud{01}{01} & A\indexud{01}{10} & 0                 \\
 0                 & A\indexud{10}{01} & A\indexud{10}{10} & 0                 \\
 A\indexud{11}{00} & 0                 & 0                 & A\indexud{11}{11}
\end{smallmatrix}\right], \qquad
\f{\mathcal{A}}_{\set{1,2}}^{-}:
\left[\begin{smallmatrix}
 0                 & A\indexud{00}{01} & A\indexud{00}{10} & 0                 \\
 A\indexud{01}{00} & 0                 & 0                 & A\indexud{01}{11} \\
 A\indexud{10}{00} & 0                 & 0                 & A\indexud{10}{11} \\
 0                 & A\indexud{11}{01} & A\indexud{11}{10} & 0                
\end{smallmatrix}\right].
\end{equation*}
Using a shorthand notation for the parity subspaces, we have the following pattern
for $Y=\set{1}$, $Y=\set{1,2}$ and $Y=\set{1,2,3}$,
\begin{equation*}
\left[\begin{smallmatrix}
+ & -  \\
- & +  \\
\end{smallmatrix}\right],\qquad
\left[\begin{smallmatrix}
++ & +- & -+ & -- \\
+- & ++ & -- & -+ \\
-+ & -- & ++ & +- \\
-- & -+ & +- & ++ 
\end{smallmatrix}\right],\qquad
\left[\begin{smallmatrix}
+++ & ++- & +-+ & +-- & -++ & -+- & --+ & --- \\
++- & +++ & +-- & +-+ & -+- & -++ & --- & --+ \\
+-+ & +-- & +++ & ++- & --+ & --- & -++ & -+- \\
+-- & +-+ & ++- & +++ & --- & --+ & -+- & -++ \\
-++ & -+- & --+ & --- & +++ & ++- & +-+ & +-- \\
-+- & -++ & --- & --+ & ++- & +++ & +-- & +-+ \\
--+ & --- & -++ & -+- & +-+ & +-- & +++ & ++- \\
--- & --+ & -+- & -++ & +-- & +-+ & ++- & +++ 
\end{smallmatrix}\right].
\end{equation*}

\subsection{\texorpdfstring{$\xi$}{xi}-locally physical subalgebra}
\label{appsec:Parity.LocPhys}

To see how the parity superselection rule affects our formalism,
note that,
for $\f{A}_X\in\f{\mathcal{A}}_X$, $\f{B}_{\bar{X}}\in\f{\mathcal{A}}_{\bar{X}}$,
\begin{equation}
\begin{split}
\bigl[\iotaf_{X,Y}(\f{A}_X), \iotaf_{\bar{X},Y}(\f{B}_{\bar{X}}) \bigr] 
&= \iotaf_{X,Y}(\f{A}_X)\iotaf_{\bar{X},Y}(\f{B}_{\bar{X}})
  -\iotaf_{\bar{X},Y}(\f{B}_{\bar{X}})\iotaf_{X,Y}(\f{A}_X)\\
&= \bigl(\f{\Lambda}_{X\bar{X}}-\f{\Lambda}_{\bar{X}X}\bigr)(\f{A}_X\otimesf\f{B}_{\bar{X}}),
\end{split}
\end{equation}
with the $\f{\Lambda}_{X\bar{X}}$ map \eqref{eq:Lambda2}, which is given in terms of the
phase factors \eqref{eq:lXXb}.
If $\f{A}_X\in\f{\mathcal{A}}_X^+$, $\f{B}_{\bar{X}}\in\f{\mathcal{A}}_{\bar{X}}^+$,
we have the restriction for the indices
\begin{equation}
(-1)^{\esum_{k\in X} (\nu_k+\nu_k')} = (-1)^{\esum_{i\in \bar{X}} (\nu_i+\nu_i')} =  +1,
\end{equation}
since no other basis element can have nonzero coefficient, see \eqref{eq:AYpmSpan}.
From this, it quickly follows that
\begin{equation*}
1 = (-1)^{\esum_{i\in \bar{X}}(\nu_i+\nu_i')\esum_{k\in X}    (\nu_k+\nu_k')}
  = (-1)^{\esum_{i\in \bar{X}}(\nu_i+\nu_i')\esum_{k\in X,i<k}(\nu_k+\nu_k')
  +       \esum_{i\in \bar{X}}(\nu_i+\nu_i')\esum_{k\in X,i>k}(\nu_k+\nu_k')}, 
\end{equation*}
that is,
\begin{equation}
\label{eq:allevenswap}
 (-1)^{\esum_{i\in \bar{X}}(\nu_i+\nu_i')\esum_{k\in X,i<k}(\nu_k+\nu_k')}
=(-1)^{\esum_{k\in X}(\nu_k+\nu_k')\esum_{i\in \bar{X},k<i}(\nu_i+\nu_i')},
\end{equation}
which is $l\indexddu{X\bar{X}}{\vs{\nu}}{\vs{\nu}'}=l\indexddu{\bar{X}X}{\vs{\nu}}{\vs{\nu}'}$ by \eqref{eq:lXXb},
which leads to that $\f{\Lambda}_{X\bar{X}}=\f{\Lambda}_{\bar{X}X}$ by \eqref{eq:Lambda2def} 
over the $\set{X,\bar{X}}$-even subalgebra $\f{\mathcal{A}}_{X\bar{X}}^{++}$.
Also, if we consider the general case for the ordered partition $\ord{\xi}$,
we have that the phase factor $l\indexddu{\ord{\xi}}{\vs{\nu}}{\vs{\nu}'}$, given in \eqref{eq:lups},
can be written as a product of bipartite phase factors,
for which we can apply the above result.

To write down the phase factors $l\indexddu{\ord{\xi}}{\vs{\nu}}{\vs{\nu}'}$,
there is a need for a fixed ordering of the parts,
however, this is artificial, all choices lead to the same phase factors
for the $\xi$-locally physical subalgebra.

\subsection{Tensor product structure on the \texorpdfstring{$\xi$}{xi}-locally physical subalgebra}
\label{appsec:Parity.LocPhysTPS}

Here we show the derivation of \eqref{eq:TPSA}, and give the unitary $U_\xi$ explicitly.
First, note that 
\begin{equation}
\bigotimesfp_{X\in\ord{\xi}}\f{A}_X
\equalsref{eq:TPfp} \prodord_{X\in\ord{\xi}} \iotaf_{X,Y}(\f{A}_X)
\equalsref{eq:Lambda} \f{\Lambda}_{\ord{\xi}}\Bigl(\bigotimesf_{X\in\xi}\f{A}_X\Bigr)
\equalsref{eq:TPfPsi} \f{\Lambda}_\xi\Bigl(\f{\Psi}_\xi\Bigl(\bigotimes_{X\in\xi}\f{A}_X\Bigr)\Bigr),
\end{equation}
so we have to derive the action of $\f{\Lambda}_{\ord{\xi}}\circ\f{\Psi}_\xi$
on the $\xi$-locally physical subalgebra $\f{\mathcal{A}}_X^\alleven$.
This is the elementwise product with the phase factors \eqref{eq:lups} and \eqref{eq:hups} as
\begin{equation*}
\begin{split}
&l\indexddu{\ord{\xi}}{\vs{\nu}}{\vs{\nu}'}
h\indexddu{\xi}{\vs{\nu}}{\vs{\nu}'}\\
&\equals (-1)^{
 \esum_{s<r}  \esum_{i\in X_r}(\nu_i+\nu_i')\esum_{k\in X_s,i<k}(\nu_k+\nu_k')
+\esum_{X,X'\in\xi, X\neq X'} \esum_{i\in X}\nu_i'\esum_{k\in X',i<k}(\nu_k+\nu_k') }\\
&\equals (-1)^{
 \esum_{s<r} \esum_{i\in X_r}(\nu_i+\nu_i')\esum_{k\in X_s,i<k}(\nu_k+\nu_k')
+\esum_{s<r} \esum_{i\in X_r}\nu_i'\esum_{k\in X_s,i<k}(\nu_k+\nu_k')
+\esum_{s>r} \esum_{i\in X_r}\nu_i'\esum_{k\in X_s,i<k}(\nu_k+\nu_k')
}\\
&\equals (-1)^{
 \esum_{s<r} \esum_{i\in X_r}\nu_i \esum_{k\in X_s,i<k}(\nu_k+\nu_k')
+\esum_{s>r} \esum_{i\in X_r}\nu_i'\esum_{k\in X_s,i<k}(\nu_k+\nu_k')
}\\
&\equals (-1)^{
 \esum_{s<r} \esum_{i\in X_r}\nu_i \esum_{k\in X_s,i<k}(\nu_k+\nu_k')
+\esum_{s>r} \esum_{i\in X_r}\nu_i'\esum_{k\in X_s,i>k}(\nu_k+\nu_k')
}\\
&\equals (-1)^{
 \esum_{s<r} \esum_{i\in X_r}\nu_i \esum_{k\in X_s,i<k}\nu_k
+\esum_{s<r} \esum_{i\in X_r}\nu_i \esum_{k\in X_s,i<k}\nu_k'
+\esum_{s>r} \esum_{i\in X_r}\nu_i'\esum_{k\in X_s,i>k}\nu_k
+\esum_{s>r} \esum_{i\in X_r}\nu_i'\esum_{k\in X_s,i>k}\nu_k'
}\\
&\equals (-1)^{
 \esum_{s<r} \esum_{i\in X_r}\nu_i \esum_{k\in X_s,i<k}\nu_k
}(-1)^{
 \esum_{s<r} \esum_{i\in X_r}\nu_i'\esum_{k\in X_s,i<k}\nu_k'
}.
\end{split}
\end{equation*}
(The fourth equality is where the superselection is used,
namely, $\sum_{k\in X} (\nu_k+\nu_k')$ is even for all $X\in\xi$ in the $\xi$-locally physical subalgebra \eqref{eq:AYpmSpan};
the last equality is by noting that, after the fifth equation sign,
the second and third terms are the same,
by relabeling the indices $i \leftrightarrow k$ and $r \leftrightarrow s$,
and by the same relabeling in the fourth term.)
That is, we managed to write 
$l\indexddu{\ord{\xi}}{\vs{\nu}}{\vs{\nu}'}
 h\indexddu{\xi}{\vs{\nu}}{\vs{\nu}'}$
for the indices corresponding to the $\xi$-locally physical subalgebra
as a product of two factors, 
depending on the unprimed or primed indices only.
Now let us have the operator
\begin{equation}
\label{eq:uupsilon}
\f{U}_{\ord{\xi}} 
= \sum_{\vs{\nu}}u_{\ord{\xi}}^{\vs{\nu}} \f{E}\indexddu{Y}{\vs{\nu}}{\vs{\nu}}
= \sum_{\vs{\nu}}u_{\ord{\xi}}^{\vs{\nu}}   {E}\indexddu{Y}{\vs{\nu}}{\vs{\nu}}, \qquad
u_{\ord{\xi}}^{\vs{\nu}} = (-1)^{\esum_{1\leq s<r\leq\abs{\ord{\xi}}} \esum_{i\in X_r}\nu_i\esum_{k\in X_s,i<k}\nu_k},
\end{equation}
which is diagonal in the standard basis \eqref{eq:phiY}, see \eqref{eq:EYexpl}, with entries $\pm1$, so it is unitary.
Then, we have
\begin{equation*}
\begin{split}
\f{\Lambda}_{\ord{\xi}}\Bigl(\f{\Psi}_\xi\Bigl(\bigotimes_{X\in\xi}{E}\indexddu{X}{\vs{\nu}_X}{\vs{\nu}'_X}\Bigr)\Bigr)
&= \f{\Lambda}_{\ord{\xi}}\Bigl(\f{\Psi}_\xi({E}\indexddu{Y}{\vs{\nu}}{\vs{\nu}'})\Bigr)
 = l\indexddu{\ord{\xi}}{\vs{\nu}}{\vs{\nu}'}
   h\indexddu{\xi}{\vs{\nu}}{\vs{\nu}'} {E}\indexddu{Y}{\vs{\nu}}{\vs{\nu}'}\\
&= \f{U}_{\ord{\xi}} {E}\indexddu{Y}{\vs{\nu}}{\vs{\nu}'} \f{U}_{\ord{\xi}}^\dagger
 = \f{U}_{\ord{\xi}} \Bigl(\bigotimes_{X\in\xi}{E}\indexddu{X}{\vs{\nu}_X}{\vs{\nu}'_X}\Bigr) \f{U}_{\ord{\xi}}^\dagger,
\end{split}
\end{equation*}
leading to \eqref{eq:TPSA} by linearity.

Notice that $u_{X\bar{X}}^{\vs{\nu}}$ in \eqref{eq:uXXb} 
is a special case of $u_{\ord{\xi}}^{\vs{\nu}}$ in \eqref{eq:uupsilon}.
Although this holds for the explicit formulas, this does not hold for the notions represented by them:
the definition \eqref{eq:iotaU} leading to \eqref{eq:uXXb} is meaningful for the whole algebra,
but the definition \eqref{eq:TPSA} leading to \eqref{eq:uupsilon} is meaningful only in the $\xi$-locally physical subalgebra.

Note that the ordering of the partition $\ord{\xi}$ has to be fixed
for writing $\f{U}_{\ord{\xi}}$ (this ordering is a ``parameter'' of the tensor product structure in the Hilbert space),
however, $\f{A}_Y \mapsto \f{U}_{\ord{\xi}} \f{A}_Y \f{U}_{\ord{\xi}}^\dagger$ acts on the $\xi$-locally physical subalgebra
in the same way, without respect to the ordering. 
To see this on the level of indices, it is enough to notice that
$l\indexddu{\ord{\xi}}{\vs{\nu}}{\vs{\nu}'}$ is ordering independent in the $\xi$-locally physical subalgebra (see Appendix~\ref{appsec:Parity.LocPhys}),
while $h\indexddu{\xi}{\vs{\nu}}{\vs{\nu}'}$ is ordering independent in general.

\subsection{Phase factors for few modes}
\label{appsec:Parity.Explu}
Here we show the phase factors $u_{\ord{\xi}}^{\vs{\nu}}$, given in \eqref{eq:uupsilon}, for small mode subsets $Y \subseteq M$.
Note that $u_{X\bar{X}}^{\vs{\nu}}$, given in \eqref{eq:uXXb}, is a special case of this.
These are written in matrices indexed with multi-indices
$\vs{\nu}_Y=\tuple{\nu_1,\nu_2,\dots}$
ordered lexicographically with respect to the Jordan--Wigner ordering of the modes.
(Note that we use a simplified notation,
$\set{1}\set{2}\set{3}:=\tuple{\set{1},\set{2},\set{3}}$,
omitting the round brackets and colons in the writing of ordered partitions, since this does not cause confusion.)
For $\abs{Y}=2$, we have
\begin{equation*}
u_{\set{1}\set{2}} = \left[\begin{smallmatrix}+&+&+&+\end{smallmatrix}\right],\qquad 
u_{\set{2}\set{1}} = \left[\begin{smallmatrix}+&+&+&-\end{smallmatrix}\right].
\end{equation*}
For $\abs{Y}=3$, we have
\begin{align*}
u_{\set{1}\set{2,3}}      &= \left[\begin{smallmatrix}+&+&+&+&+&+&+&+\end{smallmatrix}\right],\qquad &
u_{\set{2,3}\set{1}}      &= \left[\begin{smallmatrix}+&+&+&+&+&-&-&+\end{smallmatrix}\right], \\
u_{\set{2}\set{1,3}}      &= \left[\begin{smallmatrix}+&+&+&+&+&+&-&-\end{smallmatrix}\right],\qquad &
u_{\set{1,3}\set{2}}      &= \left[\begin{smallmatrix}+&+&+&-&+&+&+&-\end{smallmatrix}\right], \\
u_{\set{3}\set{1,2}}      &= \left[\begin{smallmatrix}+&+&+&-&+&-&+&+\end{smallmatrix}\right],\qquad &
u_{\set{1,2}\set{3}}      &= \left[\begin{smallmatrix}+&+&+&+&+&+&+&+\end{smallmatrix}\right], \\
u_{\set{1}\set{2}\set{3}} &= \left[\begin{smallmatrix}+&+&+&+&+&+&+&+\end{smallmatrix}\right],\qquad &
u_{\set{1}\set{3}\set{2}} &= \left[\begin{smallmatrix}+&+&+&-&+&+&+&-\end{smallmatrix}\right], \\
u_{\set{2}\set{1}\set{3}} &= \left[\begin{smallmatrix}+&+&+&+&+&+&-&-\end{smallmatrix}\right],\qquad &
u_{\set{3}\set{1}\set{2}} &= \left[\begin{smallmatrix}+&+&+&-&+&-&+&+\end{smallmatrix}\right], \\
u_{\set{2}\set{3}\set{1}} &= \left[\begin{smallmatrix}+&+&+&+&+&-&-&+\end{smallmatrix}\right],\qquad &
u_{\set{3}\set{2}\set{1}} &= \left[\begin{smallmatrix}+&+&+&-&+&-&-&-\end{smallmatrix}\right].
\end{align*}

We can also illustrate that 
although the writing of the matrix elements of $\f{U}_{\ord{\xi}}$, given in \eqref{eq:uupsilon}, is ordering dependent,
the effect of the $\f{U}_{\ord{\xi}}$-adjoint is ordering independent.
For example, the adjoint actions of $\f{U}_{\set{2}\set{1,3}}$ and $\f{U}_{\set{1,3}\set{2}}$
are just the elementwise multiplications with the phase factors
\begin{equation*}
\left[\begin{smallmatrix}
+&+&+&+&+&+&-&-\\
+&+&+&+&+&+&-&-\\
+&+&+&+&+&+&-&-\\
+&+&+&+&+&+&-&-\\
+&+&+&+&+&+&-&-\\
+&+&+&+&+&+&-&-\\
-&-&-&-&-&-&+&+\\
-&-&-&-&-&-&+&+
\end{smallmatrix}\right],\qquad
\left[\begin{smallmatrix}
+&+&+&-&+&+&+&-\\
+&+&+&-&+&+&+&-\\
+&+&+&-&+&+&+&-\\
-&-&-&+&-&-&-&+\\
+&+&+&-&+&+&+&-\\
+&+&+&-&+&+&+&-\\
+&+&+&-&+&+&+&-\\
-&-&-&+&-&-&-&+
\end{smallmatrix}\right],
\end{equation*}
which are coinciding for the indices corresponding to the $\set{\set{2},\set{1,3}}$-locally physical subalgebra,
$\f{\mathcal{A}}_{\{1,3\}}^{+}\otimesf\f{\mathcal{A}}_{\{2\}}^{+}=
\bigl(\f{\mathcal{A}}_{\{1,3\}}^{++}\oplus\f{\mathcal{A}}_{\{1,3\}}^{--}\bigr)\otimesf\f{\mathcal{A}}_{\{2\}}^{+}=
\f{\mathcal{A}}_{\{1,2,3\}}^{+++}\oplus\f{\mathcal{A}}_{\{1,2,3\}}^{-+-}$,
see at the end of Section~\ref{appsec:Parity.ExplSpaces}.

\subsection{Examples}
\label{appsec:Parity.examples}

Here we show illustrations for properties
which do not hold without the parity superselection.

First, considering the fermionic canonical embedding \eqref{eq:cEmbf},
we show that
\begin{align*}
\iotaf_{X,Y}(\f{A}_X)U_{X\bar{X}}\bigl(\cket{\psi_X}\otimes\cket{\psi_{\bar{X}}}\bigr) 
&= U_{X\bar{X}} \bigl(\f{A}_X\cket{\psi_X}\otimes\cket{\psi_{\bar{X}}}\bigr),\\
\intertext{but there exist operators $\f{B}_{\bar{X}}\in\f{\mathcal{A}}_{\bar{X}}$
and vectors $\cket{\psi_X}\in\mathcal{H}_X$ and
$\cket{\psi_{\bar{X}}}\in\mathcal{H}_{\bar{X}}$ such that}
\iotaf_{\bar{X},Y}(\f{B}_{\bar{X}})U_{X\bar{X}}\bigl(\cket{\psi_X}\otimes\cket{\psi_{\bar{X}}}\bigr) 
&\neq U_{X\bar{X}} \bigl(\cket{\psi_X}\otimes\f{B}_{\bar{X}}\cket{\psi_{\bar{X}}}\bigr),
\end{align*}
where $U_{X\bar{X}}$ is the unitary by which $\iotaf_{X,Y}$ is given, see \eqref{eq:iotaU},
so the equality follows directly from \eqref{eq:iotaU}.
To see the non-equality, using \eqref{eq:iotaU}, we have
\begin{equation*}
U_{X\bar{X}}^\dagger U_{\bar{X}X} \bigl(\Idf_X\otimes\f{B}_{\bar{X}}\bigr) U_{\bar{X}X}^\dagger U_{X\bar{X}}
\neq \Idf_X\otimes\f{B}_{\bar{X}}.
\end{equation*}
Since the unitaries $U_{\bar{X}X}$ and $U_{\bar{X}X}$ are diagonal,
we can use that multiplication by diagonal matrices from the left or right
means multiplication of rows or columns by the respective diagonal elements.
Considering the simplest case, when $X=\set{1}$, $\bar{X}=\set{2}$, $Y=X\cup\bar{X}=\set{1,2}$,
the effect of the unitaries on the left-hand side is equivalent to the
elementwise multiplication with the phase factors
\begin{equation*}
\left[\begin{smallmatrix}
 + & + & + & - \\
 + & + & + & - \\
 + & + & + & - \\
 - & - & - & +
\end{smallmatrix}\right]
\end{equation*}
in the standard basis \eqref{eq:EYf}.
(For the matrix elements of $U_{\ord{\xi}}$, see Appendix~\ref{appsec:Parity.Explu}.)
Since $\Idf_X\otimes\f{B}_{\bar{X}}$ is block-diagonal, 
we can see that $\f{B}_{\bar{X}}$ with nonvanishing offdiagonal element provides a good example.
So let $\f{B}_{\set{2}}:=\f{E}\indexddu{\set{2}}{0}{1}:\left[\begin{smallmatrix}0&1\\0&0\end{smallmatrix}\right]$,
then 
\begin{equation*}
\left[\begin{smallmatrix}
 0 & 1 & 0 & 0 \\
 0 & 0 & 0 & 0 \\
 0 & 0 & 0 &-1 \\
 0 & 0 & 0 & 0
\end{smallmatrix}\right]\neq
\left[\begin{smallmatrix}
 0 & 1 & 0 & 0 \\
 0 & 0 & 0 & 0 \\
 0 & 0 & 0 & 1 \\
 0 & 0 & 0 & 0
\end{smallmatrix}\right].
\end{equation*}

Second, we consider fermionic elementary products \eqref{eq:TPfp}.
Again, let $X=\set{1}$, $\bar{X}=\set{2}$ and $Y=X\cup\bar{X}=\set{1,2}$,
and let us have the operators given in the standard basis as
$\f{A}_X:\left[\begin{smallmatrix}1&a\\a^*&1\end{smallmatrix}\right]$ and
$\f{B}_{\bar{X}}:\left[\begin{smallmatrix}1&b\\b^*&1\end{smallmatrix}\right]$.
These are self-adjoint, moreover, positive semidefinite if and only if $\abs{a}=\abs{b}\leq1$.
Then the 
$\iotaf_{\set{1},\set{1,2}}(\f{A}_{\set{1}})$ and
$\iotaf_{\set{2},\set{1,2}}(\f{B}_{\set{2}})$
fermionic canonical embeddings \eqref{eq:cEmbf} are
\begin{equation*}
\f{A}_{\set{1}} \otimesf \Idf_{\set{2}}
: \left[\begin{smallmatrix}
1 & 0 & a & 0 \\
0 & 1 & 0 & a \\
a^*&0 & 1 & 0 \\
0 &a^*& 0 & 1
\end{smallmatrix}\right],\qquad
\Idf_{\set{1}} \otimesf \f{B}_{\set{2}}
: \left[\begin{smallmatrix}
1 & b & 0 & 0 \\
b^*&1 & 0 & 0 \\
0 & 0 & 1 &-b \\
0 & 0 &-b^*&1
\end{smallmatrix}\right],
\end{equation*}
by which
\begin{align*}
\f{A}_{\set{1}} \otimesfp \f{B}_{\set{2}} \equiv
\bigl(\f{A}_{\set{1}} \otimesf \Idf_{\set{2}}\bigr)\bigl(\Idf_{\set{1}} \otimesf \f{B}_{\set{2}}\bigr)
&: \left[\begin{smallmatrix}
 1   & b   & a   &-ab  \\
 b^* & 1   &-ab^*& a   \\
 a^* & a^*b& 1   &-b   \\
 a^*b^*&a^*&-b^* & 1   
\end{smallmatrix}\right],\\
\f{B}_{\set{2}} \otimesfp \f{A}_{\set{1}} \equiv
\bigl(\Idf_{\set{1}} \otimesf \f{B}_{\set{2}}\bigr)\bigl(\f{A}_{\set{1}} \otimesf \Idf_{\set{2}}\bigr)
&: \left[\begin{smallmatrix}
 1   & b   & a   & ab  \\
 b^* & 1   & ab^*& a   \\
 a^* &-a^*b& 1   &-b   \\
-a^*b^*&a^*&-b^* & 1   
\end{smallmatrix}\right].
\end{align*}
It can clearly be seen that 
these are not self-adjoint (therefore cannot be positive), 
if and only if $a\neq0$ and $b\neq0$,
providing example for the violation of the second part of \eqref{eq:allevenPHom} and of \eqref{eq:allevenPPos} without superselection.
On the other hand, setting
$\f{A}_{\set{2}} = \Idf_{\set{2}}$ and
$\f{B}_{\set{1}} = \Idf_{\set{1}}$,
we have 
\begin{equation*}
\bigl(\f{B}_{\set{1}}\otimesfp\f{B}_{\set{2}}\bigr)
\bigl(\f{A}_{\set{1}}\otimesfp\f{A}_{\set{2}}\bigr)
\neq \bigl(\f{B}_{\set{1}}\f{A}_{\set{1}}\bigr)\otimesfp\bigl(\f{B}_{\set{2}}\f{A}_{\set{2}}\bigr),
\end{equation*}
since, using \eqref{eq:ffpiotaf},
\begin{equation*}
\bigl(\Idf_{\set{1}} \otimesf \f{B}_{\set{2}}\bigr)\bigl(\f{A}_{\set{1}} \otimesf \Idf_{\set{2}}\bigr)
\equiv \f{B}_{\set{2}}\otimesfp\f{A}_{\set{1}}
\neq\f{A}_{\set{1}}\otimesfp\f{B}_{\set{2}},
\end{equation*}
providing example for the violation of the first part of \eqref{eq:allevenPHom} without superselection.
(We note that, because of \eqref{eq:ffpiotaf} and \eqref{eq:specialsingle}, 
these provide examples also for the violation of the similar identities with $\otimesf$ instead of $\otimesfp$,
see the end of Section~\ref{sec:Tensors.TP}.
The differences between the two kinds of product is apparent for less simple partitions.)
On the other hand, $\frac12\f{A}_{\set{1}}$ and $\frac12\f{B}_{\set{2}}$ are rank-1 projectors, representing pure states,
if and only if $\abs{a}=\abs{b}=1$,
the products of their fermionic canonical embeddings are not self-adjoint either, therefore not rank-1 projectors,
providing example for the violation of the \eqref{eq:allevenPpures} without superselection.
The matrices of 
$\f{B}_{\set{2}}\otimesfp\f{A}_{\set{1}}$ and
$\f{A}_{\set{1}}\otimesfp\f{B}_{\set{2}}$ above also illustrate that
neither of these are Kronecker products of the matrices of $\f{A}_{\set{1}}$ and $\f{B}_{\set{2}}$
if $a\neq0$ and $b\neq0$, that is, without superselection.

\newpage
\bibliographystyle{unsrtnat}
\bibliography{fermiphase}

\end{document}